\documentclass[journal]{IEEEtran}

\usepackage{cite}

\ifCLASSINFOpdf
  \usepackage[pdftex]{graphicx}

\else

\usepackage[dvips]{graphicx}

\fi

\usepackage{multirow}
\usepackage{xcolor}
\usepackage{pgf,tikz}
\usetikzlibrary{arrows}

\usepackage{amsmath, amsfonts}

\newtheorem{theorem}{Theorem}
\newtheorem{proposition}{Proposition}

\usepackage{array}

\usepackage{fixltx2e}

\usepackage{url}

\usepackage[normalem]{ulem}
\usepackage[utf8]{inputenc} 
\usepackage[T1]{fontenc}
\begin{document}

\title{The Covering-Assignment Problem for Swarm-powered Ad-hoc Clouds: A Distributed 3D Mapping Use-case}

%

\author{Leandro~R.~Costa,
        Daniel~Aloise,~\IEEEmembership{Member,~IEEE,}
        Luca~G.~Gianoli,~\IEEEmembership{Member,~IEEE,} 
        and~Andrea~Lodi. 
\thanks{Digital Object Identifier: 10.1109/JIOT.2020.3039261}
\thanks{
This work was financial supported by the Natural Sciences and Engineering Research Council of Canada through the Canada Excellence Research Chair in data science for real-time decision-making, and by Prompt. The computing resources were made available by Compute Canada. 
}
\thanks{Leandro R. Costa and Andrea Lodi are with the 
				Department of Mathematical and Industrial Engineering,
				Polytechnique Montréal, Montreal, Canada.}
\thanks{Daniel Aloise is with the 
				Computer and Software Engineering Department,
				Polytechnique Montréal, Montreal, Canada, 
				(e-mail: daniel.aloise@polymtl.ca)}
\thanks{Luca G. Gianoli is with the Humanitas Solutions, Montreal, Canada}
\thanks{2327-4662~\copyright~2020 IEEE.  Personal use of this material is permitted.  Permission from IEEE must be obtained for all other uses, in any current or future media, including reprinting/republishing this material for advertising or promotional purposes, creating new collective works, for resale or redistribution to servers or lists, or reuse of any copyrighted component of this work in other works.}
}

\markboth{IEEE Internet of Things Journal~2020}%
{The Covering-Assignment Problem for Swarm-powered Ad-hoc Clouds: A Distributed 3D Mapping Use-case for IEEE Internet of Things Journal}

%

\IEEEpubid{2327-4662~\copyright~2020 IEEE. }


\maketitle

\begin{abstract}
The popularity of drones is rapidly increasing across the different sectors of the economy. 
Aerial capabilities and relatively low costs make drones the perfect solution to improve the efficiency of operations that are typically carried out by humans.
Besides automating field operations, drones acting de facto as a swarm can serve as an ad-hoc cloud infrastructure built on top of computing and storage resources available across the swarm members and other elements. 
Even in the absence of Internet connectivity, this cloud can serve the workloads generated by the swarm members and the field agents. 
By considering the practical example of a swarm-powered 3D reconstruction application on top of such cloud infrastructure, we present a new optimization problem for the efficient generation and execution of multi-node computing workloads subject to data geolocation and clustering constraints.
The objective is the minimization of the overall computing times, including both networking delays caused by the inter-drone data transmission and computation delays. 
We prove that the problem is NP-hard and present two combinatorial formulations to model it.
Computational results on the solution of the formulations show that one of them can be used to solve, within the configured time-limit, more than 50\% of the considered real-world instances involving up to two hundred images and six drones.
\end{abstract}


%
\IEEEpeerreviewmaketitle

\section{Introduction}
\label{sec:introduction}

\IEEEPARstart{A}n Unmanned Aerial Vehicle (UAV) --- otherwise commonly known as drone --- is a flying vehicle whose weight can vary, according to the targeted applications, from a few hundreds grams to hundreds of kilos.  
The popularity of drones is rapidly increasing across the different sectors of the economy. 
Aerial capabilities and relatively low CAPEX/OPEX costs make UAVs the perfect solution to improve the efficiency of those operations that are typically carried out by humans, e.g., building inspection, photo collection, area surveillance, etc.

In normal operations, drones are remotely controlled by human pilots through wireless remote controls. 
However, by setting the UAV autopilot in \emph{auto off-board} mode, a drone can operate in a fully autonomous manner by following the inputs generated by an on-board flight computer directly connected to the autopilot. 
This capability can be leveraged to create fleets of autonomous UAVs collaborating to fulfill the desired missions. 
This is achieved by installing a collaborative drone application on each on-board flight computer of the fleet and by connecting these latter on the same wireless network.

\IEEEpubidadjcol

The possibility of organizing drones in fleets of autonomous and collaborating entities naturally attracted the attention of swarm robotics scientists \cite{tan2013research}.
Swarm robotics studies how to reproduce, with the help of artificial agents, those swarming behaviors typically observed in nature  --- in ant colonies, bee swarms, bird flocks, etc. 
Swarm behaviors have the potential to revolutionize the world of robotized applications --- including UAV applications --- because of their promise of jointly achieving maximum performance and maximum resilience through the power of distributed interactions that do not require any form of centralized supervision.

Swarming UAVs can be deployed to support operations in a long list of domains~\cite{nex2014uav}, including forestry \cite{berni2009thermal, grenzdorffer2008photogrammetric, martinez2006experimental}, archaeology and architecture \cite{chiabrando2011uav, lambers2007combining, oczipka2009small, rinaudo2012archaeological, verhoeven2009providing}, environment monitoring \cite{hartmann2012determination, manyoky2011unmanned, niethammer2010uav, smith2009volcano, zhang2008uav}, emergency management \cite{chou2010disaster, haarbrink2006helicopter, molina2012searching} and precision agriculture \cite{zarco_tejada2014, bendig2014, diaz_varela2015}.

Accordingly, the operations research community has been investigating approaches to improve the efficiency of UAV-powered applications~\cite{alena2018optm_uav, coutinho2018}.
In particular, decentralized optimization methods have fostered search problems \cite{Sujit2004search, oh2014road_search, oh2015road_relay, lanillos2014, gan2011multi, ji2015cooperative}, target assignment problems \cite{Tang2005MMMT, KARAMAN2008CG, niccolini2010multiple, viguria2010distributed, choi2011genetic, barrientos2011aerial, Moon2013, moon2015, turpin2014capt,enright2015handbook, sadeghi2017heterogeneous}, node covering problems \cite{ZORBAS201616, Ladosz2018}, scheduling problems \cite{CARABALLO2017725}, and other cases \cite{grancharova2015uavs, koulali2016green, xu2017distributed}.

In a complex mission, the UAV swarming component is typically dedicated to data collection duties, e.g., taking pictures, producing video feeds, sniffing wireless signals. 
Other technologies are then involved in processing the collected data and producing the desired output. 
For instance, digital photogrammetry algorithms \cite{Gomarasca2009, Linder2016}, can be leveraged to elaborate the images collected by the swarming UAVs, by extracting and displaying the relevant 2D/3D geometric information.

The data-processing phase --- 3D processing phase in the photogrammetry use-case \cite{doherty2016collaborative, loianno2016swarm, schmiemann2017distributed, kobayashi2017method, golodetz2018collaborative, milani2016impact}  --- is typically executed in the cloud and in a centralized fashion~\cite{zanella2014, abbas2018, chen2019}. 
However, to mitigate Internet connectivity and network latency issues, the distributed power of the UAV swarm can be leveraged to establish an ad-hoc cloud infrastructure~\cite{yu2020, asheralieva2019, mei2020, wei2019, wan2020}.
In fact, such infrastructure can be exploited to execute the data processing phases of the considered mission \cite{chen2019,hamdaqa2015,malhotra2014,yaqoob2016, hu2019, zhang2019}.
This approach aims to exploit the \emph{power of the many} --- many embedded microcomputers of limited power are installed on the swarming UAVs --- to replace the computer power typically available in a powerful work station reserved in the cloud.

This paper addresses the problem of optimizing the exploitation of a swarm-powered ad-hoc cloud, by jointly dealing with two interrelated aspects of the data-processing stage:
\begin{IEEEitemize}
    \item the workload generation, i.e., definition of the computing application elements and of the corresponding set of data input.
    \item the workload scheduling/assignment, i.e., mapping of computing application elements and physical swarm members.
\end{IEEEitemize}
In particular, we consider the generation and placement of workloads whose input data are subject to geolocation/clustering constraints 
(e.g.~\cite{tan_qu2020, yu_wang2019}).

Practically speaking, the collected data are bound to a location and must be processed in batch of neighboring samples: in the 3D reconstruction use-case, this means that groups of neighboring pictures have to be processed by the same computing element. 
This additional constraint has a non-negligible impact when dealing with a swarm powered ad-hoc cloud: due to the distributed nature of the swarm-powered data collection stage, the whole input data-set may end up being completely scattered across the UAVs of the swarm. 
If not properly dealt with during the workload generation/scheduling processes, this aspect may severely deteriorate the whole process performance due to unnecessary data transmission delays (input data must be received by the corresponding computing application element).

For the purpose of illustrating the applicability of our approach to a real-life application, we adjust the proposed solution to a relevant use-case scenario from the emergency response field. 
Our use-case is a perfect example of a real-life application subject to geolocation constraints that highly benefits from swarm-powered ad-hoc cloud infrastructure~\cite{Humanitas}.
In fact,  the drone swarm is able to perform 3D reconstruction of a region of interest --- comprising five hundred photos --- on top of twenty Raspberry Pis~\cite{RaspberryPi} microcomputers~\cite{luca2020}.

In such context, the 3D map of a given region of interest is used to improve the decision making process and the operator situational awareness through the availability of a 3D digital twin of the operation area, where the elements of interest can be even selectively highlighted, e.g., building or road damages, risky areas, etc.
Producing the relevant 3D maps in a timely manner (near real-time), even when the cloud connectivity is not available, is crucial to increase the chances of success of an operation.
To this purpose, we introduce a new optimization problem, namely the Covering-Assignment Problem for swarm-powered ad-hoc clouds (CAPsac). 
Given a set of geo-positioned aerial pictures (data) that are physically distributed across a set of UAVs (stored on the embedded microcomputers on the drones), CAPsac minimizes the 3D mapping (data-processing phase) completion time by jointly computing: 
\begin{IEEEitemize}
    \item the optimal workload configuration/the optimal covering of photos, i.e., splitting the overall photographed region across multiple convex sub-regions, and
    \item the optimal workload scheduling/the optimal assignment of photographed sub-regions to UAVs, i.e., deciding which drone (its embedded microcomputer) is responsible for the 3D reconstruction of a photographed sub-region. 
\end{IEEEitemize}
It is worth pointing out that, differently from the \textit{decompose-then-allocate} and the \textit{allocate-then-decompose} paradigms \cite{Khamis2015} broadly adopted in (both the cloud computing optimization and) the multi-robot task allocation literature, CAPsac is an integrated decision model that handles workload generation (photo covering or sub-region splitting) and workload assignment (sub-region to UAV assignment) at the same time. 
Other works employ similar principle by jointly optimizing UAVs deployment and task/position allocation~\cite{huang2020, wang2020}. However, they propose evolutionary heuristics which, by definition, cannot certify the optimality of the obtained solutions.

The remainder of the paper is organized as follows.
The next section formally introduces the CAPsac problem, by clearly highlighting how the general problem designed for swarm-powered ad-hoc clouds is naturally applied to optimize the execution of a distributed 3D mapping application. 
Once the relationship between the general problem and the specific 3D mapping use-case will have been clearly proved through Section \ref{sec:proposed_problem}, it will be possible to present the remainder of the paper by directly referring to the latter. 
We believe that this editing approach will allow the reader to better grasp the details and the added value of the proposed solution. 
Two mathematical programming formulations to solve the CAPsac problem are described in Section~\ref{sec:problem_formulation} while Section~\ref{sec:np_hardness_proof} presents the NP-hardness proof for the problem. 
Finally, Section~\ref{sec:experiments} presents and discusses the computational results obtained by experimenting with realistic 3D reconstruction instances, while Section~\ref{sec:conclusions} summarizes our concluding remarks.

\section{Covering-Assignment Problem for Swarm-powered Ad-hoc Clouds - CAPsac}
\label{sec:proposed_problem}
A swarm-powered mission can be typically decomposed in two phases:
\begin{IEEEenumerate}[\IEEEsetlabelwidth{ii}]
    \item[i] \textbf{Data collection}: the UAVs of the swarm dynamically collaborate to collect all the necessary information within the area of interest. 
    In a swarm-powered 3D mapping mission, this phase corresponds to the photo-collection process meant to shoot the required aerial photos of the selected area. 
    Note that the set of required pictures is typically computed by a dedicated mapping software and is merely an input of the mapping mission.
    
    \item[ii] \textbf{Data processing}: the collected data are collaboratively processed by the ad-hoc cloud built on top of the microcomputers installed on the UAVs to produce the desired output.
    During this process, thanks to the swarm-powered ad-hoc cloud, the computing workload can be parallelized over the available computing units. 
    Furthermore, the collected data can be transferred over the inter-drone wireless network to satisfy the input requirements of the distributed processing tasks.
    In a swarm-powered 3D mapping mission, this phase corresponds to the 3D-processing process meant to compute a 3D point cloud and/or a 3D mesh of the selected area.
\end{IEEEenumerate}
The proposed CAPsac problem deals with the optimization of the data (3D) processing phase and has no direct control on the data (photo) collection strategy. 
Given the set of data (aerial pictures) just collected by the UAVs, CAPsac aims at minimizing the overall processing time required to compute the desired output (3D map).

\begin{figure}[!t]
    \centering
\begin{tikzpicture}[line cap=round,line join=round,>=triangle 45,x=0.6cm,y=0.6cm]
\clip(0,-1.04) rectangle (12,8);
\draw node at (6.5,3.5) {\includegraphics[scale=0.135]{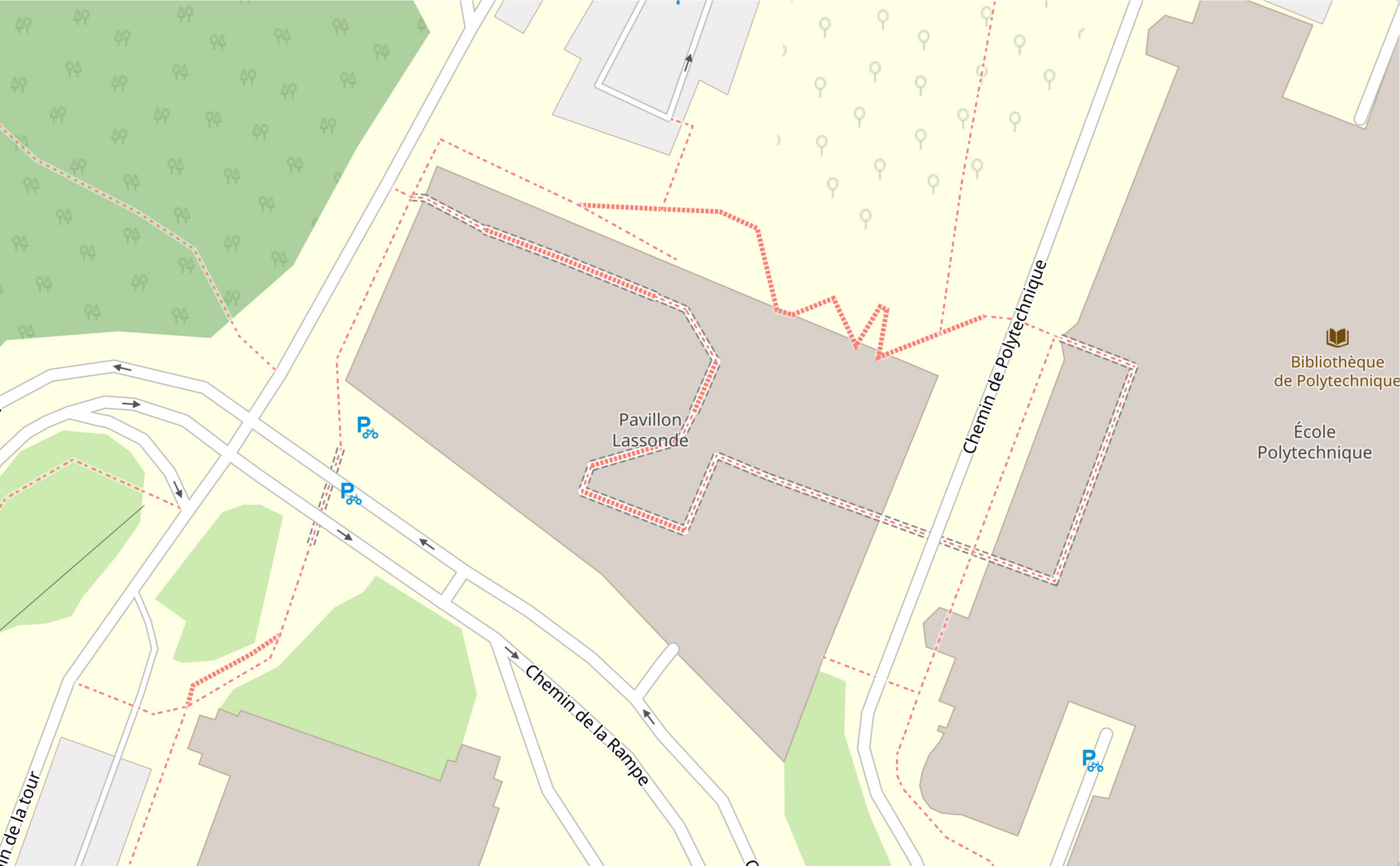} };
\draw (2.82,5.92)-- (2.58,1.22);
\draw (2.58,1.22)-- (3.75,0.76);
\draw (3.75,0.76)-- (9.2,0.93);
\draw (9.2,0.93)-- (9.12,5.74);
\draw (9.12,5.74)-- (2.82,5.92);
\draw [line width=1.2pt,loosely dotted] (3.88,1.83)-- (6.23,3.76);
\draw [line width=1.2pt,loosely dotted] (6.23,3.76)-- (8.08,1.69);
\draw [line width=1.2pt,loosely dotted] (6.23,3.76)-- (6.38,5.08);
\begin{scriptsize}
\draw [color=black] (3,3.98)-- ++(-2.5pt,0 pt) -- ++(5.0pt,0 pt) ++(-2.5pt,-2.5pt) -- ++(0 pt,5.0pt);
\draw [color=black] (5,3.98) circle (2.5pt);
\draw [color=black] (3,2.48)-- ++(-2.5pt,0 pt) -- ++(5.0pt,0 pt) ++(-2.5pt,-2.5pt) -- ++(0 pt,5.0pt);
\draw [color=black] (5,2.48) circle (2.5pt);
\draw [color=black] (7,2.48)-- ++(-2.5pt,-2.5pt) -- ++(5.0pt,5.0pt) ++(-5.0pt,0) -- ++(5.0pt,-5.0pt);
\draw [color=black] (7,3.98)-- ++(-2.5pt,-2.5pt) -- ++(5.0pt,5.0pt) ++(-5.0pt,0) -- ++(5.0pt,-5.0pt);
\draw [color=black] (9,3.98) ++(-2.5pt,0 pt) -- ++(2.5pt,2.5pt)--++(2.5pt,-2.5pt)--++(-2.5pt,-2.5pt)--++(-2.5pt,2.5pt);
\draw [color=black] (9,2.48) ++(-2.5pt,0 pt) -- ++(2.5pt,2.5pt)--++(2.5pt,-2.5pt)--++(-2.5pt,-2.5pt)--++(-2.5pt,2.5pt);
\draw [color=black] (3,1.14)-- ++(-2.5pt,0 pt) -- ++(5.0pt,0 pt) ++(-2.5pt,-2.5pt) -- ++(0 pt,5.0pt);
\draw [color=black] (5,1.16) circle (2.5pt);
\draw [color=black] (7,1.14)-- ++(-2.5pt,-2.5pt) -- ++(5.0pt,5.0pt) ++(-5.0pt,0) -- ++(5.0pt,-5.0pt);
\draw [color=black] (9,1.14) ++(-2.5pt,0 pt) -- ++(2.5pt,2.5pt)--++(2.5pt,-2.5pt)--++(-2.5pt,-2.5pt)--++(-2.5pt,2.5pt);
\draw [color=black] (3,5.58)-- ++(-2.5pt,0 pt) -- ++(5.0pt,0 pt) ++(-2.5pt,-2.5pt) -- ++(0 pt,5.0pt);
\draw [color=black] (5,5.58)-- ++(-2.5pt,0 pt) -- ++(5.0pt,0 pt) ++(-2.5pt,-2.5pt) -- ++(0 pt,5.0pt);
\draw [color=black] (7,5.58) ++(-2.5pt,0 pt) -- ++(2.5pt,2.5pt)--++(2.5pt,-2.5pt)--++(-2.5pt,-2.5pt)--++(-2.5pt,2.5pt);
\draw [color=black] (9,5.58)-- ++(-2.5pt,-2.5pt) -- ++(5.0pt,5.0pt) ++(-5.0pt,0) -- ++(5.0pt,-5.0pt);
\draw [color=black, line width=1.2pt] (3.88,1.83)-- ++(-4.5pt,0 pt) -- ++(9.0pt,0 pt) ++(-4.5pt,-4.5pt) -- ++(0 pt,9.0pt);
\draw [color=black, line width=1.2pt] (6.38,5.08) ++(-4.5pt,0 pt) -- ++(4.5pt,4.5pt)--++(4.5pt,-4.5pt)--++(-4.5pt,-4.5pt)--++(-4.5pt,4.5pt);
\draw [color=black, line width=1.2pt] (8.08,1.69)-- ++(-4.5pt,-4.5pt) -- ++(9.0pt,9.0pt) ++(-9.0pt,0) -- ++(9.0pt,-9.0pt);
\draw [color=black, line width=1.2pt] (6.23,3.76) circle (4.5pt);

\end{scriptsize}
\end{tikzpicture}
\begin{tikzpicture}[line cap=round,line join=round,>=triangle 45,x=0.6cm,y=0.6cm]
\begin{scriptsize}
\matrix [draw,above right, fill=white] at (current bounding box.south west) {
	\draw [color=black, line width=1.2pt] (0, 0.15)-- ++(-2.5pt,0 pt) -- ++(5.0pt,0 pt) ++(-2.5pt,-2.5pt) -- ++(0 pt,5.0pt);		
	\draw [color=black, line width=1.2pt] (0.5, 0.15)-- ++(-2.5pt,-2.5pt) -- ++(5.0pt,5.0pt) ++(-5.0pt,0) -- ++(5.0pt,-5.0pt); 
	\draw node at (0.4, 0)[label=right:3D-capable drones] {}; 
	
	\draw [color=black, line width=1.2pt] (4.5, 0.15) circle (2.5pt); 	
	\draw [color=black, line width=1.2pt] (5, 0.15) ++(-2.5pt,0 pt) -- ++(2.5pt,2.5pt)--++(2.5pt,-2.5pt)--++(-2.5pt,-2.5pt)--++(-2.5pt,2.5pt); 
	\draw node at (4.9, 0)[label=right:Ordinary drones] {}; 
	
	\draw [color=black, line width=1.2pt,loosely dotted] (8.6, 0.15) -- (9.5, 0.15); 
	\draw node at (9.1, 0)[label=right:Network] {}; \\

	\draw [color=black] (0, 0.15)-- ++(-2.5pt,0 pt) -- ++(5.0pt,0 pt) ++(-2.5pt,-2.5pt) -- ++(0 pt,5.0pt);	
	\draw [color=black] (0.4, 0.15) circle (2.5pt); 	
	\draw [color=black] (0.8, 0.15)-- ++(-2.5pt,-2.5pt) -- ++(5.0pt,5.0pt) ++(-5.0pt,0) -- ++(5.0pt,-5.0pt); 
	\draw [color=black] (1.2, 0.15) ++(-2.5pt,0 pt) -- ++(2.5pt,2.5pt)--++(2.5pt,-2.5pt)--++(-2.5pt,-2.5pt)--++(-2.5pt,2.5pt); 
	\draw node at (1, 0)[label=right:Photos(shot by respective larger symbols) ] {};
	
	\draw [color=black] (8.6, 0.15) -- (9.4, 0.15) ; 
	\draw node at (9.05, 0)[label=right: Area selec.] {}; \\
};
\end{scriptsize}
\end{tikzpicture}
    \caption{Explanatory instance of CAPsac problem accordingly to the 3D mapping use-case.}
   	\label{fig:intro_entry}
\end{figure}
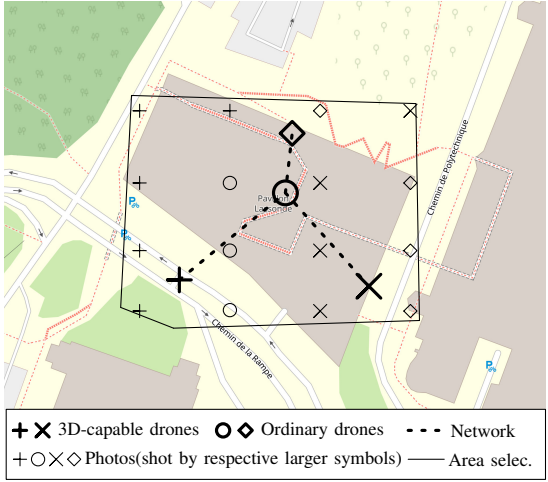

An explanatory CAPsac problem instance involving a swarm of four drones performing a 3D mapping mission is represented in Fig. \ref{fig:intro_entry}. 
The full lines delimit the area of interest represented by the set $P$ of aerial pictures just captured by the four drones during the photo collection phase.
Each photo $p \in P$ was taken by a specific drone (which also stores it in memory). Furthermore, each picture must be processed during the 3D processing phase to guarantee a proper reconstruction. 

In Fig.~\ref{fig:intro_entry}, the positions of the four drones are represented by the large symbols “$\times$”, “$+$”, “$\diamond$”, and “$\circ$”. 
Not all drones may be equipped with microcomputers powerful enough to support the 3D processing workload. 
In the example of Fig.~\ref{fig:intro_entry}, only two drones are considered \emph{3D-capable}, those represented by the $+$ and the $\times$ symbols. 

Each photo is characterized by its shooting location, which is denoted by the small versions of the symbols previously used to represent the UAVs: 
the pictures represented by a small $+$ were shot by the drone represented by the large $+$, and so on. 
Note that, given the dynamic nature of the decentralized decision-making process employed by the swarm of drones \cite{Khamis2015}, it is impossible to know a-priori which UAV will shoot which picture. 

A solution of the CAPsac problem describes how to: 
\begin{IEEEitemize}
    \item Split the processing workload into multiple processing (application) components, each responsible for dealing with a specific subset of the collected data, e.g., of the aerial pictures. 
    Note that in the 3D reconstruction use-case each 3D reconstruction sub-task corresponds to a specific sub-region and requires as input all the aerial pictures that belong to that sub-region.
    
    \item Assign each processing component and all its corresponding input data to at least one of the computing elements available within the swarm-powered ad-hoc cloud, i.e., the microcomputers installed on the swarming-UAVs or on any other ground element connected to the swarm itself.
\end{IEEEitemize}

The optimal solution of CAPsac minimizes the latest processing time among all the involved computing elements, which corresponds to minimizing the makespan of the whole 3D reconstruction process.
Three main issues cannot be ignored when assigning the photos (and thus the sub-regions) to the optimal 3D processing drones.
Note that for sake of simplicity, we consider the case of not more than one computing device available on each drone. 

\begin{IEEEenumerate}[\IEEEsetlabelwidth{iii}]
 
    \item[i] A feasible region (workload) subdivision is characterized by the creation of a \emph{spatial-convex covering}: 
    the union of the sub-regions corresponds to the whole region and the photos associated to each sub-region must be a \emph{spatial-convex set}.
    Accordingly, a photo can be assigned to a drone if and only if it lies inside the convex hull of all the photos assigned to that drone.
    Fig.~\ref{fig:realistic_bad_set} illustrates a set of photos which is \textbf{not} spatial-convex. 
    The assigned photos are represented by colored “$\bullet$” symbols. The “$\circ$” symbols represent photos that do not belong to the set, which are hence assigned to other drones.
    Spatial-convex sets are crucial to perform the 3D mapping procedure since the presence of non-overlapping photo footprints (represented by the dashed colored rectangles in Fig. \ref{fig:realistic_bad_set}) makes the 3D reconstruction of the associated region impossible. 
    Fig.~\ref{fig:realistic_good_set} shows an example of a photo spatial-convex set assigned to one 3D-capable drone.
    It is worth pointing out that the creation of a spatial-convex covering is required by any workload operating over geo-located/clustered data to be processed in neighboring batches.
      
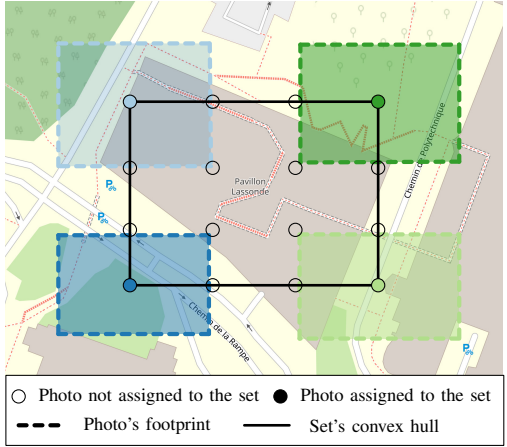
\begin{figure}[!t]
    \centering
\definecolor{esrzrz}{rgb}{0.89,0.1,0.1}
\definecolor{fzzbzz}{rgb}{0.98,0.61,0.6}
\definecolor{afddyz}{rgb}{0.69,0.87,0.54}
\definecolor{rexxbs}{rgb}{0.12,0.47,0.7}
\definecolor{ttzesb}{rgb}{0.2,0.62,0.17}
\definecolor{avcces}{rgb}{0.65,0.8,0.89}
\begin{tikzpicture}[line cap=round,line join=round,>=triangle 45,x=0.55cm,y=0.55cm]
\clip(0,-1.04) rectangle (12,8);
\draw node at (6.5,3.5) {\includegraphics[scale=0.135]{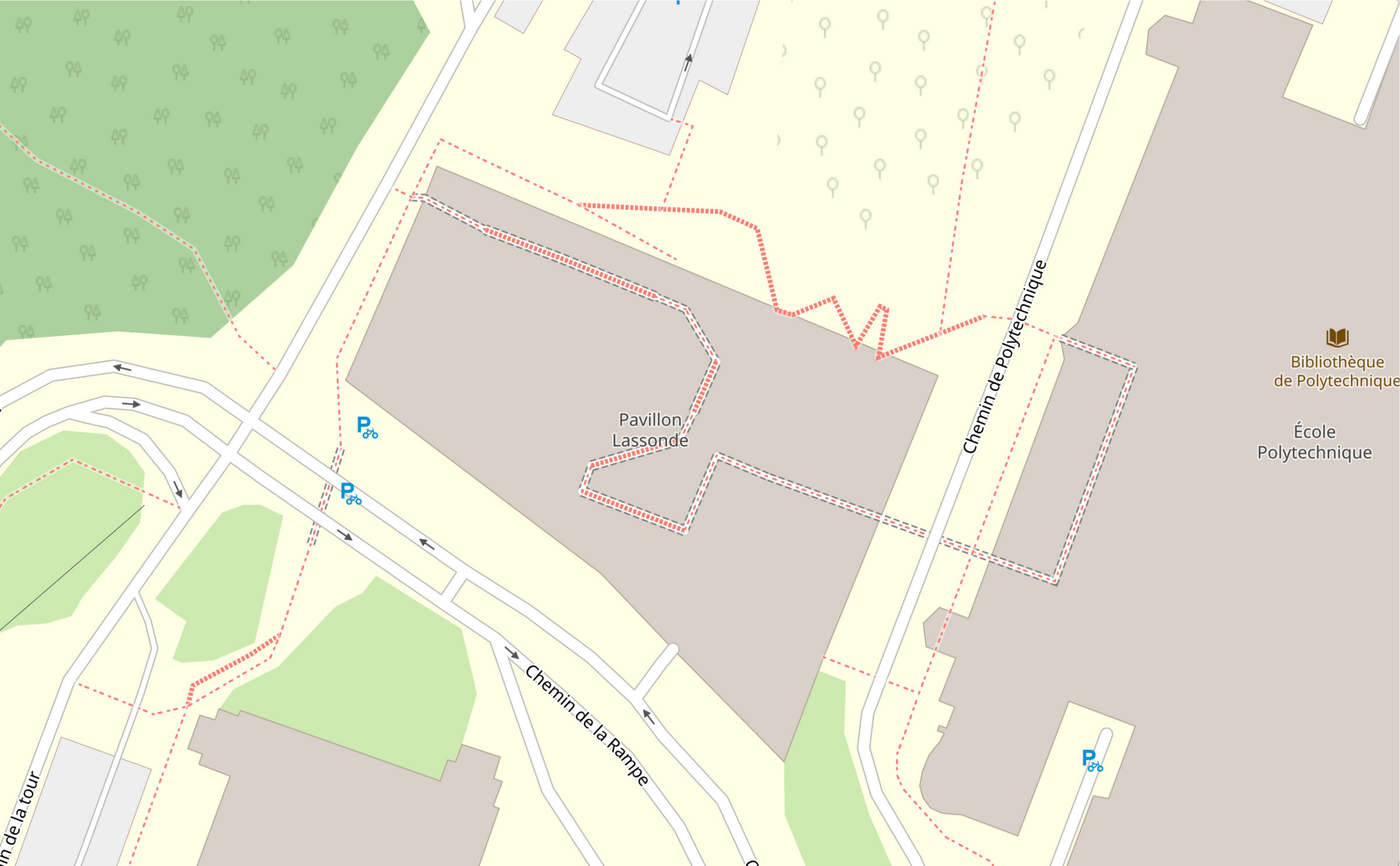}};
\fill[line width=2.8pt,color=rexxbs,fill=rexxbs,fill opacity=0.2] (1.27,4.02) -- (1.27,6.99) -- (4.93,6.99) -- (4.96,4.02) -- cycle;
\fill[line width=2.8pt,color=ttzesb,fill=ttzesb,fill opacity=0.5] (7.12,4.11) -- (7.12,6.96) -- (10.96,6.96) -- (10.96,4.13) -- cycle;
\fill[line width=2.8pt,color=afddyz,fill=afddyz,fill opacity=0.5] (7.1,-0.15) -- (7.1,2.37) -- (10.96,2.37) -- (10.96,-0.15) -- cycle;
\fill[line width=2.8pt,color=rexxbs,fill=rexxbs,fill opacity=0.5] (1.27,-0.04) -- (1.27,2.35) -- (4.91,2.35) -- (4.91,-0.04) -- cycle;
\draw [line width=1.6pt,color=avcces, dashed] (1.27,4.02)-- (1.27,6.99);
\draw [line width=1.6pt,color=avcces, dashed] (1.27,6.99)-- (4.93,6.99);
\draw [line width=1.6pt,color=avcces, dashed] (4.93,6.99)-- (4.96,4.02);
\draw [line width=1.6pt,color=avcces, dashed] (4.96,4.02)-- (1.27,4.02);
\draw [line width=1.6pt,color=ttzesb, dashed] (7.12,4.11)-- (7.12,6.96);
\draw [line width=1.6pt,color=ttzesb, dashed] (7.12,6.96)-- (10.96,6.96);
\draw [line width=1.6pt,color=ttzesb, dashed] (10.96,6.96)-- (10.96,4.13);
\draw [line width=1.6pt,color=ttzesb, dashed] (10.96,4.13)-- (7.12,4.11);
\draw [line width=1.6pt,color=afddyz, dashed] (7.1,-0.15)-- (7.1,2.37);
\draw [line width=1.6pt,color=afddyz, dashed] (7.1,2.37)-- (10.96,2.37);
\draw [line width=1.6pt,color=afddyz, dashed] (10.96,2.37)-- (10.96,-0.15);
\draw [line width=1.6pt,color=afddyz, dashed] (10.96,-0.15)-- (7.1,-0.15);
\draw [line width=1.6pt,color=rexxbs, dashed] (1.27,-0.04)-- (1.27,2.35);
\draw [line width=1.6pt,color=rexxbs, dashed] (1.27,2.35)-- (4.91,2.35);
\draw [line width=1.6pt,color=rexxbs, dashed] (4.91,2.35)-- (4.91,-0.04);
\draw [line width=1.6pt,color=rexxbs, dashed] (4.91,-0.04)-- (1.27,-0.04);
\draw [line width=1.0pt] (3,1.14)-- (9,1.14);
\draw [line width=1.0pt] (9,1.14)-- (9,5.58);
\draw [line width=1.0pt] (3,5.58)-- (9,5.58);
\draw [line width=1.0pt] (3,5.58)-- (3,1.14);
\begin{scriptsize}
\draw [color=black] (3,3.98) circle (2.5pt);
\draw [color=black] (5,3.98) circle (2.5pt);
\draw [color=black] (3,2.48) circle (2.5pt);
\draw [color=black] (5,2.48) circle (2.5pt);
\draw [color=black] (7,2.48) circle (2.5pt);
\draw [color=black] (7,3.98) circle (2.5pt);
\draw [color=black] (9,3.98) circle (2.5pt);
\draw [color=black] (9,2.48) circle (2.5pt);
\fill [color=rexxbs] (3,1.14) circle (2.5pt);
\draw [color=black] (5,1.14) circle (2.5pt);
\draw [color=black] (7,1.14) circle (2.5pt);
\fill [color=afddyz] (9,1.14) circle (2.5pt);
\fill [color=avcces] (3,5.58) circle (2.5pt);
\draw [color=black] (5,5.58) circle (2.5pt);
\draw [color=black] (7,5.58) circle (2.5pt);
\fill [color=ttzesb] (9,5.58) circle (2.5pt);
\draw [color=black] (3,1.14) circle (2.5pt);
\draw [color=black] (9,1.14) circle (2.5pt);
\draw [color=black] (3,5.58) circle (2.5pt);
\draw [color=black] (9,5.58) circle (2.5pt);
\end{scriptsize}
\end{tikzpicture} 
\begin{tikzpicture}[line cap=round,line join=round,>=triangle 45,x=0.6cm,y=0.6cm]
\begin{scriptsize}
\matrix [draw,above right, fill=white] at (current bounding box.south west) {
	\draw [color=black] (0,0.15) circle (2.5pt);
	\draw node at (0.0, 0)[label=right:Photo not assigned to the set] {};

	\draw [color=black, fill=black] (5.8,0.15) circle (2.5pt);
	\draw node at (5.8, 0)[label=right:Photo assigned to the set] {};\\
	
	\draw [line width=1.6pt, dashed] (0.0, 0.15)-- (1.0, 0.15);
	\draw node at (1.0, 0)[label=right:Photo's footprint] {};

	\draw [line width=1.0pt] (5.0, 0.15)-- (6.0, 0.15);
	\draw node at (6.0, 0)[label=right:Set's convex hull] {};\\
};
\end{scriptsize}
\end{tikzpicture}
	\caption{Ordinary set and the respective convex hull.}
	\label{fig:realistic_bad_set}              
\end{figure}    

\begin{figure}[!t]
    \centering
\definecolor{esrzrz}{rgb}{0.89,0.1,0.1}
\definecolor{fzzbzz}{rgb}{0.98,0.61,0.6}
\definecolor{afddyz}{rgb}{0.69,0.87,0.54}
\definecolor{rexxbs}{rgb}{0.12,0.47,0.7}
\definecolor{ttzesb}{rgb}{0.2,0.62,0.17}
\definecolor{avcces}{rgb}{0.65,0.8,0.89}
\centering
\begin{tikzpicture}[line cap=round,line join=round,>=triangle 45,x=0.55cm,y=0.55cm]
\clip(0,-1.04) rectangle (12,8);
\draw node at (6.5,3.5) {\includegraphics[scale=0.135]{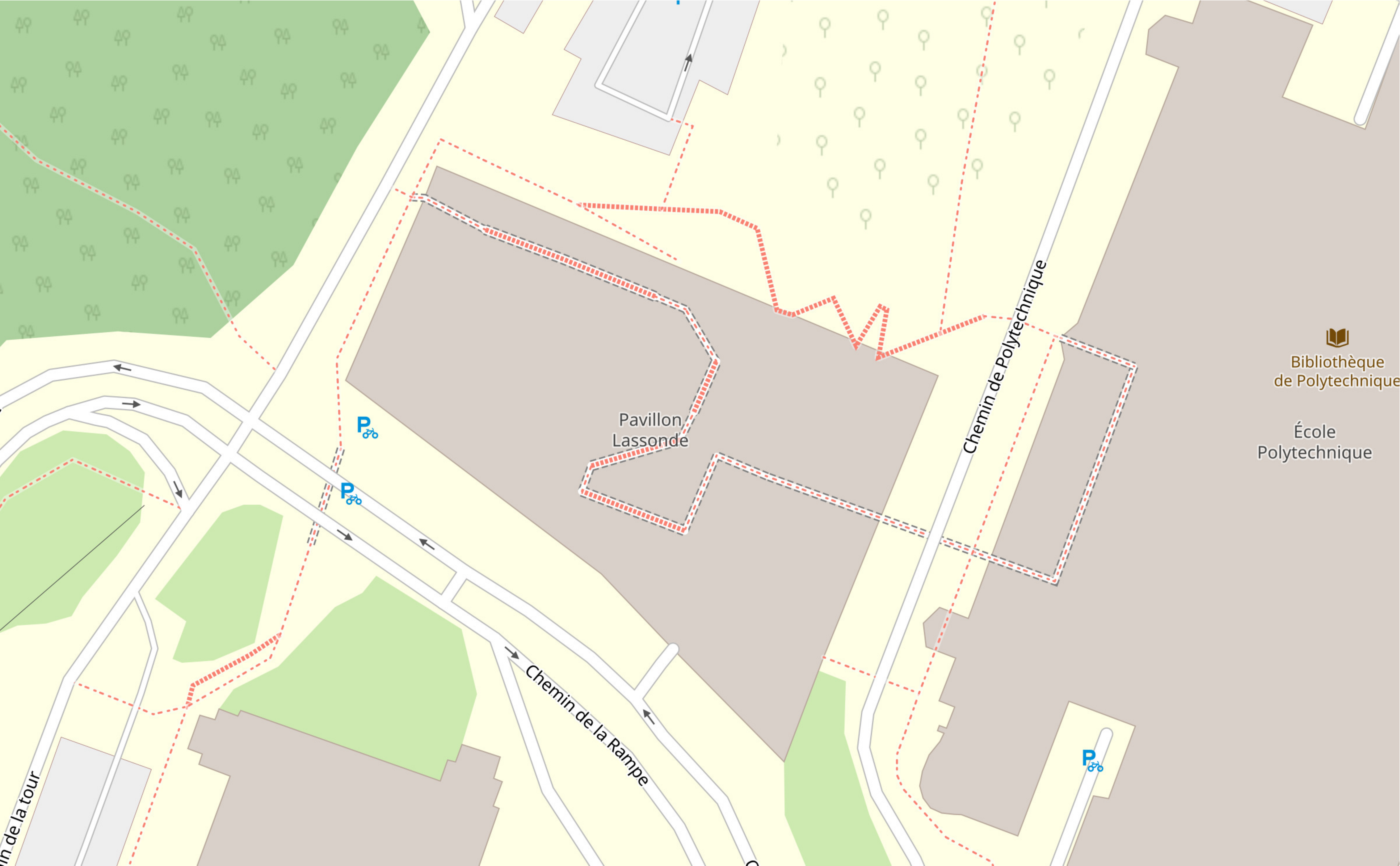}};
\fill[line width=2.8pt,color=rexxbs,fill=rexxbs,fill opacity=0.2] (0.86,2.55) -- (0.87,5.55) -- (4.95,5.55) -- (4.93,2.48) -- cycle;
\fill[line width=2.8pt,color=ttzesb,fill=ttzesb,fill opacity=0.5] (3.05,2.61) -- (3.1,5.44) -- (6.99,5.43) -- (6.97,2.59) -- cycle;
\fill[line width=2.8pt,color=afddyz,fill=afddyz,fill opacity=0.5] (3.11,1.26) -- (3.14,3.95) -- (6.88,3.92) -- (6.88,1.21) -- cycle;
\fill[line width=2.8pt,color=rexxbs,fill=rexxbs,fill opacity=0.5] (1,1.36) -- (1.01,3.81) -- (4.79,3.79) -- (4.76,1.35) -- cycle;
\fill[line width=2.8pt,color=fzzbzz,fill=fzzbzz,fill opacity=0.2] (0.92,4.1) -- (0.93,7.11) -- (5.01,7.11) -- (5,4.04) -- cycle;
\fill[line width=2.8pt,color=esrzrz,fill=esrzrz,fill opacity=0.2] (2.96,4.13) -- (3,6.97) -- (6.89,6.95) -- (6.88,4.12) -- cycle;
\draw [line width=1.6pt,color=avcces, dashed] (0.86,2.55)-- (0.87,5.55);
\draw [line width=1.6pt,color=avcces, dashed] (0.87,5.55)-- (4.95,5.55);
\draw [line width=1.6pt,color=avcces, dashed] (4.95,5.55)-- (4.93,2.48);
\draw [line width=1.6pt,color=avcces, dashed] (4.93,2.48)-- (0.86,2.55);
\draw [line width=1.6pt,color=ttzesb, dashed] (3.05,2.61)-- (3.1,5.44);
\draw [line width=1.6pt,color=ttzesb, dashed] (3.1,5.44)-- (6.99,5.43);
\draw [line width=1.6pt,color=ttzesb, dashed] (6.99,5.43)-- (6.97,2.59);
\draw [line width=1.6pt,color=ttzesb, dashed] (6.97,2.59)-- (3.05,2.61);
\draw [line width=1.6pt,color=afddyz, dashed] (3.11,1.26)-- (3.14,3.95);
\draw [line width=1.6pt,color=afddyz, dashed] (3.14,3.95)-- (6.88,3.92);
\draw [line width=1.6pt,color=afddyz, dashed] (6.88,3.92)-- (6.88,1.21);
\draw [line width=1.6pt,color=afddyz, dashed] (6.88,1.21)-- (3.11,1.26);
\draw [line width=1.6pt,color=rexxbs, dashed] (1,1.36)-- (1.01,3.81);
\draw [line width=1.6pt,color=rexxbs, dashed] (1.01,3.81)-- (4.79,3.79);
\draw [line width=1.6pt,color=rexxbs, dashed] (4.79,3.79)-- (4.76,1.35);
\draw [line width=1.6pt,color=rexxbs, dashed] (4.76,1.35)-- (1,1.36);
\draw [line width=1.6pt,color=fzzbzz, dashed] (0.92,4.1)-- (0.93,7.11);
\draw [line width=1.6pt,color=fzzbzz, dashed] (0.93,7.11)-- (5.01,7.11);
\draw [line width=1.6pt,color=fzzbzz, dashed] (5.01,7.11)-- (5,4.04);
\draw [line width=1.6pt,color=fzzbzz, dashed] (5,4.04)-- (0.92,4.1);
\draw [line width=1.6pt,color=esrzrz, dashed] (2.96,4.13)-- (3,6.97);
\draw [line width=1.6pt,color=esrzrz, dashed] (3,6.97)-- (6.89,6.95);
\draw [line width=1.6pt,color=esrzrz, dashed] (6.89,6.95)-- (6.88,4.12);
\draw [line width=1.6pt,color=esrzrz, dashed] (6.88,4.12)-- (2.96,4.13);
\draw [line width=1.0pt] (3,5.58)-- (5,5.58);
\draw [line width=1.0pt] (5,5.58)-- (5,3.98);
\draw [line width=1.0pt] (5,3.98)-- (5,2.48);
\draw [line width=1.0pt] (3,5.58)-- (3,3.98);
\draw [line width=1.0pt] (3,3.98)-- (3,2.48);
\draw [line width=1.0pt] (3,2.48)-- (5,2.48);
\begin{scriptsize}
\fill [color=avcces] (3,3.98) circle (2.5pt);
\fill [color=ttzesb] (5,3.98) circle (2.5pt);
\fill [color=rexxbs] (3,2.48) circle (2.5pt);
\fill [color=afddyz] (5,2.48) circle (2.5pt);
\draw [color=black] (7,2.48) circle (2.5pt);
\draw [color=black] (7,3.98) circle (2.5pt);
\draw [color=black] (9,3.98) circle (2.5pt);
\draw [color=black] (9,2.48) circle (2.5pt);
\draw [color=black] (3,1.14) circle (2.5pt);
\draw [color=black] (5,1.16) circle (2.5pt);
\draw [color=black] (7,1.14) circle (2.5pt);
\draw [color=black] (9,1.14) circle (2.5pt);
\fill [color=fzzbzz] (3,5.58) circle (2.5pt);
\fill [color=esrzrz] (5,5.58) circle (2.5pt);
\draw [color=black] (7,5.58) circle (2.5pt);
\draw [color=black] (9,5.58) circle (2.5pt);
\draw [color=black] (3,3.98) circle (2.5pt);
\draw [color=black] (5,3.98) circle (2.5pt);
\draw [color=black] (3,2.48) circle (2.5pt);
\draw [color=black] (5,2.48) circle (2.5pt);
\draw [color=black] (3,5.58) circle (2.5pt);
\draw [color=black] (5,5.58) circle (2.5pt);
\end{scriptsize}
\end{tikzpicture}
\begin{tikzpicture}[line cap=round,line join=round,>=triangle 45,x=0.6cm,y=0.6cm]
\begin{scriptsize}
\matrix [draw,above right, fill=white] at (current bounding box.south west) {
	\draw [color=black] (0,0.15) circle (2.5pt);
	\draw node at (0.0, 0)[label=right:Photo not assigned to the set] {};

	\draw [color=black, fill=black] (5.8,0.15) circle (2.5pt);
	\draw node at (5.8, 0)[label=right:Photo assigned to the set] {};\\
	
	\draw [line width=1.6pt, dashed] (0.0, 0.15)-- (1.0, 0.15);
	\draw node at (1.0, 0)[label=right:Photo's footprint] {};

	\draw [line width=1.0pt] (5.0, 0.15)-- (6.0, 0.15);
	\draw node at (6.0, 0)[label=right:Set's convex hull] {};\\
};
\end{scriptsize}
\end{tikzpicture}
	\caption{Spatial-convex set and the respective convex hull.}
	\label{fig:realistic_good_set} 
\end{figure}
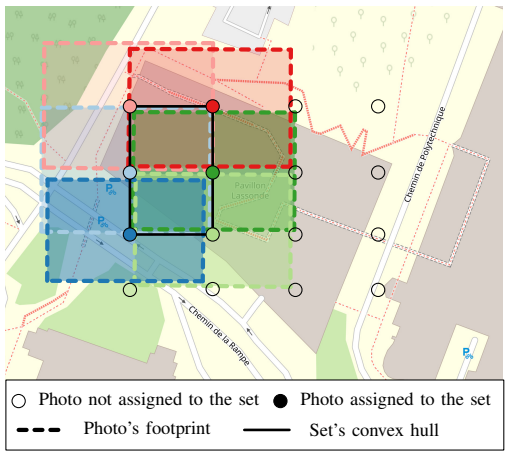    
    
    \item[ii] As the sub-regions (and their corresponding pictures) are assigned to the 3D-capable drones, a drone may need to  require some input pictures (data) from the other swarm members. 
    The CAPsac considers a pre-defined single-tree network topology built by a networking middleware running on the swarming drones \cite{Bekmezci2013}. 
    Note that in Fig.~\ref{fig:intro_result}, the network links are represented by the pointed lines. 
    Besides, in a single-tree network topology, only one routing path exists to connect each pair of UAVs.
    A drone cannot start the 3D reconstruction of the assigned sub-region until all the required photos are received. 
    The TCP communication protocol is widely applied in engineering to achieve reliable transmissions and flow control, and it is adopted to model the swarm communications in the CAPsac problem.
    According to \cite{massoulie1999bandwidth}, a good way to approximate the transmission behavior concerning the TCP protocol is to assume that the transmission rate allocation follows the Max-Min Fairness - MMF paradigm \cite{bertsekas1992data}. Thus, minimizing the makespan of the 3D reconstruction requires that all the transmission rates of the network follow the MMF paradigm. It is well known that, in multi-hop wireless networks, the allocation behavior of TCP may deviate from the ideal MMF paradigm \cite{raniwala2007}. However, when a traffic engineering problem involves elastic traffic demands, i.e., the transmission rate is autonomously determined by a distributed end-to-end congestion control scheme, the choice of considering even just an approximated form of fairness plays a crucial role to avoid those poor routing solutions that would be otherwise obtained by approximating the traffic demands as inelastic \cite{coniglio2020}. Furthermore, note that sources of MMF deviations, such as round-trip-time variance and multi-connection schemes \cite{raniwala2007}, can be explicitly accounted for with the help of simple multiplicative parameters added to a specific group of constraints \cite{coniglio2020}. Another more complex source of  deviation, i.e., the hidden node phenomenon, could be addressed by means of robust optimization techniques \cite{bertsimas2003, delage2018} that fall beyond the scope for this work.

    \item[iii] Finally, a \textit{reliability factor} should be considered to immunize the CAPsac assignment with respect to drone malfunctions. 
    The reliability factor defines the minimum number of drones (computing elements) that should process each sub-region.
\end{IEEEenumerate}   

Fig. \ref{fig:intro_result} shows a feasible solution to the CAPsac problem optimizing the makespan of a 3D mapping mission and considering a reliability factor equal to one.  
The dashed and the dashed-and-pointed lines define a feasible spatial-convex covering. 
The number of sets comprising the covering is equal to the number of 3D-capable drones.
For instance, the covering in Fig. \ref{fig:intro_result} has only two sets. 
The photos lying within the left sub-region are processed by drone $+$, whereas those in the right sub-region are elaborated by drone $\times$. 

\begin{figure}[!t]
    \centering
\begin{tikzpicture}[line cap=round,line join=round,>=triangle 45,x=0.6cm,y=0.6cm]
\clip(0,-1.04) rectangle (12,8);
\draw node at (6.5,3.5) {\includegraphics[scale=0.135]{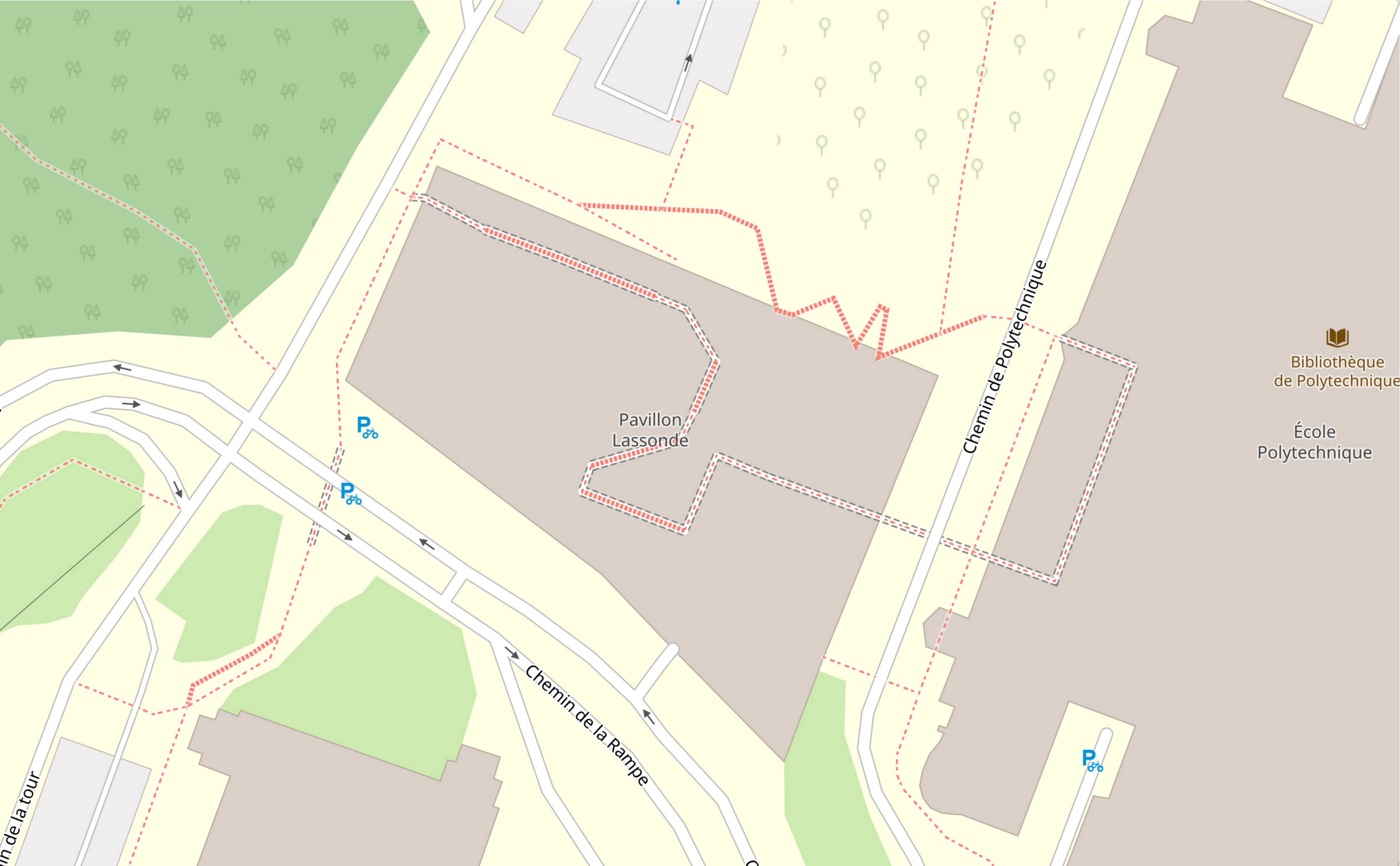}};
\draw (2.82,5.92)-- (2.58,1.22);
\draw (2.58,1.22)-- (3.75,0.76);
\draw (3.75,0.76)-- (9.2,0.93);
\draw (9.2,0.93)-- (9.12,5.74);
\draw (9.12,5.74)-- (2.82,5.92);
\draw [line width=1.2pt,loosely dotted] (3.88,1.83)-- (6.23,3.76);
\draw [line width=1.2pt,loosely dotted] (6.23,3.76)-- (8.08,1.69);
\draw [line width=1.2pt,loosely dotted] (6.23,3.76)-- (6.38,5.08);
\draw [line width=1.4pt, dashed] (5.74,5.75)-- (5.82,0.93);
\draw [line width=1.4pt, dashed] (5.82,0.93)-- (3.77,0.82);
\draw [line width=1.4pt, dashed] (3.77,0.82)-- (2.68,1.29);
\draw [line width=1.4pt, dashed] (2.68,1.29)-- (2.89,5.83);
\draw [line width=1.4pt, dashed] (2.89,5.83)-- (5.74,5.75);
\draw [line width=1.0pt, dashdotted] (5.87,5.74)-- (5.96,0.93);
\draw [line width=1.0pt, dashdotted] (5.96,0.93)-- (9.14,1.01);
\draw [line width=1.0pt, dashdotted] (9.14,1.01)-- (9.06,5.66);
\draw [line width=1.0pt, dashdotted] (9.06,5.66)-- (5.87,5.74);
\begin{scriptsize}
\draw [color=black] (3,3.98)-- ++(-2.5pt,0 pt) -- ++(5.0pt,0 pt) ++(-2.5pt,-2.5pt) -- ++(0 pt,5.0pt);
\draw [color=black] (5,3.98) circle (2.5pt);
\draw [color=black] (3,2.48)-- ++(-2.5pt,0 pt) -- ++(5.0pt,0 pt) ++(-2.5pt,-2.5pt) -- ++(0 pt,5.0pt);
\draw [color=black] (5,2.48) circle (2.5pt);
\draw [color=black] (7,2.48)-- ++(-2.5pt,-2.5pt) -- ++(5.0pt,5.0pt) ++(-5.0pt,0) -- ++(5.0pt,-5.0pt);
\draw [color=black] (7,3.98)-- ++(-2.5pt,-2.5pt) -- ++(5.0pt,5.0pt) ++(-5.0pt,0) -- ++(5.0pt,-5.0pt);
\draw [color=black] (9,3.98) ++(-2.5pt,0 pt) -- ++(2.5pt,2.5pt)--++(2.5pt,-2.5pt)--++(-2.5pt,-2.5pt)--++(-2.5pt,2.5pt);
\draw [color=black] (9,2.48) ++(-2.5pt,0 pt) -- ++(2.5pt,2.5pt)--++(2.5pt,-2.5pt)--++(-2.5pt,-2.5pt)--++(-2.5pt,2.5pt);
\draw [color=black] (3,1.14)-- ++(-2.5pt,0 pt) -- ++(5.0pt,0 pt) ++(-2.5pt,-2.5pt) -- ++(0 pt,5.0pt);
\draw [color=black] (5,1.16) circle (2.5pt);
\draw [color=black] (7,1.14)-- ++(-2.5pt,-2.5pt) -- ++(5.0pt,5.0pt) ++(-5.0pt,0) -- ++(5.0pt,-5.0pt);
\draw [color=black] (9,1.14) ++(-2.5pt,0 pt) -- ++(2.5pt,2.5pt)--++(2.5pt,-2.5pt)--++(-2.5pt,-2.5pt)--++(-2.5pt,2.5pt);
\draw [color=black] (3,5.58)-- ++(-2.5pt,0 pt) -- ++(5.0pt,0 pt) ++(-2.5pt,-2.5pt) -- ++(0 pt,5.0pt);
\draw [color=black] (5,5.58)-- ++(-2.5pt,0 pt) -- ++(5.0pt,0 pt) ++(-2.5pt,-2.5pt) -- ++(0 pt,5.0pt);
\draw [color=black] (7,5.58) ++(-2.5pt,0 pt) -- ++(2.5pt,2.5pt)--++(2.5pt,-2.5pt)--++(-2.5pt,-2.5pt)--++(-2.5pt,2.5pt);
\draw [color=black] (9,5.58)-- ++(-2.5pt,-2.5pt) -- ++(5.0pt,5.0pt) ++(-5.0pt,0) -- ++(5.0pt,-5.0pt);
\draw [color=black, line width=1.2pt] (3.88,1.83)-- ++(-4.5pt,0 pt) -- ++(9.0pt,0 pt) ++(-4.5pt,-4.5pt) -- ++(0 pt,9.0pt);
\draw [color=black, line width=1.2pt] (6.38,5.08) ++(-4.5pt,0 pt) -- ++(4.5pt,4.5pt)--++(4.5pt,-4.5pt)--++(-4.5pt,-4.5pt)--++(-4.5pt,4.5pt);
\draw [color=black, line width=1.2pt] (8.08,1.69)-- ++(-4.5pt,-4.5pt) -- ++(9.0pt,9.0pt) ++(-9.0pt,0) -- ++(9.0pt,-9.0pt);
\draw [color=black, line width=1.2pt] (6.23,3.76) circle (4.5pt);
\end{scriptsize}
\end{tikzpicture}
\begin{tikzpicture}[line cap=round,line join=round,>=triangle 45,x=0.6cm,y=0.6cm]
\begin{scriptsize}
\matrix [draw,above right, fill=white] at (current bounding box.south west) {
	\draw [color=black, line width=1.2pt] (0, 0.15)-- ++(-2.5pt,0 pt) -- ++(5.0pt,0 pt) ++(-2.5pt,-2.5pt) -- ++(0 pt,5.0pt);		
	\draw [color=black, line width=1.2pt] (0.5, 0.15)-- ++(-2.5pt,-2.5pt) -- ++(5.0pt,5.0pt) ++(-5.0pt,0) -- ++(5.0pt,-5.0pt); 
	\draw node at (0.4, 0)[label=right:3D-capable drones] {}; 
	
	\draw [color=black, line width=1.2pt] (4.5, 0.15) circle (2.5pt); 	
	\draw [color=black, line width=1.2pt] (5, 0.15) ++(-2.5pt,0 pt) -- ++(2.5pt,2.5pt)--++(2.5pt,-2.5pt)--++(-2.5pt,-2.5pt)--++(-2.5pt,2.5pt); 
	\draw node at (4.9, 0)[label=right:Ordinary drones] {}; 
	
	\draw [color=black, line width=1.2pt,loosely dotted] (8.6, 0.15) -- (9.5, 0.15); 
	\draw node at (9.1, 0)[label=right:Network] {}; \\

	\draw [color=black] (0, 0.15)-- ++(-2.5pt,0 pt) -- ++(5.0pt,0 pt) ++(-2.5pt,-2.5pt) -- ++(0 pt,5.0pt);	
	\draw [color=black] (0.4, 0.15) circle (2.5pt); 	
	\draw [color=black] (0.8, 0.15)-- ++(-2.5pt,-2.5pt) -- ++(5.0pt,5.0pt) ++(-5.0pt,0) -- ++(5.0pt,-5.0pt); 
	\draw [color=black] (1.2, 0.15) ++(-2.5pt,0 pt) -- ++(2.5pt,2.5pt)--++(2.5pt,-2.5pt)--++(-2.5pt,-2.5pt)--++(-2.5pt,2.5pt); 
	\draw node at (1, 0)[label=right:Photos(shot by respective larger symbols) ] {};
	
	\draw [color=black] (8.6, 0.15) -- (9.4, 0.15) ; 
	\draw node at (9.05, 0)[label=right: Area selec.] {}; \\
	
	\draw [line width=1.4pt, dashed] (0,0.15)-- (0.8,0.15);
	\draw node at (0.6, 0)[label=right:Sub-region assigned to“] {};
	\draw [color=black, line width=1.2pt] (5.25, 0.15)-- ++(-2.5pt,0 pt) -- ++(5.0pt,0 pt) ++(-2.5pt,-2.5pt) -- ++(0 pt,5.0pt);
	\draw node at (5.05, 0.05)[label=right:"] {};
	
	\draw [line width=1.0pt, dashdotted] (5.8,0.15)-- (6.6,0.15);
	\draw node at (6.35, 0)[label=right:Sub-region assigned to“] {};
	\draw [color=black, line width=1.2pt] (11.0, 0.15)-- ++(-2.5pt,-2.5pt) -- ++(5.0pt,5.0pt) ++(-5.0pt,0) -- ++(5.0pt,-5.0pt);
	\draw node at (10.8, 0.05)[label=right:"] {}; \\
};
\end{scriptsize}
\end{tikzpicture}
   	\caption{Spatial-convex covering and its assignment optimizing the makespan of a 3D mapping mission.}
   	\label{fig:intro_result}
\end{figure}
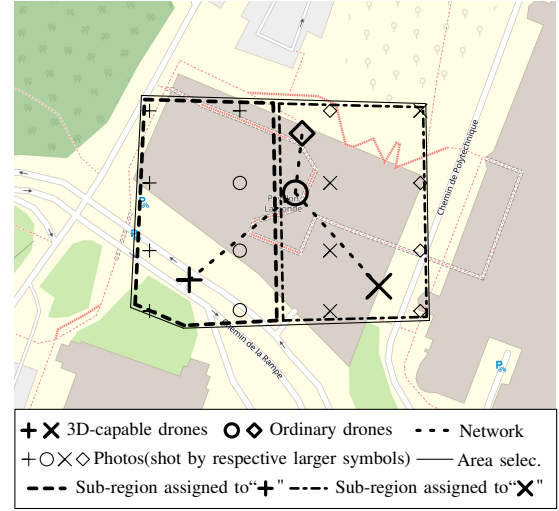

Given that the relationship between the general CAPsac and the swarm-powered 3D mission use-case has been clearly established, we will present the remainder of the paper referring directly to the specific use-case.
Thus, it will be possible to the reader to better grasp the details and the added value of our proposed optimization problem.

\section{Mathematical programming formulations}
\label{sec:problem_formulation}

\subsection{Common definitions}

Let us consider a swarm of drones $D$ of different types which is responsible for the 3D reconstruction of a region described by the set $P$ of photos. 
Let $\bar{D} \subseteq D$, with $|\bar{D}| = m$, be the set of 3D-capable drones that have enough computing power to support the 3D reconstruction workloads. 
The location of all photos $p \in P$ are also known. 

Given a photo $p \in P$, let $\lambda_{p}$ and $\mu_{p}$ be the non-negative real parameters representing, respectively, the estimated processing time of $p$ and the storage space occupied by $p$ (expressed in megabytes). 
Also, for each drone $d \in D$, let $\theta_{dp}$ be the binary parameter equal to 1 if photo $p \in P$ is stored on drone $d$. 
The subset of photos processed by a drone directly corresponds to a specific sub-region. 
Therefore, note that the number of sub-regions are hence equal to the number of 3D-capable drones available in the swarm. 

Some pictures may have to be transferred among different drones to respect the computed sub-region assignment configuration.
The picture transmission is supported by an undirected transmission tree $T {=} (N,A)$, where the nodes of set $N$ correspond to the swarming drones and the arcs of set $A$ represent the device-to-device communication links (e.g., Wi-Fi links) between the drone themselves. 
Furthermore, let $F$ be the set of traffic demands defined for each pair of drones $(h,d) \in D \times \bar{D}$, where  $f^{hd}$ denotes the demand (possibly null) between drones $h$ and $d$. 
Also for each $(i,j) \in A$, denote $c_{ij}$ the transmission capacity of the link $(i,j)$.

Finally, the maximum allowed time for transmitting the traffic demands through the network is denoted by $\hat{T}$.

\begin{table}[!t]
\renewcommand{\arraystretch}{1.0}
	\caption{CAPsac parameters in the context of the 3D mapping use-case.}
	\label{tab:parameters}
	\centering
	\begin{tabular}{c | l }
	\hline
	Parameters & Description \\
	\hline	
	$\lambda_{p}$ & estimated processing time of a photo $p$\\
	$\mu_{p}$ & amount of data of a photo $p$ in Mb \\
	$\theta_{dp}$ & equal to 1 if drone $d$ has the photo $p$ stored in its memory \\
	$F$ & set of traffic demands between each pair of drones \\
	$c_{ij}$ & transmission capacity of the link $(i,j) \in A$ \\
	$\bar{c}^{hd}$ & minimum $c_{ij}$ on the sole routing path of $f^{hd}$ \\
	$\sigma$ & reliability factor \\
	$m$ & number of drones (equiv. number of sub-regions) which \\ & can perform 3D reconstruction\\
	$\hat{T}$ & maximum allowed time for exchanging photos between \\ & drones \\
	\hline
	\end{tabular}
\end{table}

With the support of the notation just introduced --- grouped in Table \ref{tab:parameters}, we present two different Mixed-Integer Linear Programming (MILP) formulations to optimize the 3D-processing phase of 3D mapping missions with UAV swarms:

\begin{itemize}
    \item The Photo-based CAPsac (pCAPsac), where each picture $p \in P$ is assigned to one sub-region, among a pre-defined set of initially empty  sub-regions.
	\item The Region-based CAPsac (rCAPsac), where all the feasible rectangular sub-regions are given; the formulation is responsible for selecting the optimal set of sub-regions among those available (presented at the Appendix).
\end{itemize}

\subsection{Photo-based CAPsac}
\label{subsec:pr_formulation} 

In the pCAPsac formulation, the decision variables will be optimized to compose $R = \{1,\ldots,m\}$ sub-regions (equiv. subsets of photos) to be reconstructed by the set of drones.
The formulation aims to jointly perform two assignment operations:

\begin{IEEEitemize}
    \item each picture $p \in P$ is assigned to one sub-region $r \in R$,
    \item each non-empty sub-region $r \in R$ is assigned to one 3D-capable drone $d \in \bar{D}$.
\end{IEEEitemize}

To this purpose, let $y^{r}_{p}$ and $x^{r}_{d}$ be the binary variables equal to 1 when, respectively, photo $p \in P$ is assigned to sub-region $r \in R$, and sub-region $r \in R$ is assigned to drone $d \in \bar{D}$. 
Furthermore, let $g^{r}_{dp}$ be the binary variables equal to 1 if drone $d \in \bar{D}$ is assigned to a sub-region $r \in R$ that contains picture $p \in P$.

\subsubsection{Assignment constraints}
\label{subsec:assignment_constraints}

To obtain a proper 3D-reconstruction, each photo must be processed at least one time, i.e., it must belong to at least one sub-region:
\begin{equation}
	\sum_{r \in R} y^{r}_{p} \geq 1 \quad \forall p \in P. \label{eq:nlm_1}
\end{equation}

\noindent Similarly, each sub-region must be assigned to at least $\sigma$ 3D-capable drones, with $\sigma$ representing the previously introduced reliability factor meant to immunize the system toward possible drone failures:  
\begin{equation}
	\sum_{d \in \bar{D}} x^{r}_{d} \geq \sigma \quad \forall r \in R. \label{eq:nlm_2}
\end{equation}
\noindent 

Finally, let us introduce the group of constraints necessary to correctly compute the $g$ variables without introducing any non-linearity:
\begin{align}
	&\quad g^{r}_{dp} \leq x^{r}_{d} &\forall p \in P, \forall r \in R, \forall d \in \bar{D} \label{eq:nlm_4} \\ 
	&\quad g^{r}_{dp} \leq y^{r}_{p} &\forall p \in P, \forall r \in R, \forall d \in \bar{D} \label{eq:nlm_5} \\	
	&\quad g^{r}_{dp} \geq y^{r}_{p} + x^{r}_{d} - 1 &\forall p \in P, \forall r \in R, \forall d \in \bar{D}. \label{eq:nlm_6} 
\end{align}
\noindent That is, constraints (\ref{eq:nlm_4})-(\ref{eq:nlm_6}) are the classical McCormick inequalities \cite{mccormick1976} such that $g^{r}_{dp} = x^{r}_{d} y^{r}_{p} \text{ } \forall p \in P, \forall r \in R, \forall d \in \bar{D}$ are represented in linear form.

\subsubsection{Spatial-convexity constraints}
\label{subsec:convex_constraints}

To properly work, state-of-the-art 3D reconstruction algorithms~\cite{tang2020} have to deal with convex regions and/or sub-regions, which is also equivalent to work with spatial-convex sets of photos.
To this purpose, we approximate the convex hull of the set of photos assigned to a drone by its smallest enclosing hyperrectangle. 
This is not a bad approximation since 3D mapping missions often considers 50\% to 80\% of photo overlapping~\cite{pepe2018planning}.

Since the GPS position of each photo shooting point is known, let $C$ be the set of distinct picture longitudes and let $L$ be the set of  distinct photo latitudes. 
Note that the following relations are always respected: $1 \leq |C| \leq |P|$ and $1 \leq |L| \leq |P|$.
For each sub-region, there exists a finite set of photo latitudes and longitudes that represents the bounding rectangular box, i.e. the approximated boundaries of the sub-region.

The boundary of a sub-region $r$ is defined by its left ($\alpha^r$), right ($\beta^r$), bottom ($\gamma^r$), and top ($\omega^r$) borders.
Binary variables $\alpha^{r}_{c}$, $\beta^{r}_{c}$, $\gamma^{r}_{\ell}$, and $\omega^{r}_{\ell}$ are used to designate the latitudes and the longitudes defining these borders:

\begin{IEEEitemize}
    \item Binary variable $\alpha^{r}_{c}$ is equal to one if longitude $c \in C$ delimits the left border of sub-region $r \in R$,
    \item Binary variable $\beta^{r}_{c}$ is equal to one if longitude $c \in C$ delimits the right border of sub-region $r \in R$,
    \item Binary variable $\gamma^{r}_{\ell}$ is equal to one if latitude $\ell \in L$ delimits the bottom (inferior) border of sub-region $r \in R$,
    \item Binary variable $\omega^{r}_{\ell}$ is equal to one if latitude $\ell \in L$ delimits the top (superior) border of sub-region $r \in R$.
\end{IEEEitemize}

Each sub-region $r\in R$ must be associated to a unique tuple of borders:
\begin{align}
	\sum_{c \in C} \alpha^{r}_{c} = 1 & & \forall r \in R \label{eq:boundary1} \\	
	\sum_{c \in C} \beta^{r}_{c} = 1 & & \forall r \in R \label{eq:boundary2} \\
	\sum_{\ell \in L} \gamma^{r}_{l} = 1 & & \forall r \in R \label{eq:boundary3} \\
	\sum_{\ell \in L} \omega^{r}_{l} = 1 & & \forall r \in R. \label{eq:boundary4} 		
\end{align}

To respect the sub-region convexity, a photo $p \in P$ can be assigned to sub-region $r \in R$ if and only if $p$ is contained within the boundary defined for $r$.
Geometrically, such constraint is fulfilled when (i) $lng_{\alpha^r} \leq lng_{p} \leq lng_{\beta^r}$ and (ii) $lat_{\gamma^r} \leq lat_{p} \leq lat_{\omega^r}$, where the $lat$ stands for the latitude and $lng$ for longitude. 
To capture this geometrical pattern, for each photo $p$, the sets $\mathcal{L}^{p}_{\alpha}$, $\mathcal{L}^{p}_{\beta}$, $\mathcal{L}^{p}_{\gamma}$, $\mathcal{L}^{p}_{\omega}$ are defined as:
 \begin{IEEEitemize}
   \item $\mathcal{L}^{p}_{\alpha} = \{c \in C| lng_c \leq lng_p \}$, i.e., $\mathcal{L}^{p}_{\alpha}$ contains the longitudes on the left of $lng_p$,
   \item $\mathcal{L}^{p}_{\beta} = \{c \in C| lng_c \geq lng_p \}$, i.e., $\mathcal{L}^{p}_{\beta}$ contains the longitudes on the right of $lng_p$,
   \item $\mathcal{L}^{p}_{\gamma} = \{\ell \in L| lat_\ell \leq lat_p \}$, i.e., $\mathcal{L}^{p}_{\gamma}$ contains the latitudes  below  $lat_p$,
   \item $\mathcal{L}^{p}_{\omega} = \{\ell \in L| lat_\ell \geq lat_p \}$, i.e., $\mathcal{L}^{p}_{\omega}$ contains the latitudes  above  $lat_p$.
 \end{IEEEitemize}
Finally, the sub-region convexity is modeled by the \textit{Boundary Constraints} - $BC_{1}$, expressed as:
\begin{align}
    (BC_{1}^{\alpha}) & & y^{r}_{p} \leq \sum_{c \in \mathcal{L}^{p}_\alpha} \alpha^{r}_{c} & & \forall p \in P, \forall r \in R \label{eq:sbc_y_one_alpha} \\	
    (BC_{1}^\beta) & & y^{r}_{p} \leq \sum_{c \in \mathcal{L}^{p}_\beta} \beta^{r}_{c} & & \forall p \in P, \forall r \in R \label{eq:sbc_y_one_beta} \\
    (BC_{1}^\gamma) & & y^{r}_{p} \leq \sum_{\ell \in \mathcal{L}^{p}_\gamma} \gamma^{r}_{\ell} & & \forall p \in P, \forall r \in R \label{eq:sbc_y_one_gamma} \\
    (BC_{1}^\omega) & & y^{r}_{p} \leq \sum_{\ell \in \mathcal{L}^{p}_\omega} \omega^{r}_{\ell} & & \forall p \in P, \forall r \in R \label{eq:sbc_y_one_omega}.
\end{align}
Constraints (\ref{eq:sbc_y_one_alpha}) restrict the longitudes which can compose the left border $\alpha^{r}$ to the left of the photo $p$'s longitude. 
Similarly, Constraints (\ref{eq:sbc_y_one_beta})-(\ref{eq:sbc_y_one_omega}) impose restrictions on the right (longitude), the bottom (latitude) and the top (latitude) borders, respectively. 

However, a photo $p$ is not assigned to sub-region $r$ if and only if it lies outside the boundary of $r$, i.e., if $lng_{p} < lng_{\alpha^r}$, or $lng_{p} > lng_{\beta^r}$, or $lat_{p} < lat_{\gamma^r}$, or $lat_{p} > lat_{\omega^r}$, which may be expressed by either 
\begin{align}
& (\overline{BC}_0) \nonumber\\
& \sum_{c \in C - \mathcal{L}^{p}_\alpha} \alpha^{r}_{c} {+} \sum_{c \in C - \mathcal{L}^{p}_\beta} \beta^{r}_{c} 
{+} \sum_{\ell \in L - \mathcal{L}^{p}_\gamma} \gamma^{r}_{\ell} {+} \sum_{\ell \in L - \mathcal{L}^{p}_\omega} \omega^{r}_{\ell} \geq 1 - y_{p}^{r} \nonumber\\
& \text{\hspace{6.5cm}} \forall p \in P, \forall r \in R \label{eq:sbc_zero_1}
\end{align}
or
\begin{align}
& (BC_0) \nonumber\\
& \sum_{c \in \mathcal{L}^{p}_\alpha} \alpha^{r}_{c} + \sum_{c \in \mathcal{L}^{p}_\beta} \beta^{r}_{c} + \sum_{\ell \in \mathcal{L}^{p}_\gamma} \gamma^{r}_{\ell} + \sum_{\ell \in \mathcal{L}^{p}_\omega} \omega^{r}_{\ell} \leq 3 + y_{p}^{r} \nonumber\\
& \text{\hspace{6.4cm}}  \forall p \in P, \forall r \in R. \label{eq:sbc_zero_2}
\end{align}

\begin{figure}[!tb]
    \centering   
	\includegraphics[width=3.2in]{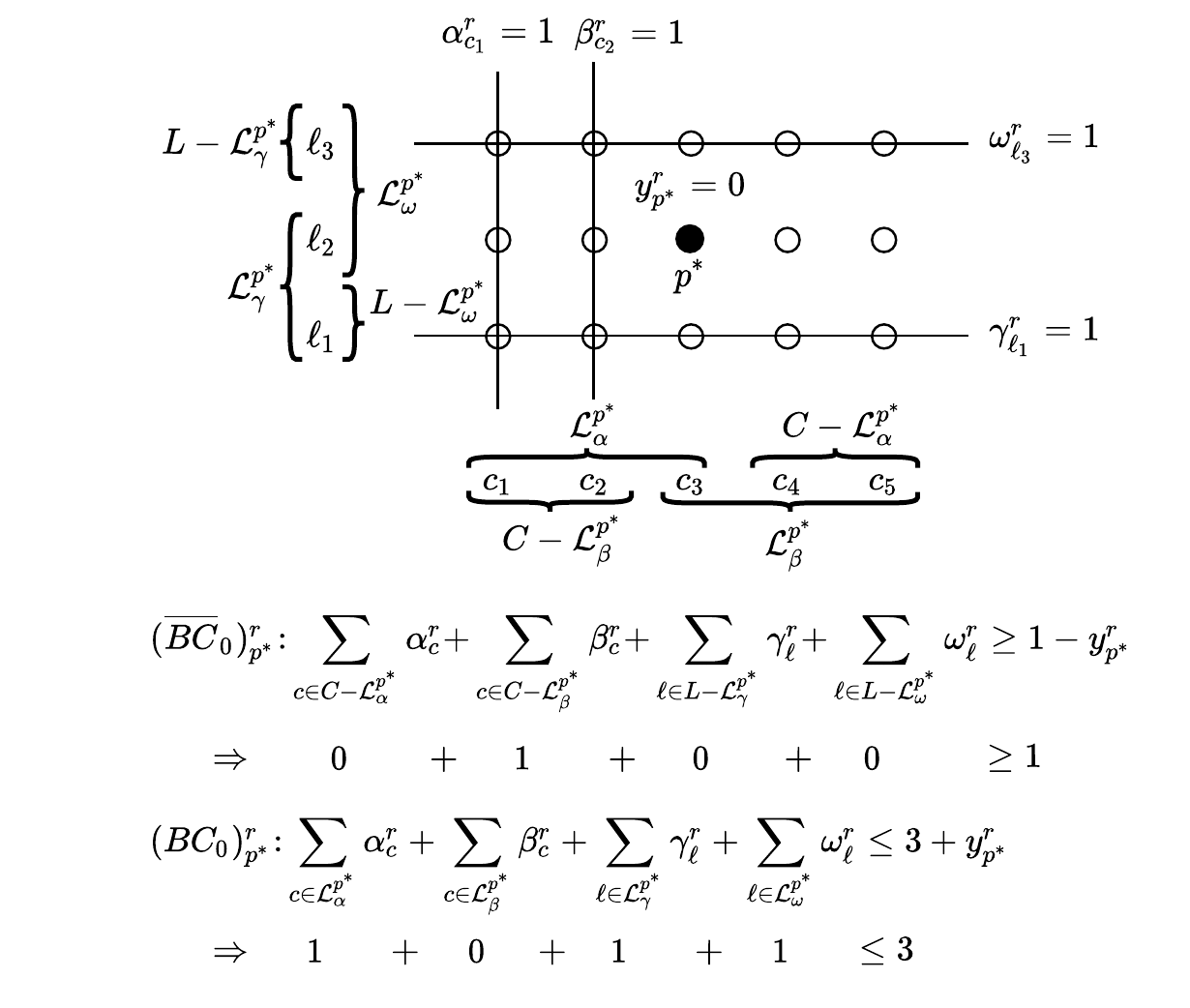}
	\caption{Behaviour of constraints (\ref{eq:sbc_zero_1}) and (\ref{eq:sbc_zero_2}) when the photo $p_8$ is \textbf{not} assigned to a sub-region $r$.}
	\label{fig:ex_spatial_const_NOT_assigned}
\end{figure}

Given constraints (\ref{eq:boundary1})-(\ref{eq:boundary4}), constraints (\ref{eq:sbc_zero_1}) guarantee that at least one of left-side summations is equal to one when $p$ is not assigned to the sub-region $r$ (i.e., $y_{p}^{r} = 0$). Consequently, at least one boundary of $r$ makes the photo $p$ to lie outside $r$. 
In a complimentary way, constraints (\ref{eq:sbc_zero_2}) force that at most three boundaries are satisfied when the photo $p$ is not assigned to the sub-region $r$. 
The Fig.~\ref{fig:ex_spatial_const_NOT_assigned} mathematically illustrates the behaviour of constraints (\ref{eq:sbc_zero_1}) and (\ref{eq:sbc_zero_2}) concerning a photo $p^{\ast}$ (represented by “$\bullet$” ) and a sub-region $r$ when the photo $p^{\ast}$ is \textbf{not} assigned to $r$.

Moreover, let us define valid inequalities (namely Ordering inequalities) to preemptively remove the infeasible boundaries in the continuous space for any possible sub-region. 
For instance, a boundary is infeasible if the  right border is placed on the left side of the left border. Such boundaries are removed through the following set of ordering inequalities:
\begin{align}
	&\alpha^{r}_{c} \leq \sum_{j \in C: lng_j > lng_c } \beta^{r}_{j} &\forall c \in C, \forall r \in R \label{eq:cut_1} \\
	&\beta^{r}_{c} \leq \sum_{j \in C: lng_j < lng_c } \alpha^{r}_{j} &\forall c \in C, \forall r \in R \label{eq:cut_2} \\
	&\gamma^{r}_{\ell} \leq \sum_{j \in L: lat_j > lat_\ell } \omega^{r}_{j} &\forall \ell \in L, \forall r \in R \label{eq:cut_3} \\
	&\omega^{r}_{\ell} \leq \sum_{j \in L: lat_j < lat_\ell } \gamma^{r}_{j} &\forall \ell \in L, \forall r \in R \label{eq:cut_4}
\end{align}

\subsubsection{Photo transmission constraints}

For the purpose of minimizing the 3D processing computation time, it cannot be ignored that an additional delay is introduced any time a picture is transmitted by the drone where it is currently stored, to the drone that is responsible for reconstructing the corresponding sub-region. 
Given a demand $f^{hd} \in F$, let $\bar{c}^{hd}$ represent the minimum link capacity on the sole routing path of $f^{hd}$ and let $z^{hd}$ be the binary variable equal to 1 if traffic demand $f^{hd} > 0$ is active, i.e., 
if at least one picture has to be transferred from drone $h \in D$ to drone $d \in \bar{D}$. 
In this case, non-negative real variables $\phi^{hd}$ are used to represent the transmission rate achieved by traffic demand $f^{hd}$ on its routing path.

Given the demands $f^{hd} \in F$, the following constraints are introduced to correctly activate binary variables $z^{hd}$:

\begin{equation}
z^{hd} \leq \sum_{r \in R, p \in P} g^{r}_{dp} \theta_{hp} \quad \forall (h,d) \in D \times \bar{D}. \label{eq:nlm_7}
\end{equation}
That is, a demand $f^{hd}$ is active if there exists a photo exchange between drones $h$ and $d$.
Besides, the flow variables $\phi$ are forced to 0 when the corresponding traffic demands are idle, or forced to the upper bound on the transmission rate otherwise. This is represented by constraints:
\begin{equation}
	\phi^{hd} \leq \bar{c}^{hd} z^{hd} \quad \forall (h,d) \in D \times \bar{D}. \label{eq:nlm_8}
\end{equation}

As mentioned in Section \ref{sec:proposed_problem}, the transmission times of the traffic demands are computed by considering the MMF paradigm for computing the traffic demand transmission rates. 
A flow (transmission rate) allocation vector is MMF if and only if there is at least one bottleneck link $(i,j) \in A$ on the routing path $V^{hd}$ of each active traffic demand $f^{hd} \in F$ \cite{nace2008max}.
In this work, we adopt the constraints proposed in~\cite{amaldi2013network,coniglio2020} to impose the MMF paradigm for all the swarm communications (i.e., the photo transmissions) in pCAPsac. 
For the sake of simplicity, the multiplicative parameters introduced in ~\cite{amaldi2013network,coniglio2020} to cope with MMF deviations were all considered equal to 1.

In pCAPsac formulation, a maximum networking/transmission latency of $\hat{T}$ seconds is imposed for each activated traffic demand: 
\begin{equation}
    \hat{T} \cdot \phi^{hd} \geq \sum_{r \in R, p \in P} g^{r}_{dp} \theta_{hp} \mu_{p} \quad \forall (h,d) \in D \times \bar{D}. \label{eq:nlm_14}
\end{equation}

\noindent The summation term $\sum_{r \in R, p \in P} g^{r}_{dp} \theta_{hp} \mu_{p}$ computes the overall amount of data to be transferred from drone $h \in D$ to the drone $d \in \bar{D}$.  
Limiting the transmission times means to ensure every drone receives all the photos belonging to the assigned sub-region within a maximum prefixed time set by the domain expert. 
This constraint can be relaxed by setting $\hat{T}$ to a suitable very high value.

\subsubsection{Symmetry breaking constraints}

Formulation pCAPsac suffers from symmetry in both photo-to-sub-region (i.e., $y^{r}_{p}$) and sub-region-to-drone (i.e., $x^{r}_{d}$) assignments.
It is possible to partially break the symmetry of the sub-region-to-drone assignments. Each sub-region can be assigned to one distinct drone in advance.
That is, $m$ variables $x^{r}_{d}$ are fixed where the fixed pairs $\{(r_1,d_1),\ldots,(r_m,d_m)\} \in \{R \times \bar{D} \}$ have distinct indexes. Considering Fig. \ref{fig:intro_entry}, the sub-region 1 could be assigned to the drone “$+$” and the sub-region 2 to the drone “$\times$” for instance. Note that setting variables $x^{r}_{d}$ does not affect the variables $y^{r}_{p}$. Consequently, just redundant integer solutions are eliminated.

\subsubsection{Complete formulation}

Let $T_{\max}$ be the 3D mapping completion time, i.e., the makespan, calculated as the maximum processing time obtained from the swarm of drones. 
The variable $T_{\max}$ is computed by the group of constraints
\begin{equation}
	T_{\max} \geq \sum_{r \in R,p \in P} g^{r}_{dp} \lambda_{p}  \quad \forall d \in \bar{D} \label{eq:nlm_15}
\end{equation}
where  $\sum_{r \in R,p \in P} (g^{r}_{dp} \lambda_{p})$ computes the required time to process all the photos assigned to drone $d \in \bar{D}$.

Finally, the pCAPsac formulation is expressed by the following MILP.
\begin{align}
   \min\limits_{x,y} &\quad T_{\max} \label{eq:nlm_obj_func} \\
    \text{s.t. } &(\ref{eq:nlm_1})-(\ref{eq:nlm_15}) \nonumber \\
   & \text{MMF constraints~\cite{amaldi2013network} }  \nonumber \\
	&x^{r}_{d}, y^{r}_{p}, g^{r}_{dp} \in \{ 0, 1 \} &&\forall r \in R, \forall p \in P, \forall d \in \bar{D} \label{eq:nlm_16} \\
	&w^{hd}_{ij} \in \{0, 1\} &&\forall (i,j) \in A, \forall (h,d) \in  D \times \bar{D} \label{eq:nlm_17} \\
	&\phi^{hd} \geq 0, z^{hd} \in \{0, 1\} &&\forall (h,d) \in  D \times \bar{D} \label{eq:nlm_18} \\
	&\alpha^{r}_{c}, \beta^{r}_{c} \in \{ 0, 1 \} &&\forall c \in C, \forall r \in R \label{eq:nlm_19} \\
 	&\gamma^{r}_{\ell}, \omega^{r}_{\ell} \in \{ 0, 1 \} &&\forall \ell \in L, \forall r \in R  \label{eq:nlm_20} \\
	&u_{ij} \geq 0 &&\forall (i,j) \in A. \label{eq:nlm_21}
\end{align}
The objective function (\ref{eq:nlm_obj_func}) minimizes the makespan of the whole 3D mapping procedure. 
The domain constraints are given by (\ref{eq:nlm_16})-(\ref{eq:nlm_21}). 
As $|R| = |\bar{D}|$, the total number of constraints in the model is $O(|P| \cdot |\bar{D}|^2)$ as well as the number of its variables. 

\section{NP-Hardness of the CAPsac}
\label{sec:np_hardness_proof}

The proof comes from a reduction of the decision version of the \textit{unweighted Geometric Set-Covering Problem} - $GSCP$, whose objective is to assert, for a finite set of points $P' = \{ p_1, p_2, ..., p_n \in \mathbb{R}^{d}\}$ and a finite collection $\mathcal{S'}$ of subsets of $P'$, if there exists a covering for the points $P'$ composed by at most $k < |\mathcal{S'}|$ sets of $\mathcal{S'}$, i.e., if there exists a $\mathcal{C} \subset \mathcal{S'}$ such that $\cup_{c \in \mathcal{C}}c = P'$ and $|\mathcal{C}| \leq k$. The collection $\mathcal{S'}$ is induced by a fixed polytope $\mathcal{T}$, that is, $\mathcal{S'}$ is formed by the points covered by the distinct placements of $\mathcal{T}$ over the coordinates of the points $P'$. The decision problem is NP-Complete even when $\mathcal{T}$ is a fixed square \cite{fowler1981boxcover} or a fixed circumference \cite{johnson1982np}.

\begin{proposition}
\label{prop:polynomial_transformation}
\emph{Given an instance $I^{\prime}$ of the $GSCP$, there exists a polynomial-time transformation from $I^{\prime}$ to an instance $I$ of the $CAPsac$}.
\end{proposition}

\begin{IEEEproof}
Consider an instance of the $GSCP$ with $P'$ points, a collection $\mathcal{S'}$ of subsets of $P'$ induced by a fixed square of length $s$, and a positive integer $k < |\mathcal{S'}|$. The instance of the $rCAPsac$ (see the Appendix) is created on polynomial-time as follows. 

Let the set of photos $P$ and their location be equal to the set of points $P'$ (i.e., $P = P'$), and consider a set of $k$ drones which can do the 3D reconstruction, i.e., $|D| = |\bar{D}| = k$. The communication network $T=(N,A)$ is a random tree whose links $(i,j) \in A$ have infinite capacity. Therefore, the transmission times, the transmission rates $\phi^{hd}$, and the photos storage location $\theta_{dp}$ can be dismissed. Thus, the latest completion time $T_{\max}$ is defined only by the photo processing times of the drones. Concerning the reliability factor, it is made equal to 1 ($\sigma = 1$). Indeed, the collection $\mathcal{S'}$ does not have all possible rectangular spatial-convex sets, being the $\mathcal{S} \setminus \mathcal{S'}$ missing spatial-convex sets  obtained in polynomial-time by inspecting the tuples in $C \times C  \times L \times L$. Finally, the photo processing times of the spatial-convex sets will be either $1$ or $+\infty$. For $S \in \mathcal{S'}$, $t^{S} = 1$, while for the remaining $S \in \mathcal{S} \setminus \mathcal{S'}$ $t^{S} = +\infty$.
\end{IEEEproof}

\begin{proposition}
\label{prop:papmap_solves_gscp}
\emph{The $CAPsac$ answers the $GSCP$}.
\end{proposition}
\begin{IEEEproof}
Consider $rCAPsac(P,D,T,\mathcal{S})$ a routine which solves the $CAPsac$ by the $rCAPsac$ formulation (see the Appendix). 
Let its optimal solution be comprised by the optimal set $Q^{*}$ of variables $q^{S}_{d}$ and the optimal completion time $T^{*}_{\max}$. 
Given a solution of an instance of the $CAPsac$ created from an instance of the $GSCP$, evaluating $T^{*}_{\max}$ is enough to answer the $GSCP$. 
If $T^{*}_{\max} = 1$, reply \textit{yes}. Otherwise, reply \textit{no}. 
For an optimal solution whose $T^{*}_{\max} = 1$, the covering $\mathcal{C}$ of the $P'$, with $|\mathcal{C}| \leq k$, can be extracted from $Q^{*}$.
\end{IEEEproof}

\begin{theorem}
\label{theor:papmap_np_hard}
\emph{The $CAPsac$ is NP-Hard}.
\end{theorem}
\begin{IEEEproof}
Given the propositions \ref{prop:polynomial_transformation} and \ref{prop:papmap_solves_gscp}, one can state that the $GSCP$ is no harder than the $CAPsac$. Since the $GSCP$ is NP-complete, the $CAPsac$ is NP-Hard.
\end{IEEEproof}

\section{Computational experiments}
\label{sec:experiments}
 
Our experimental analysis assessed 
(i) the effectiveness of the ordering inequalities and the branching strategies for the pCAPsac formulation; 
(ii) the performance of formulation pCAPsac; 
(iii) the sensitivity of the pCAPsac formulation with respect to both reliability factor $\sigma$ and maximum allowed transmission time $\hat{T}$. 
We used CPLEX v12.8 as general-purpose MILP solver.  
All experiments were carried exploiting a single core on a machine powered by an Intel E5-2683 v4 Broadwell 2.1GHz with 20Gb of RAM, and running the CentOS Linux 7.5.1804 OS.

We do not present computational experiments for formulation rCAPsac (see the Appendix). 
As demonstrated in the Appendix, the number of variables of that formulation is bounded by $O(|P|^4)$, which can rapidly increase. 
Column Generation (CG) strategy has been extensively applied to formulations with a massive number of variables \cite{column2005}.
The great advantage of employing CG is to solve the Linear Programming (LP) continuous relaxation considering only a relevant subset of variables.
Such LP with reduced number of variables is called the \emph{restricted master problem} (RMP).
The subset of relevant variables is iteratively created as needed by solving the so-called \emph{pricing subproblem} (PS).
Usually, a CG iteration comprises:
i) solving the current RMP to obtain current primal optimal solution and its associated dual variables, and
ii) optimizing the PS to find new variables with negative reduced costs (when considering minimization problems).
The CG terminates when the PS does not find any variable with negative reduced cost, i.e., when the optimality of the current RMP has been proved.
Preliminary experiments (restricted to $\sigma=1$ and $\hat{T} = +\infty$) were performed applying a vanilla column generation on the $rCAPsac$. 
Unfortunately, $rCAPsac$ has proved highly degenerate requiring several CG iterations to prove optimality. 
Therefore, its performance for solving CAPsac was largely inferior to that of using the pCAPsac formulation. Finally, note that the above degeneragy is a common CG drawback that leads the resulting algorithms (and codes) to be difficult to tune and somehow delicate, thus incompatible for the applied context we deal with.

\subsection{MILP technology}

Whenever a new problem can be described by a compact MILP formulation, it is sensible (if not necessary) to first assess the viability of solving it by a general-purpose MILP solver like CPLEX. The reason is that MILP technology is very mature, with essentially all advances in algorithmic discrete optimization now part of the arsenal of the main solvers \cite{lodi2010mixed}. In other words, even the design of an \emph{ad hoc} exact or heuristic algorithm for a new discrete optimization problem always needs to be accompanied by a thoughtful analysis of a MILP solver performance on a possibly non-trivial formulation on the problem. This is the path we have followed and in this section, we briefly describe the basic ingredients that make the MILP technology so successful and mature.\footnote{The interested reader is referred to \cite{lodi2010mixed} for more details and for all necessary references that we avoid to add to the section to keep it compact.}

MILP solvers implement a sophisticated version of the divide-and-conquer algorithm called branch and bound. The basic idea is to relax the complicated constraints of the problem, namely the integrality requirements, and solve the continuous (or LP) relaxation to compute a lower bound (in our minimization case) on the optimal solution value. If none of the removed constraints is violated (in this case, all discrete variables are integer) in the LP solution, then the solution is feasible and the problem is solved. Otherwise, one of the variables taking a fractional value in the continuous relaxation is rounded up and down and two subproblems are created by imposing either that the variable must be smaller or equal than the rounded-down value or greater or equal than the rounded-up value, respectively. This step is called branching and it is easy to see that for both subproblems the solution of the LP relaxation is infeasible and that the optimal solution of the original problem must belong to one of the two subproblems. It is easy to visualize the branching process by a tree in which the original MILP is the ``root'' node and each subproblem is a ``child" node. The algorithm iterates by the selection (according to different policies) of one of the subproblems / nodes and a node can be fathomed either because integer feasible, or because LP infeasible or, finally, because the local lower bound is provably worse than the incumbent solution, i.e., the best upper bound value found so far in the search.

This basic scheme clearly leads to potentially explore an exponentially-large search tree, which is to be expected in the worst case (MILP is theoretically NP-hard) but obviously undesirable. The quality of the formulation is measured with respect to the gap between its LP relaxation and the optimal solution values,\footnote{We specify and use the notion of gap extensively in the computational evaluation.} and MILP technology includes a number of enhancements to this basic scheme to help avoid the tree growing too large. Namely, (i) preprocessing and probing techniques are applied to simplify the formulation by fixing variables and improving coefficients, (ii) redundant linear inequalities exploiting integrality of the variables are added to improve the LP relaxation,\footnote{Those inequalities are referred to as ``cutting planes'' or ``cuts'' and this is the reason the overall algorithm is often referred to as ``branch and cut'' instead of ``branch and bound".} (iii) effective rules to select, at every node, the variable to branch on are used and (iv) heuristic methods are applied to compute better and better incumbent solutions (and, consequently, fathom more nodes).

\subsection{Tested instances and computational settings}

The instances, which were constructed from realistic data, comprise two scenarios:
\begin{IEEEenumerate}[\IEEEsetlabelwidth{ii}]
    \item[i] \textit{Unweighted}: all photos require the same amount of processing time $\lambda$.
    \item[ii] \textit{Weighted}: each photo $p \in P$ requires a certain amount of processing time $\lambda_p$.
\end{IEEEenumerate}
The $\lambda_p$ are acquired from the equivalent unweighted case: $\lceil |P| \times 0.1 \rceil$ groups of nine adjacent photos are randomly selected, and then, for each of those groups, a single $\lambda_p$ is drawn from a normal distribution ($\mu=26.72$ seconds and $\sigma=5.0$) and attributed to all photos of that group.
Changing the photo-processing time in that fashion allows to represent the 3D reconstruction of distinct complex objects in the region of interest.
The name of the instances follows the notation \textit{X}-P\textit{YY}D\textit{Z}$\%\bar{D}$\textit{WW} where “\textit{X}” is “u” for the unweighted instances and “w” for the weighted instances, “\textit{YY}” stands for the number of photos in the instance, “\textit{Z}” specifies the number of drones in the swarm, and “\textit{WW}” informs the percentage of drones that can do 3D reconstruction. 
The number of drones able to perform 3D reconstruction, called $|\bar{D}|$, is always equal to $\lfloor  Z \times \frac{WW}{100} \rfloor $. 
The characteristics of the tested instances are listed in Table \ref{tab:instances}.
Many factors directly impact the number of photos required to obtain 3D maps with good resolution, e.g., perspective, lens quality, overlap, coverage, and object geometry~\cite{tang2020}. 
In fact, the same number of photos is suitable to areas with different dimensions when varying those parameters~\cite{pepe2018planning}.
As an illustration, according to~\cite{luca2020}, an area surrounding a football field of approximate size of 160m $\times$ 80m was reconstructed with about 200 pictures.
Hence, the number of photos used in our instances is comparable to those of real-world applications.

\begin{table}[!t]
	\renewcommand{\arraystretch}{1.3}
	\centering
	\caption{Characteristics of the tested instances.}
	\label{tab:instances}
	\begin{tabular}{l | l }
  		\hline
  		Photos(P) &  200, 400 \\
  		Drones(D) & 5, 7, 10 \\ 
  		$\%$3D-capable drones($\%\bar{D}$) & 50\%, 70\%, 90\% \\
  		\hline
	\end{tabular}
\end{table}

The tables presented hereafter report 
the instance employed at each row (column “Instance”),
the corresponding formulation (column “Form.”), 
the dual gap (in percentage) wrt the optimal solution found at the root node (column “$gap_0$”),
the number of cuts added by CPLEX at the root node (column “cuts”), the
number of nodes explored by the CPLEX's branch-and-cut method (column “Nodes”),  and
the dual gap (in percentage) wrt the optimal solution (best known, see below) at the end of the branch-and-cut enumeration (column “$gap$”). 
CPU times spent in the solution of the root node and by the branch-and-cut algorithm are also reported (column “$sec.$”).

Note that dual gaps are computed with respect to the best upper bound solution found whenever the optimal solutions are not obtained by CPLEX within one day of execution. 
These situations are represented in the tables by the symbol “$\ast$”. 

\subsection{pCAPsac experiments}
\label{subsec:photo_region_exp}

This section evaluates the performance of the proposed pCAPsac formulation. 
It investigates the effectiveness of valid inequalities (\ref{eq:cut_1})-(\ref{eq:cut_4}), whose aim is, for any possible sub-region, to clean the search space from all the infeasible boundary configuration. 
Also, the experiments analyze how different branching priorities can influence the branch-and-cut method.
All the experiments of this section are for $\sigma=1$ and $\hat{T} {=} {+}\infty$.

This section also reports the performance profiles~\cite{dolan2002} concerning the CPU times of the formulations.
Let us define $I$ as the set of all instances and $\mathcal{F}$ as the set comprising all formulations to be compared.
Given an instance $i \in I$, the performance profile compares the CPU time of a formulation $form \in \mathcal{F}$, denoted $cpu_{i, form}$, with the best CPU time obtained across all formulations in $\mathcal{F}$ when solving that instance $i$.
It is done by computing the performance ratio $r_{i, form}$~\cite{dolan2002}
\begin{equation}
	r_{i, form} = \frac{ cpu_{i, form} } { \min\limits_{form \in \mathcal{F}}  \{ cpu_{i, form}  \}  }	\quad \forall i \in I, form \in \mathcal{F}. 	\nonumber
\end{equation}
Thus, the overall assessment of performance is captured by the cumulative distribution function
\begin{equation}
	\rho_{form}(\tau)  = \frac{ | \{ r_{i, form} \leq \tau | i \in I  \} | } { | I | }	.		\nonumber
\end{equation}
Finally, $\rho_{form}(\tau)$ is the probability for the formulation $form$ that the ratio $r_{i, form}$ is at most $\tau$ of the best ratio found~\cite{dolan2002}.

\subsubsection{pCAPsac performance}
\label{subsec:bc_vs_bc}

The pCAPsac formulation using the $\overline{BC_0}$ constraints (\ref{eq:sbc_zero_1}), called “$PB{:}\overline{BC_0}$”, and the pCAPsac formulation using the $BC_0$ constraints (\ref{eq:sbc_zero_2}), named “$PB{:}BC_0$”, are compared in Tables \ref{tab:bc_formulations_comparisson_unweighted} and \ref{tab:bc_formulations_comparisson_weighted}.
For those tables, the dual gaps in the column “$gap_0$” indicate how strong is the formulation.
Small dual gaps mean that the value of the optimal objective function is close to the value of the optimal objective function obtained when solving the continuous relaxation of the formulation.  
Small “$gap_0$” values lead to the exploration of fewer nodes by the branch-and-cut method, reducing its overall computing time. 
As the dual gaps “$gap_0$” are equal across the proposed formulations, one should focus on the execution time (column “sec”) and number of nodes explored (column “nodes”) of the whole branch-and-cut method.\footnote{Note that the time and nodes can vary even if the gap is initially equal because CPLEX evolves differently, for example, because it adds different cutting planes (see, column ``cuts").} 
The results related to lines “$PB{:}\overline{BC_0}{+}Ord.$” are discussed in the following section. 

\begin{table}[!t]
\renewcommand{\arraystretch}{1.3}
\setlength{\tabcolsep}{1.2pt}
\centering
\caption{CPLEX results when solving unweighted instances for the “$PB{:}\overline{BC_0}$”, the “$PB{:}BC_0$”, and “$PB{:}\overline{BC_0}{+}Ord.$” formulations.} 
\label{tab:bc_formulations_comparisson_unweighted}
\begin{tabular}{ c | r || r r r || r | r r}
	\hline
	\multicolumn{1}{ c |}{ \multirow{2}{*}{\textbf{Instance}} } & \multicolumn{1}{ c ||}{ \multirow{2}{*}{\textbf{Form.}} } & \multicolumn{3}{ c ||}{\textbf{Root Node}} & \multicolumn{3}{ c }{\textbf{Branch-and-Cut}} \\
	\cline{3-8}
    & & $\mathbf{gap_0}$ & \textbf{cuts} & \textbf{sec.} & \textbf{Nodes} & $\mathbf{gap}$ & \textbf{sec.} \\				
	\hline
\multirow{3}{*}{u-P200D5$\%\bar{D}$70} & $PB{:}\overline{BC}_0$ & 4.76 & 28 & 1.47 & 182 & 0.00 & 15.81 \\ 
 & $PB{:}BC_0$ & 4.76 & 110 & 2.84 & 300 & 0.00 & 33.83 \\ 
 & $PB{:}\overline{BC}_0$+Ord. & 4.76 & 242 & 1.15 & 345 & 0.00 & 19.02 \\ 
\hline 
\multirow{3}{*}{u-P200D7$\%\bar{D}$50} & $PB{:}\overline{BC}_0$ & 4.76 & 17 & 1.25 & 395 & 0.00 & 27.48 \\ 
 & $PB{:}BC_0$ & 4.76 & 62 & 1.14 & 383 & 0.00 & 19.70 \\ 
  & $PB{:}\overline{BC}_0$+Ord. & 4.76 & 189 & 1.46 & 239 & 0.00 & 11.45 \\ 
\hline 
\multirow{3}{*}{u-P400D5$\%\bar{D}$70} & $PB{:}\overline{BC}_0$ & 1.23 & 28 & 3.33 & 386 & 0.00 & 87.16 \\ 
 & $PB{:}BC_0$ & 1.23 & 74 & 3.57 & 264 & 0.00 & 57.94 \\ 
  & $PB{:}\overline{BC}_0$+Ord. & 1.23 & 183 & 3.38 & 500 & 0.00 & 91.56 \\ 
\hline 
\multirow{3}{*}{u-P400D7$\%\bar{D}$50} & $PB{:}\overline{BC}_0$ & 1.23 & 27 & 3.42 & 556 & 0.00 & 156.15 \\ 
 & $PB{:}BC_0$ & 1.23 & 50 & 3.16 & 1178 & 0.00 & 181.14 \\ 
  & $PB{:}\overline{BC}_0$+Ord. & 1.23 & 188 & 4.31 & 707 & 0.00 & 133.70 \\ 
\hline 
\hline 
\multirow{3}{*}{u-P200D5$\%\bar{D}$90} & $PB{:}\overline{BC}_0$ & 0.00 & 9 & 2.16 & 433 & 0.00 & 24.73 \\ 
 & $PB{:}BC_0$ & 0.00 & 31 & 3.21 & 41 & 0.00 & 13.98 \\ 
  & $PB{:}\overline{BC}_0$+Ord. & 0.00 & 227 & 1.99 & 359 & 0.00 & 22.60 \\ 
\hline 
\multirow{3}{*}{u-P200D7$\%\bar{D}$70} & $PB{:}\overline{BC}_0$ & 0.00 & 23 & 2.14 & 371 & 0.00 & 27.49 \\ 
 & $PB{:}BC_0$ & 0.00 & 22 & 4.07 & 175 & 0.00 & 24.74 \\ 
  & $PB{:}\overline{BC}_0$+Ord. & 0.00 & 226 & 2.33 & 699 & 0.00 & 45.39 \\ 
\hline 
\multirow{3}{*}{u-P400D5$\%\bar{D}$90} & $PB{:}\overline{BC}_0$ & 0.00 & 4 & 5.94 & 631 & 0.00 & 440.53 \\ 
 & $PB{:}BC_0$ & 0.00 & 81 & 4.09 & 121 & 0.00 & 44.60 \\ 
  & $PB{:}\overline{BC}_0$+Ord. & 0.00 & 267 & 5.00 & 49 & 0.00 & 35.02 \\ 
\hline 
\multirow{3}{*}{u-P400D7$\%\bar{D}$70} & $PB{:}\overline{BC}_0$ & 0.00 & 71 & 5.04 & 790 & 0.00 & 141.43 \\ 
 & $PB{:}BC_0$ & 0.00 & 84 & 5.06 & 1631 & 0.00 & 696.21 \\ 
  & $PB{:}\overline{BC}_0$+Ord. & 0.00 & 175 & 6.23 & 7011 & 0.00 & 6665.71 \\ 
\hline 
\hline 
\multirow{3}{*}{u-P200D10$\%\bar{D}$50} & $PB{:}\overline{BC}_0$ & 0.00 & 81 & 3.63 & 6591 & 0.00 & 1036.81 \\ 
 & $PB{:}BC_0$ & 0.00 & 95 & 2.72 & 18840 & 0.00 & 3141.16 \\ 
  & $PB{:}\overline{BC}_0$+Ord. & 0.00 & 312 & 4.99 & 3546 & 0.00 & 556.97 \\ 
\hline 
\multirow{3}{*}{u-P400D10$\%\bar{D}$50} & $PB{:}\overline{BC}_0$ & 0.00 & 38 & 9.99 & 4363 & 0.00 & 1737.28 \\ 
 & $PB{:}BC_0$ & 0.00 & 127 & 10.64 & 16216 & 0.00 & 7200.00 \\ 
  & $PB{:}\overline{BC}_0$+Ord. & 0.00 & 46 & 6.35 & 990 & 0.00 & 219.40 \\ 
\hline 
\hline 
\multirow{3}{*}{u-P200D7$\%\bar{D}$90} & $PB{:}\overline{BC}_0$ & 1.96 & 24 & 5.17 & 8448 & 0.00 & 1354.41 \\ 
 & $PB{:}BC_0$ & 1.96 & 47 & 9.60 & 13835 & 1.96 & 7200.00 \\ 
  & $PB{:}\overline{BC}_0$+Ord. & 1.96 & 130 & 8.80 & 521 & 0.00 & 49.52 \\ 
\hline 
\multirow{3}{*}{u-P400D7$\%\bar{D}$90} & $PB{:}\overline{BC}_0$ & *1.96 & 130 & 12.28 & 12362 & *1.96 & 7200.00 \\ 
 & $PB{:}BC_0$ & *1.96 & 84 & 13.56 & 9063 & *1.96 & 7200.00 \\ 
  & $PB{:}\overline{BC}_0$+Ord. & *1.96 & 50 & 17.94 & 14897 & *1.96 & 7200.00 \\ 
\hline 
\end{tabular}
\end{table}

\begin{table}[!t]
\renewcommand{\arraystretch}{1.3}
\setlength{\tabcolsep}{1.2pt}
\centering
\caption{CPLEX results when solving weighted instances for the “$PB{:}\overline{BC_0}$”, the “$PB{:}BC_0$”, “$PB{:}\overline{BC_0}{+}Ord.$” formulations.} 
\label{tab:bc_formulations_comparisson_weighted}
\begin{tabular}{ c | r || r r r || r | r r}
	\hline
	\multicolumn{1}{ c |}{ \multirow{2}{*}{\textbf{Instance}} } & \multicolumn{1}{ c ||}{ \multirow{2}{*}{\textbf{Form.}} } & \multicolumn{3}{ c ||}{\textbf{Root Node}} & \multicolumn{3}{ c }{\textbf{Branch-and-Cut}} \\
	\cline{3-8}
    & & $\mathbf{gap_0}$ & \textbf{cuts} & \textbf{sec.} & \textbf{Nodes} & $\mathbf{gap}$ & \textbf{sec.} \\				
	\hline
\multirow{3}{*}{w-P200D5$\%\bar{D}$70} & $PB{:}\overline{BC}_0$ & 3.36 & 28 & 1.40 & 226 & 0.00 & 19.13 \\ 
 & $PB{:}BC_0$ & 3.36 & 19 & 2.14 & 450 & 0.00 & 45.84 \\ 
  & $PB{:}\overline{BC}_0$+Ord. & 3.36 & 93 & 1.25 & 300 & 0.00 & 16.87 \\ 
\hline 
\multirow{3}{*}{w-P200D7$\%\bar{D}$50} & $PB{:}\overline{BC}_0$ & 3.67 & 22 & 1.34 & 961 & 0.00 & 50.73 \\ 
 & $PB{:}BC_0$ & 3.67 & 40 & 1.27 & 1057 & 0.00 & 65.62 \\ 
  & $PB{:}\overline{BC}_0$+Ord. & 3.67 & 111 & 1.49 & 549 & 0.00 & 31.52 \\ 
\hline 
\multirow{3}{*}{w-P400D5$\%\bar{D}$70} & $PB{:}\overline{BC}_0$ & 0.86 & 9 & 2.57 & 483 & 0.00 & 97.08 \\ 
 & $PB{:}BC_0$ & 0.86 & 45 & 3.19 & 910 & 0.00 & 148.94 \\
  & $PB{:}\overline{BC}_0$+Ord. & 0.86 & 190 & 3.48 & 414 & 0.00 & 79.99 \\  
\hline 
\multirow{3}{*}{w-P400D7$\%\bar{D}$50} & $PB{:}\overline{BC}_0$ & 2.02 & 44 & 2.92 & 1008 & 0.00 & 191.33 \\ 
 & $PB{:}BC_0$ & 2.02 & 22 & 3.47 & 930 & 0.00 & 121.78 \\ 
  & $PB{:}\overline{BC}_0$+Ord. & 2.02 & 264 & 3.40 & 974 & 0.00 & 241.08 \\ 
\hline 
\hline 
\multirow{3}{*}{w-P200D5$\%\bar{D}$90} & $PB{:}\overline{BC}_0$ & 2.96 & 13 & 2.37 & 5080 & 0.00 & 748.98 \\ 
 & $PB{:}BC_0$ & 2.96 & 42 & 4.76 & 24186 & 0.00 & 5199.65 \\ 
  & $PB{:}\overline{BC}_0$+Ord. & 2.96 & 84 & 2.30 & 3427 & 0.00 & 427.78 \\ 
\hline 
\multirow{3}{*}{w-P200D7$\%\bar{D}$70} & $PB{:}\overline{BC}_0$ & 2.69 & 58 & 2.55 & 4227 & 0.00 & 453.18 \\ 
 & $PB{:}BC_0$ & 2.69 & 21 & 3.95 & 18204 & 0.00 & 4191.87 \\ 
  & $PB{:}\overline{BC}_0$+Ord. & 2.69 & 87 & 3.01 & 3736 & 0.00 & 514.63 \\ 
\hline 
\multirow{3}{*}{w-P400D5$\%\bar{D}$90} & $PB{:}\overline{BC}_0$ & 1.26 & 19 & 4.85 & 3161 & 0.00 & 1425.80 \\ 
 & $PB{:}BC_0$ & 1.26 & 51 & 5.29 & 18195 & 1.24 & 7200.00 \\ 
  & $PB{:}\overline{BC}_0$+Ord. & 1.26 & 442 & 6.14 & 3048 & 0.00 & 1532.36 \\ 
\hline 
\multirow{3}{*}{w-P400D7$\%\bar{D}$70} & $PB{:}\overline{BC}_0$ & 0.68 & 34 & 4.97 & 4529 & 0.00 & 1805.83 \\ 
 & $PB{:}BC_0$ & 0.68 & 13 & 4.63 & 19444 & 0.68 & 7200.00 \\ 
  & $PB{:}\overline{BC}_0$+Ord. & 0.68 & 628 & 6.68 & 9429 & 0.00 & 6031.09 \\ 
\hline 
\hline 
\multirow{3}{*}{w-P200D10$\%\bar{D}$50} & $PB{:}\overline{BC}_0$ & 1.65 & 15 & 5.42 & 32327 & 1.65 & 7200.00 \\ 
 & $PB{:}BC_0$ & 1.65 & 63 & 3.62 & 42035 & 1.65 & 7200.00 \\ 
  & $PB{:}\overline{BC}_0$+Ord. & 1.65 & 80 & 5.55 & 19491 & 1.65 & 7200.00 \\ 
\hline 
\multirow{3}{*}{w-P400D10$\%\bar{D}$50} & $PB{:}\overline{BC}_0$ & *3.38 & 129 & 10.59 & 9416 & *3.38 & 7200.00 \\ 
 & $PB{:}BC_0$ & *3.38 & 82 & 8.88 & 7364 & *3.38 & 7200.00 \\ 
  & $PB{:}\overline{BC}_0$+Ord. & *3.38 & 194 & 7.03 & 10625 & *3.38 & 7200.00 \\ 
\hline 
\hline 
\multirow{3}{*}{w-P200D7$\%\bar{D}$90} & $PB{:}\overline{BC}_0$ & *3.81 & 1 & 8.28 & 54078 & *3.81 & 7200.00 \\ 
 & $PB{:}BC_0$ & *3.81 & 71 & 6.17 & 19819 & *3.81 & 7200.00 \\ 
  & $PB{:}\overline{BC}_0$+Ord. & *3.81 & 57 & 10.48 & 23206 & *3.81 & 7200.00 \\ 
\hline 
\multirow{3}{*}{w-P400D7$\%\bar{D}$90} & $PB{:}\overline{BC}_0$ & *3.18 & 15 & 9.42 & 13316 & *3.18 & 7200.00 \\ 
 & $PB{:}BC_0$ & *3.18 & 83 & 17.06 & 10581 & *3.18 & 7200.00 \\ 
  & $PB{:}\overline{BC}_0$+Ord. & *3.18 & 226 & 13.09 & 9753 & *3.18 & 7200.00 \\ 
\hline 
\end{tabular}
\end{table}

\begin{figure}[!tb]
    \centering   
	\includegraphics[width=3.2in]{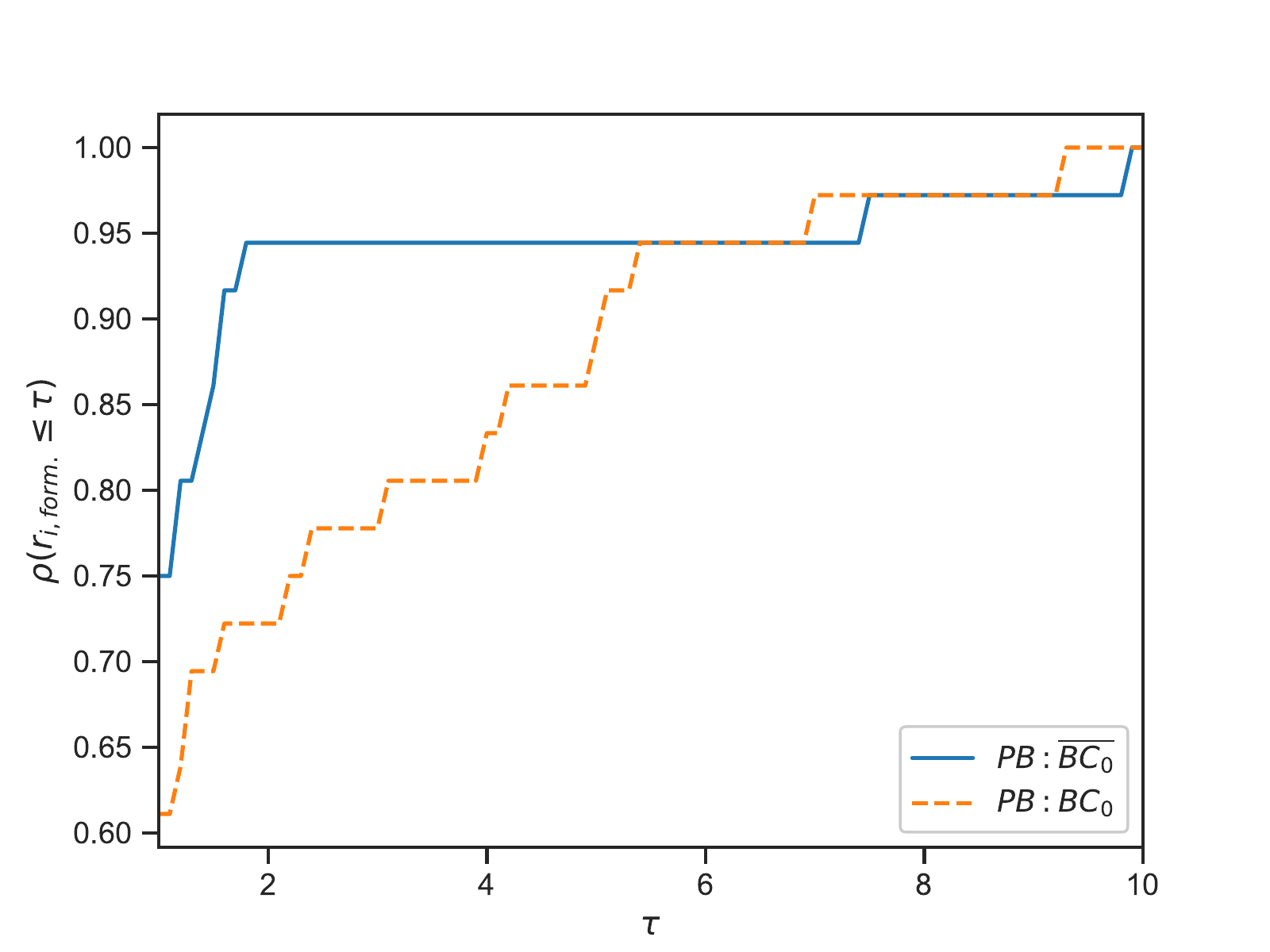}
	\caption{Performance profile w.r.t. the CPU times of the “$PB{:}\overline{BC_0}$” and “$PB{:}BC_0$”.}
	\label{fig:form_performance}
\end{figure}

The results in both Tables \ref{tab:bc_formulations_comparisson_unweighted}  and \ref{tab:bc_formulations_comparisson_weighted} clearly show that the “$PB{:}\overline{BC_0}$” solves faster than “$PB{:}BC_0$” formulation (T=-2.877 and p-val=0.008 via “$PB{:}\overline{BC_0}$” vs. “$PB{:}BC_0$” paired t-test \cite{jain1990art}).
The performance profile in Fig.~\ref{fig:form_performance} confirms the better performance of the “$PB{:}\overline{BC_0}$”.
In fact, “$PB{:}\overline{BC_0}$” has the largest probability ($0.75$) to have the best performance ratio (point when $\tau = 1$).

We can observe that the dual gaps at the root node are equal to zero for the unweighted instances whenever the number of photos is divisible by the number of drones that can do the 3D reconstruction. 
Consequently, the objective function value of the optimum solution coincides with the dual bound already at  the root node for instances “u-P200D5$\%\bar{D}$90”, “u-P200D7$\%\bar{D}$70”, “u-P400D5$\%\bar{D}$90”, “u-P400D7$\%\bar{D}$70”, “u-P200D10$\%\bar{D}$50” and “u-P400D10$\%\bar{D}$50”.

\begin{figure*}[!t]
    \centering
    \begin{minipage}{0.45\textwidth}
       \centering
		\includegraphics[width=3.2in]{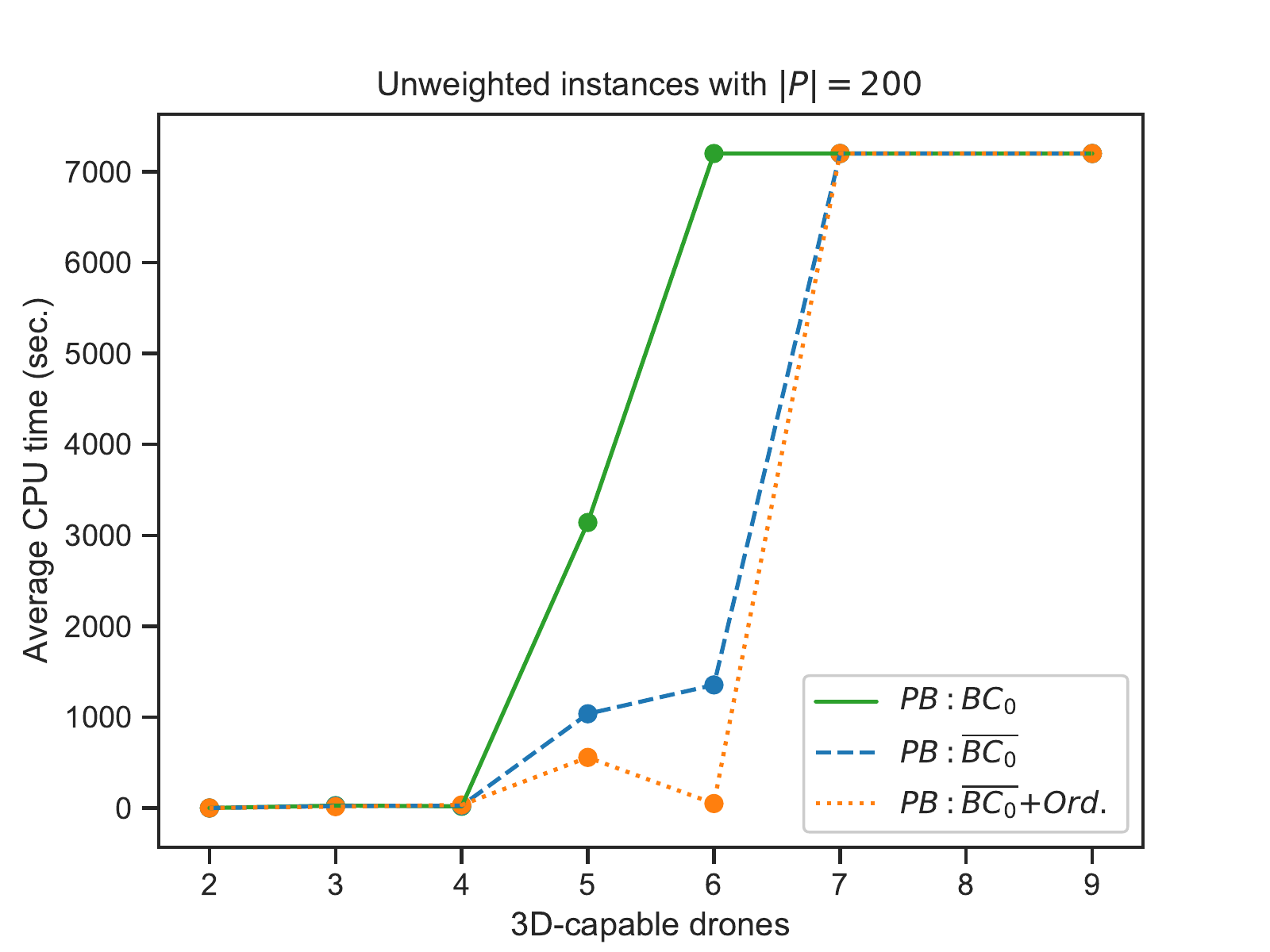}
		\caption{Average CPU times when solving unweighted instances with 200 photos according to “$PB{:}BC_0$”, “$PB{:}\overline{BC_0}$”, and “$PB{:}\overline{BC_0}{+}Ord.$”.}
		\label{fig:avg_time_unweighted_p200}       
    \end{minipage}
	\begin{minipage}{0.45\textwidth}
		\centering		
		\includegraphics[width=3.2in]{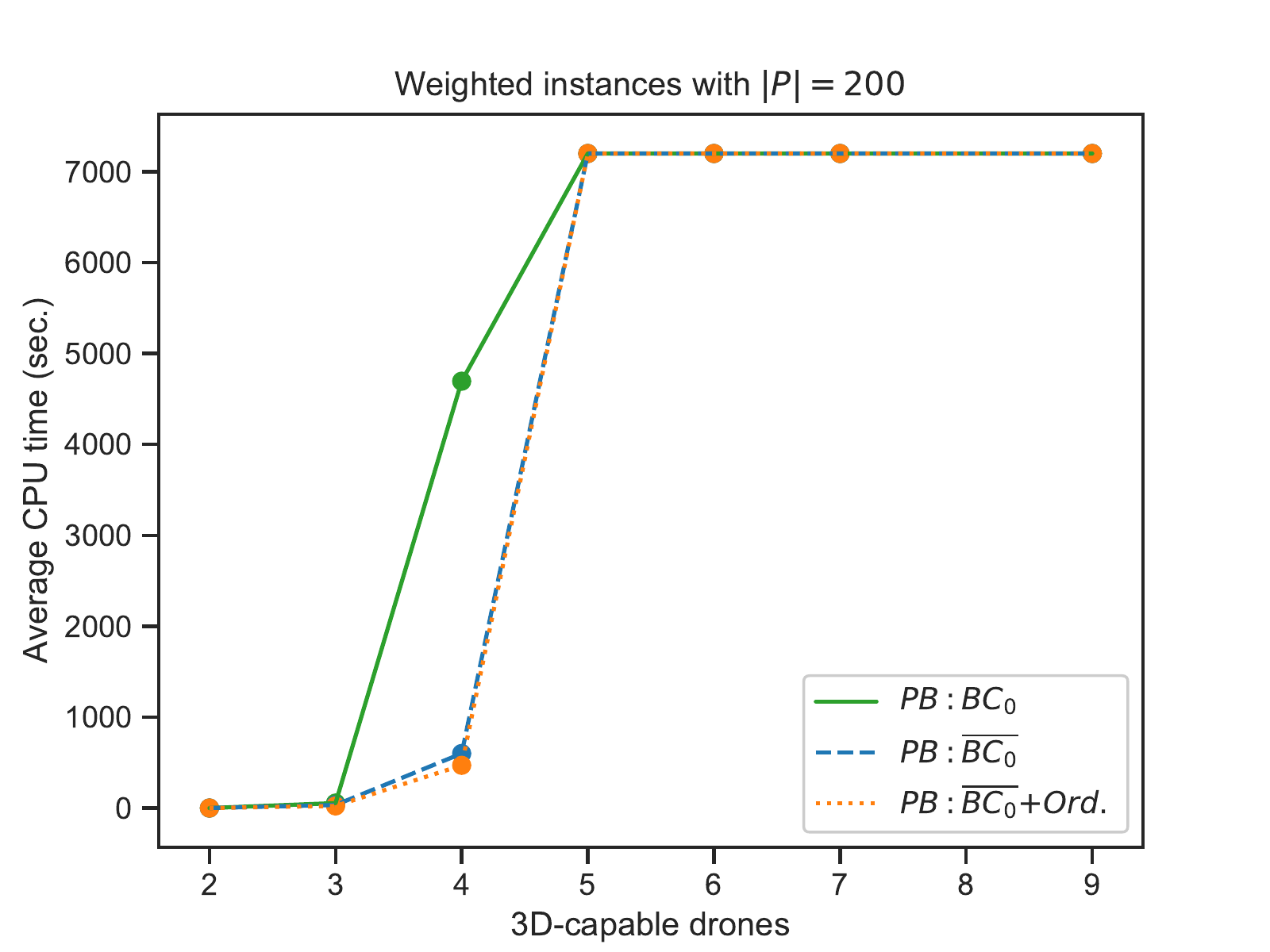}
		\caption{Average CPU times when solving weighted instances with 200 photos according to “$PB{:}BC_0$”, “$PB{:}\overline{BC_0}$”, and “$PB{:}\overline{BC_0}{+}Ord.$”.}
		\label{fig:avg_time_weighted_p200}       
    \end{minipage}
\end{figure*}

\begin{figure*}[!t]
    \centering
    \begin{minipage}{0.45\textwidth}
       \centering
		\includegraphics[width=3.2in]{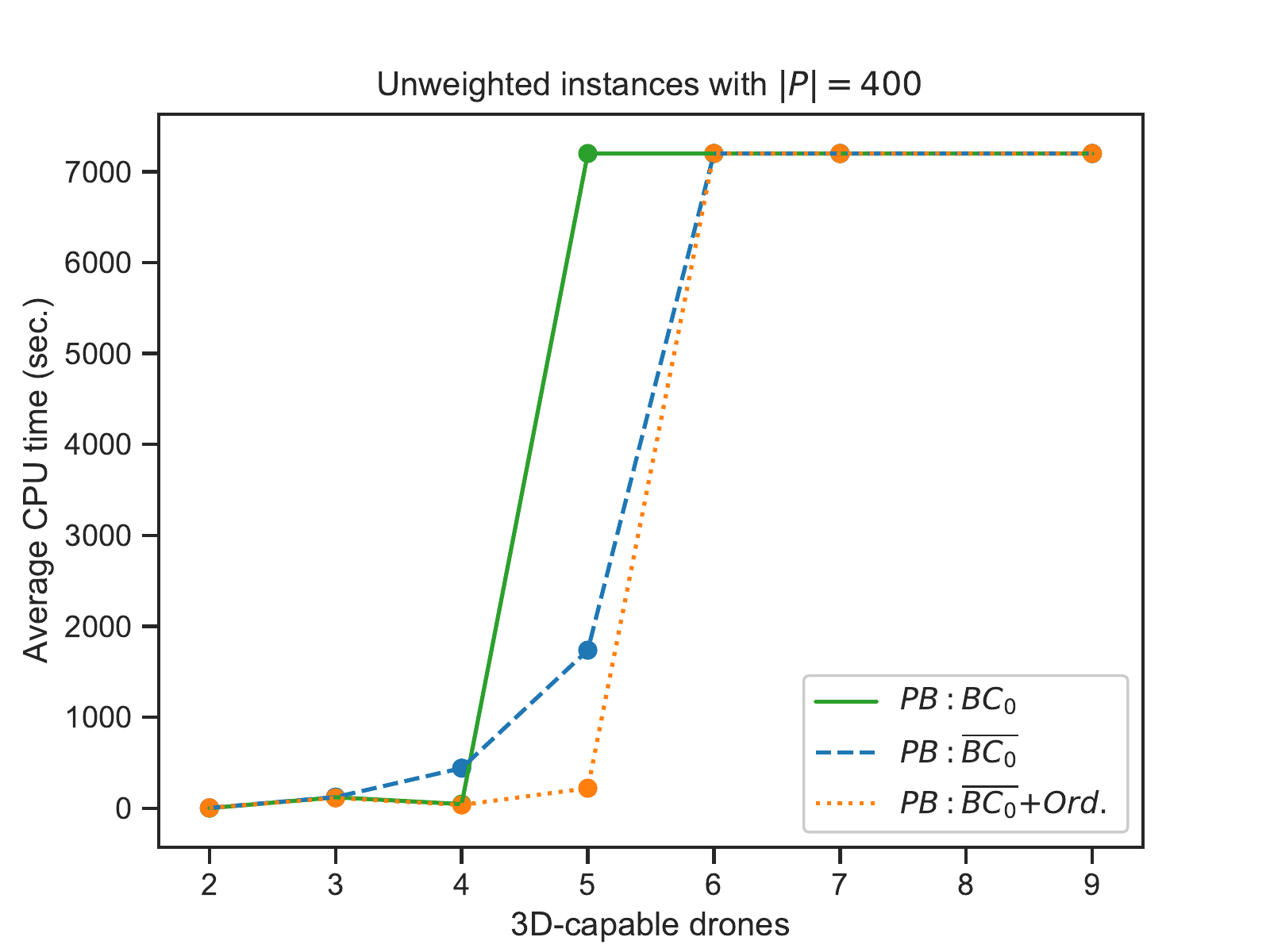}
		\caption{Average CPU times when solving unweighted instances with 400 photos according to “$PB{:}BC_0$”, “$PB{:}\overline{BC_0}$”, and “$PB{:}\overline{BC_0}{+}Ord.$”.}
		\label{fig:avg_time_unweighted_p400}       
    \end{minipage}
	\begin{minipage}{0.45\textwidth}
		\centering		
		\includegraphics[width=3.2in]{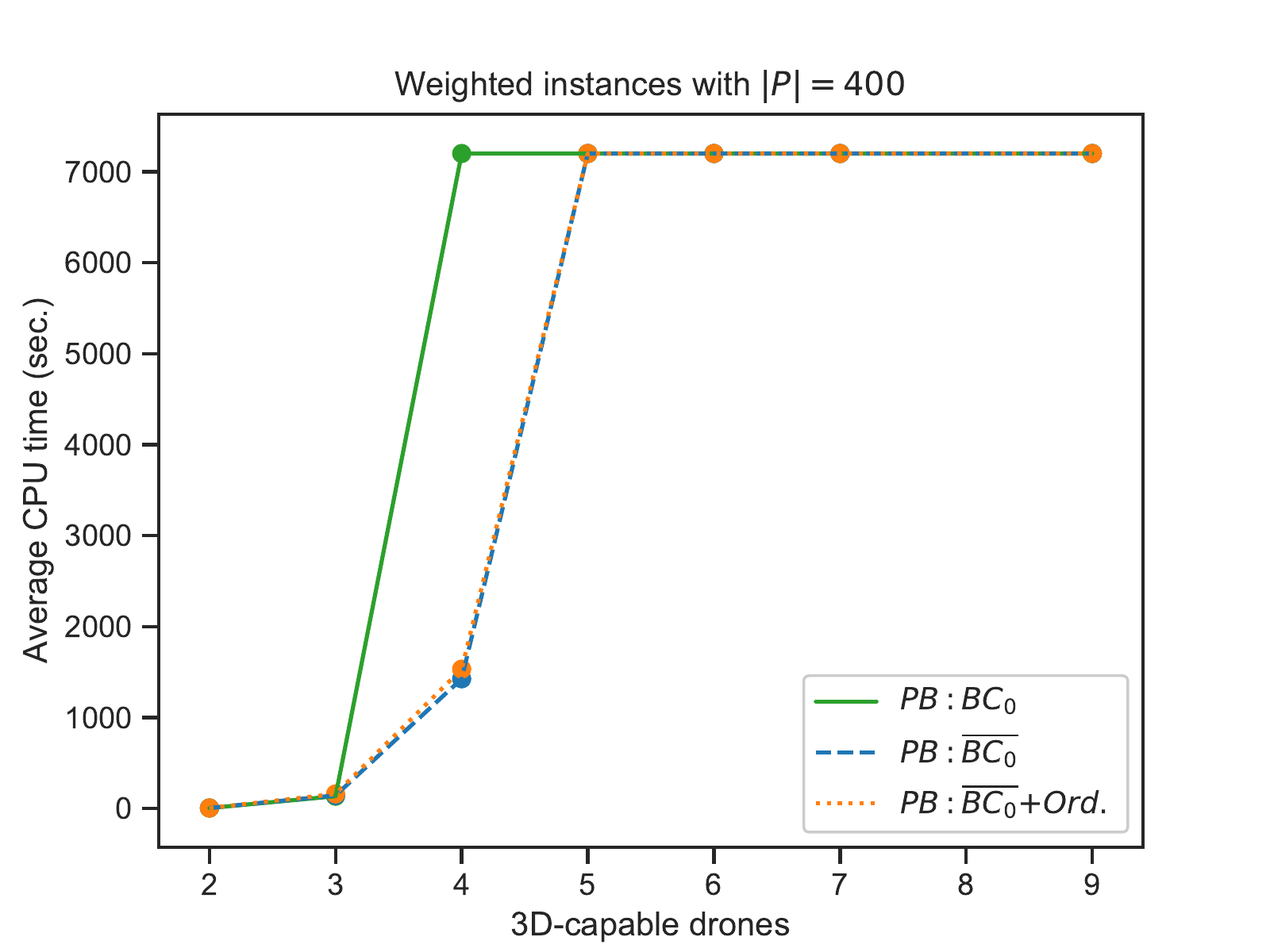}
		\caption{Average CPU times when solving weighted instances with 400 photos according to “$PB{:}BC_0$”, “$PB{:}\overline{BC_0}$”, and “$PB{:}\overline{BC_0}{+}Ord.$”.}
		\label{fig:avg_time_weighted_p400}       
    \end{minipage}
\end{figure*}

Finally, based on tables \ref{tab:bc_formulations_comparisson_unweighted} and \ref{tab:bc_formulations_comparisson_weighted},  Figs.~\ref{fig:avg_time_unweighted_p200},~\ref{fig:avg_time_weighted_p200},~\ref{fig:avg_time_unweighted_p400}, and~\ref{fig:avg_time_weighted_p400}
 present how the CPU time (column “sec.”) is affected (on average) with the increase of the number of 3D-capable drones in the unweighted and weighted instances.
Figs.~\ref{fig:avg_time_unweighted_p200},~\ref{fig:avg_time_weighted_p200},~\ref{fig:avg_time_unweighted_p400}, and~\ref{fig:avg_time_weighted_p400} group the average CPU times --- computed w.r.t. the collection of instances with the same number of 3D-capable drones and  number of photos --- obtained by  formulations “$PB{:}BC_0$” (green lines), “$PB{:}\overline{BC_0}$” (blue dashed lines), and “$PB{:}\overline{BC_0}{+}Ord.$” (orange dotted lines).
We observe that the average CPU time increases when the instances have more 3D-capable drones.
Also, the average CPU time of the instances with 200 photos tends to be shorter than the average CPU time of the instances with 400 photos.

\subsubsection{Ordering inequalities effectiveness}
\label{subsec:ordering_inequalities_experiments}

The effect of adding all the ordering inequalities (\ref{eq:cut_1})-(\ref{eq:cut_4}) into the “$PB{:}\overline{BC_0}$” formulation is analyzed in Tables \ref{tab:bc_formulations_comparisson_unweighted} and \ref{tab:bc_formulations_comparisson_weighted}. 
The inclusion of the ordering inequalities is identified by the “${+}Ord.$” in the formulation's name.

\begin{figure}[!tb]
    \centering   
	\includegraphics[width=3.2in]{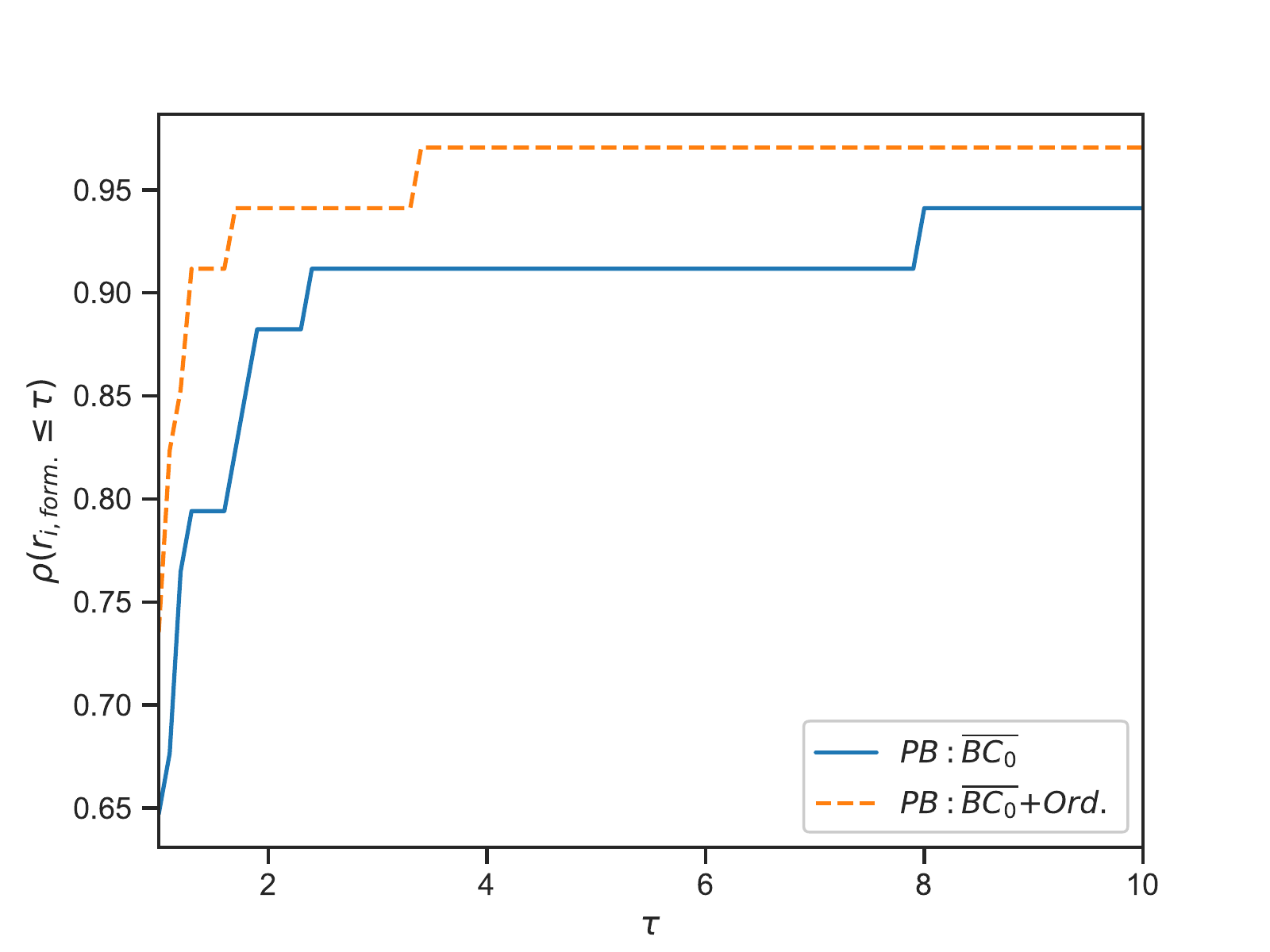}
	\caption{Performance profile w.r.t. the CPU times of the “$PB{:}\overline{BC_0}$” and “$PB{:}\overline{BC_0}{+}Ord.$”.}
	\label{fig:cuts_performance}
\end{figure}

The valid inequalities (\ref{eq:cut_1})-(\ref{eq:cut_4}) eliminate infeasible boundaries in the continuous solution space whereas not necessarily excluding the continuous optimum solution.
Consequently, these inequalities are not guaranteed to increase the dual bound obtained. 
In fact, $gap_0$ was never improved in our experiments after adding the ordering inequalities. 
Nevertheless, the insertion of (\ref{eq:cut_1})-(\ref{eq:cut_4}) improved the CPU time required to reach the optimum solution of 11 out to 19 instances solved to optimality (considering 2h of execution). 
In fact, paired t-tests show significant improvements (T= 1.907 p-val= 0.034) by adding them into the “$PB{:}\overline{BC_0}$” formulation, except for instances “P400D7$\%\bar{D}$70”. 
This is also confirmed by the performance profile in Fig.~\ref{fig:cuts_performance} in which “$PB{:}\overline{BC_0}{+}Ord.$” has the best CPU times approximately 75\% of the times ($\tau = 1$).
However, when limited to weighted cases, there is no significant improvement on reducing the enumeration CPU time. 
For these cases, 11 seconds of improvement is obtained when comparing the average computing CPU time of non-inserting against inserting constraints (\ref{eq:cut_1})-(\ref{eq:cut_4}). 
Finally, we observed that the number of cuts added by CPLEX at the root node increased considerably when the ordering inequalities were employed.

\subsubsection{Branching priority}
\label{subsec:branching_experiments}

Different branching priorities for the selection of the boundary assignment (i.e., $\alpha^{r}_{c}$, $\beta^{r}_{c}$, $\gamma^{r}_{\ell}$, and $\omega^{r}_{\ell}$) and the photo assignment (i.e., $y^{r}_{p}$) variables were also explored in formulation “$PB{:}\overline{BC_0}{+}Ord.$” (simply denoted “$PB{:}\overline{BC_0}$” at this section) and the results reported in Tables \ref{tab:branching_unweighted} and \ref{tab:branching_weighted}. 
The distinct branching priorities are denoted as “$b {>} y$” and “$y {>} b$”.
 The default option of CPLEX is identified by the absence of those notations. 
 The “$b {>} y$” is used when the boundary assignment variables are given higher priority over the photo assignment variables, which still have higher priority over the remaining variables.
 The “$y {>} b$” refers to the opposite case, that is, when photo assignment variables are rather branched over boundary assignment variables.
Since the values in the column “gap” are equal in most of the cases, the number of nodes explored (column “nodes”) and execution times (column “gap”) should be used as the comparison metric.
This means that smaller values in those columns stand for more efficient enumeration performed by CPLEX. 

\begin{table}[!t]
\renewcommand{\arraystretch}{1.3}
\setlength{\tabcolsep}{1.3pt}
\centering
\caption{CPLEX results when solving unweighted instances for the “$PB{:}\overline{BC_0}$”, “$PB{:}\overline{BC_0}{-}b{>}y$” and “$PB{:}\overline{BC_0}{-}y{>}b$” branching priority strategies.}
 \label{tab:branching_unweighted}
\begin{tabular}{ c | r || r | r r}	
	\hline
	\multicolumn{1}{ c |}{ \multirow{2}{*}{\textbf{Instance}} } & \multicolumn{1}{ c ||}{ \multirow{2}{*}{\textbf{Form.}} } & \multicolumn{3}{ c }{\textbf{Branch-and-Cut}} \\
	\cline{3-5}
    & & \textbf{Nodes} & $\mathbf{gap}$ & \textbf{sec.} \\				
	\hline
\multirow{1}{*}{u-P200D5$\%\bar{D}$70} & $PB{:}\overline{BC}_0$ & 345 & 0.00 & 19.02 \\ 
 & $PB{:}\overline{BC}_0{-}b{>}y$ & 529 & 0.00 & 25.46 \\ 
 & $PB{:}\overline{BC}_0{-}y{>}b$ & 178 & 0.00 & 15.05 \\ 
\hline 
\multirow{1}{*}{u-P200D7$\%\bar{D}$50} & $PB{:}\overline{BC}_0$ & 239 & 0.00 & 11.45 \\ 
 & $PB{:}\overline{BC}_0{-}b{>}y$ & 744 & 0.00 & 33.73 \\ 
 & $PB{:}\overline{BC}_0{-}y{>}b$ & 232 & 0.00 & 11.47 \\ 
\hline 
\multirow{1}{*}{u-P400D5$\%\bar{D}$70} & $PB{:}\overline{BC}_0$ & 500 & 0.00 & 91.56 \\ 
 & $PB{:}\overline{BC}_0{-}b{>}y$ & 1993 & 0.00 & 255.79 \\ 
 & $PB{:}\overline{BC}_0{-}y{>}b$ & 179 & 0.00 & 50.86 \\ 
\hline 
\multirow{1}{*}{u-P400D7$\%\bar{D}$50} & $PB{:}\overline{BC}_0$ & 707 & 0.00 & 133.70 \\ 
 & $PB{:}\overline{BC}_0{-}b{>}y$ & 1011 & 0.00 & 207.03 \\ 
 & $PB{:}\overline{BC}_0{-}y{>}b$ & 202 & 0.00 & 68.28 \\ 
\hline 
\hline 
\multirow{1}{*}{u-P200D5$\%\bar{D}$90} & $PB{:}\overline{BC}_0$ & 359 & 0.00 & 22.60 \\ 
 & $PB{:}\overline{BC}_0{-}b{>}y$ & 471 & 0.00 & 41.63 \\ 
 & $PB{:}\overline{BC}_0{-}y{>}b$ & 72 & 0.00 & 7.06 \\ 
\hline 
\multirow{1}{*}{u-P200D7$\%\bar{D}$70} & $PB{:}\overline{BC}_0$ & 699 & 0.00 & 45.39 \\ 
 & $PB{:}\overline{BC}_0{-}b{>}y$ & 1096 & 0.00 & 110.24 \\ 
 & $PB{:}\overline{BC}_0{-}y{>}b$ & 393 & 0.00 & 25.89 \\ 
\hline 
\multirow{1}{*}{u-P400D5$\%\bar{D}$90} & $PB{:}\overline{BC}_0$ & 49 & 0.00 & 35.02 \\ 
 & $PB{:}\overline{BC}_0{-}b{>}y$ & 812 & 0.00 & 215.49 \\ 
 & $PB{:}\overline{BC}_0{-}y{>}b$ & 49 & 0.00 & 35.16 \\ 
\hline 
\multirow{1}{*}{u-P400D7$\%\bar{D}$70} & $PB{:}\overline{BC}_0$ & 7011 & 0.00 & 6665.71 \\ 
 & $PB{:}\overline{BC}_0{-}b{>}y$ & 1330 & 0.00 & 235.25 \\ 
 & $PB{:}\overline{BC}_0{-}y{>}b$ & 430 & 0.00 & 59.71 \\ 
\hline 
\hline 
\multirow{1}{*}{u-P200D10$\%\bar{D}$50} & $PB{:}\overline{BC}_0$ & 3546 & 0.00 & 556.97 \\ 
 & $PB{:}\overline{BC}_0{-}b{>}y$ & 2231 & 0.00 & 217.96 \\ 
 & $PB{:}\overline{BC}_0{-}y{>}b$ & 1328 & 0.00 & 144.06 \\ 
\hline 
\multirow{1}{*}{u-P400D10$\%\bar{D}$50} & $PB{:}\overline{BC}_0$ & 990 & 0.00 & 219.40 \\ 
 & $PB{:}\overline{BC}_0{-}b{>}y$ & 1022 & 0.00 & 107.68 \\ 
 & $PB{:}\overline{BC}_0{-}y{>}b$ & 922 & 0.00 & 199.24 \\ 
\hline 
\hline 
\multirow{1}{*}{u-P200D7$\%\bar{D}$90} & $PB{:}\overline{BC}_0$ & 521 & 0.00 & 49.52 \\ 
 & $PB{:}\overline{BC}_0{-}b{>}y$ & 18419 & 0.00 & 3891.44 \\ 
 & $PB{:}\overline{BC}_0{-}y{>}b$ & 2466 & 0.00 & 557.22 \\ 
\hline 
\multirow{1}{*}{u-P400D7$\%\bar{D}$90} & $PB{:}\overline{BC}_0$ & 14897 & *1.96 & 7200.00 \\ 
 & $PB{:}\overline{BC}_0{-}b{>}y$ & 30016 & *1.96 & 7200.00 \\ 
 & $PB{:}\overline{BC}_0{-}y{>}b$ & 7749 & *1.96 & 7200.00 \\ 
\hline 
\end{tabular}
\end{table}

\begin{table}[!t]
\renewcommand{\arraystretch}{1.3}
\setlength{\tabcolsep}{1.3pt}
\centering
\caption{CPLEX results when solving weighted instances for the “$PB{:}\overline{BC_0}$”, “$PB{:}\overline{BC_0}-b{>}y$” and “$PB{:}\overline{BC_0}-y{>}b$” branching priority strategies.} 
\label{tab:branching_weighted}
\begin{tabular}{ c | r || r | r r}
	\hline
	\multicolumn{1}{ c |}{ \multirow{2}{*}{\textbf{Instance}} } & \multicolumn{1}{ c ||}{ \multirow{2}{*}{\textbf{Form.}} } & \multicolumn{3}{ c }{\textbf{Branch-and-Cut}} \\
	\cline{3-5}
    & & \textbf{Nodes} & $\mathbf{gap}$ & \textbf{sec.} \\				
	\hline
\multirow{1}{*}{w-P200D5$\%\bar{D}$70} & $PB{:}\overline{BC}_0$ & 300 & 0.00 & 16.87 \\ 
 & $PB{:}\overline{BC}_0{-}b{>}y$ & 620 & 0.00 & 38.68 \\ 
 & $PB{:}\overline{BC}_0{-}y{>}b$ & 279 & 0.00 & 20.40 \\ 
\hline 
\multirow{1}{*}{w-P200D7$\%\bar{D}$50} & $PB{:}\overline{BC}_0$ & 549 & 0.00 & 31.52 \\ 
 & $PB{:}\overline{BC}_0{-}b{>}y$ & 600 & 0.00 & 28.76 \\ 
 & $PB{:}\overline{BC}_0{-}y{>}b$ & 327 & 0.00 & 29.57 \\ 
\hline 
\multirow{1}{*}{w-P400D5$\%\bar{D}$70} & $PB{:}\overline{BC}_0$ & 414 & 0.00 & 79.99 \\ 
 & $PB{:}\overline{BC}_0{-}b{>}y$ & 1918 & 0.00 & 330.83 \\ 
 & $PB{:}\overline{BC}_0{-}y{>}b$ & 299 & 0.00 & 80.89 \\ 
\hline 
\multirow{1}{*}{w-P400D7$\%\bar{D}$50} & $PB{:}\overline{BC}_0$ & 974 & 0.00 & 241.08 \\ 
 & $PB{:}\overline{BC}_0{-}b{>}y$ & 602 & 0.00 & 94.42 \\ 
 & $PB{:}\overline{BC}_0{-}y{>}b$ & 315 & 0.00 & 126.37 \\ 
\hline 
\hline 
\multirow{1}{*}{w-P200D5$\%\bar{D}$90} & $PB{:}\overline{BC}_0$ & 3427 & 0.00 & 427.78 \\ 
 & $PB{:}\overline{BC}_0{-}b{>}y$ & 4281 & 0.00 & 409.38 \\ 
 & $PB{:}\overline{BC}_0{-}y{>}b$ & 3833 & 0.00 & 560.64 \\ 
\hline 
\multirow{1}{*}{w-P200D7$\%\bar{D}$70} & $PB{:}\overline{BC}_0$ & 3736 & 0.00 & 514.63 \\ 
 & $PB{:}\overline{BC}_0{-}b{>}y$ & 5879 & 0.00 & 663.39 \\ 
 & $PB{:}\overline{BC}_0{-}y{>}b$ & 4680 & 0.00 & 679.89 \\ 
\hline 
\multirow{1}{*}{w-P400D5$\%\bar{D}$90} & $PB{:}\overline{BC}_0$ & 3048 & 0.00 & 1532.36 \\ 
 & $PB{:}\overline{BC}_0{-}b{>}y$ & 2189 & 0.00 & 925.85 \\ 
 & $PB{:}\overline{BC}_0{-}y{>}b$ & 3146 & 0.00 & 2052.07 \\ 
\hline 
\multirow{1}{*}{w-P400D7$\%\bar{D}$70} & $PB{:}\overline{BC}_0$ & 9429 & 0.00 & 6031.09 \\ 
 & $PB{:}\overline{BC}_0{-}b{>}y$ & 1397 & 0.00 & 605.48 \\ 
 & $PB{:}\overline{BC}_0{-}y{>}b$ & 3341 & 0.00 & 2097.93 \\ 
\hline 
\hline 
\multirow{1}{*}{w-P200D10$\%\bar{D}$50} & $PB{:}\overline{BC}_0$ & 19491 & 1.65 & 7200.00 \\ 
 & $PB{:}\overline{BC}_0{-}b{>}y$ & 46502 & 1.08 & 7200.00 \\ 
 & $PB{:}\overline{BC}_0{-}y{>}b$ & 25588 & 1.65 & 7200.00 \\ 
\hline 
\multirow{1}{*}{w-P400D10$\%\bar{D}$50} & $PB{:}\overline{BC}_0$ & 10625 & *3.38 & 7200.00 \\ 
 & $PB{:}\overline{BC}_0{-}b{>}y$ & 10981 & *3.38 & 7200.00 \\ 
 & $PB{:}\overline{BC}_0{-}y{>}b$ & 8520 & *3.38 & 7200.00 \\ 
\hline 
\hline 
\multirow{1}{*}{w-P200D7$\%\bar{D}$90} & $PB{:}\overline{BC}_0$ & 23206 & *3.81 & 7200.00 \\ 
 & $PB{:}\overline{BC}_0{-}b{>}y$ & 33427 & *3.81 & 7200.00 \\ 
 & $PB{:}\overline{BC}_0{-}y{>}b$ & 29215 & *3.81 & 7200.00 \\ 
\hline 
\multirow{1}{*}{w-P400D7$\%\bar{D}$90} & $PB{:}\overline{BC}_0$ & 9753 & *3.18 & 7200.00 \\ 
 & $PB{:}\overline{BC}_0{-}b{>}y$ & 20758 & *3.18 & 7200.00 \\ 
 & $PB{:}\overline{BC}_0{-}y{>}b$ & 9417 & *3.18 & 7200.00 \\ 
\hline 
\end{tabular}
\end{table}

The adoption of different branching priorities improves the CPU times to solve some instances. 
In fact, the “$y{>}b$” strategy achieves better or equivalent  CPU times in 10 out to 11 unweighted instances solved to optimality (within 2h of execution).
Regarding weighted cases, the “$b{>}y$” strategy results in better or equivalent CPU times in 5 out to 8 instances solved to optimality (considering 2h of execution).
However, paired t-tests do not show significant improvement for both “$b {>} y$” (T= 0.889 p-val= 0.191) or “$y {>} b$” (T=1.295 p-val= 0.103) strategies considering overall cases.

 \subsection{Sensitivity of the pCAPsac formulation}
 \label{subsec:sensitivity}

This section evaluates the sensitivity of pCAPsac formulation with respect to both reliability factor $\sigma$ and to maximum transmission time allowed $\hat{T}$.

\subsubsection{Reliability factor sensitivity}
\label{subsec:sensitivity_sigma}

Tables \ref{tab:sensitivity_sigma_unweighted} and \ref{tab:sensitivity_sigma_weighted} report results for various values of $\sigma$, ranging from 1 to $|\bar{D}|-1$. 
The subset of instances used in this experiment consists of those for which CPLEX was able to solve within 2h of execution the associated problem  with $\sigma=1$. 
Besides, the communication constraints concerning $\hat{T}$ were relaxed, i.e. $\hat{T} {=} {+}\infty$.
For this analysis, one should concentrate on how much varying $\sigma$ affects the initial dual gap (column “$gap_0$”), which might lead to a large number of explored nodes and execution times.

\begin{table}[!t]
\renewcommand{\arraystretch}{1.3}
\setlength{\tabcolsep}{1.3pt}
\centering
\caption{CPLEX results when solving unweighted instances for the $PB$ formulation with $\sigma \in \{1,\ldots,|\bar{D}|-1\}$.} \label{tab:sensitivity_sigma_unweighted} 
\begin{tabular}{ c | c || r r r || r | r r}
	\hline
	\multicolumn{1}{ c |}{ \multirow{2}{*}{\textbf{Instance}} } & \multicolumn{1}{ c ||}{ \multirow{2}{*}{$\mathbf{\sigma}$} } & \multicolumn{3}{ c ||}{\textbf{Root Node}} & \multicolumn{3}{ c }{\textbf{Branch-and-Cut}} \\
	\cline{3-8}
    & & $\mathbf{gap_0}$ & \textbf{cuts} & \textbf{sec.} & \textbf{Nodes} & $\mathbf{gap}$ & \textbf{sec.} \\
	\hline
 \multirow{1}{*}{u-P200D5$\%\bar{D}$70} & 1 & 4.76 & 242 & 1.15 & 345 & 0.00 & 19.02 \\ 
 & 2 & 50.62 & 752 & 10.21 & 1113 & 0.00 & 138.18 \\ 
\hline 
 \multirow{1}{*}{u-P200D7$\%\bar{D}$50} & 1 & 4.76 & 189 & 1.46 & 239 & 0.00 & 11.45 \\ 
 & 2 & 50.62 & 1053 & 9.16 & 902 & 0.00 & 101.71 \\ 
\hline 
 \multirow{1}{*}{u-P400D5$\%\bar{D}$70} & 1 & 1.23 & 183 & 3.38 & 500 & 0.00 & 91.56 \\ 
 & 2 & 50.25 & 667 & 2.90 & 1912 & 0.00 & 539.52 \\ 
\hline 
 \multirow{1}{*}{u-P400D7$\%\bar{D}$50} & 1 & 1.23 & 188 & 4.31 & 707 & 0.00 & 133.70 \\ 
 & 2 & 50.25 & 366 & 2.80 & 3070 & 0.00 & 1723.51 \\ 
\hline 
\hline 
 \multirow{1}{*}{u-P200D5$\%\bar{D}$90} & 1 & 0.00 & 227 & 1.99 & 359 & 0.00 & 22.60 \\ 
 & 2 & 50.00 & 228 & 3.47 & 14578 & 0.00 & 5685.71 \\ 
 & 3 & 66.67 & 194 & 2.46 & 2392 & 0.00 & 1014.65 \\ 
\hline 
 \multirow{1}{*}{u-P200D7$\%\bar{D}$70} & 1 & 0.00 & 226 & 2.33 & 699 & 0.00 & 45.39 \\ 
 & 2 & *50.00 & 119 & 3.71 & 16579 & *33.33 & 7200.00 \\ 
 & 3 & 66.67 & 402 & 3.52 & 2209 & 0.00 & 1150.97 \\ 
\hline 
 \multirow{1}{*}{u-P400D5$\%\bar{D}$90} & 1 & 0.00 & 267 & 5.00 & 49 & 0.00 & 35.02 \\ 
 & 2 & *50.00 & 349 & 6.82 & 3573 & *33.33 & 7200.00 \\ 
 & 3 & 66.67 & 656 & 5.65 & 2631 & 0.00 & 4593.69 \\ 
\hline 
 \multirow{1}{*}{u-P400D7$\%\bar{D}$70} & 1 & 0.00 & 175 & 6.23 & 7011 & 0.00 & 6665.71 \\ 
 & 2 & *50.00 & 403 & 6.59 & 3019 & *49.50 & 7200.00 \\ 
 & 3 & *66.67 & 749 & 7.69 & 3275 & *11.11 & 7200.00 \\ 
\hline 
\hline 
 \multirow{1}{*}{u-P200D10$\%\bar{D}$50} & 1 & 0.00 & 312 & 4.99 & 3546 & 0.00 & 556.97 \\ 
 & 2 & 50.00 & 522 & 8.13 & 11659 & 16.67 & 7200.00 \\ 
 & 3 & *67.48 & 202 & 9.37 & 9759 & *59.32 & 7200.00 \\ 
 & 4 & 75.00 & 2419 & 967.58 & 1145 & 26.00 & 7200.00 \\ 
\hline 
\end{tabular}
\end{table}

\begin{table}[!t]
\renewcommand{\arraystretch}{1.3}
\setlength{\tabcolsep}{1.3pt}
\centering
\caption{CPLEX results when solving weighted instances for the $PB$ formulation with $\sigma \in \{1,\ldots,|\bar{D}|-1\}$.} \label{tab:sensitivity_sigma_weighted} 
\begin{tabular}{ c | c || r r r || r | r r}
	\hline
	\multicolumn{1}{ c |}{ \multirow{2}{*}{\textbf{Instance}} } & \multicolumn{1}{ c ||}{ \multirow{2}{*}{$\mathbf{\sigma}$} } & \multicolumn{3}{ c ||}{\textbf{Root Node}} & \multicolumn{3}{ c }{\textbf{Branch-and-Cut}} \\
	\cline{3-8}
    & & $\mathbf{gap_0}$ & \textbf{cuts} & \textbf{sec.} & \textbf{Nodes} & $\mathbf{gap}$ & \textbf{sec.} \\
	\hline
 \multirow{1}{*}{w-P200D5$\%\bar{D}$70} & 1 & 3.36 & 93 & 1.25 & 300 & 0.00 & 16.87 \\ 
 & 2 & 50.71 & 831 & 9.81 & 1663 & 0.00 & 172.46 \\ 
\hline 
 \multirow{1}{*}{w-P200D7$\%\bar{D}$50} & 1 & 3.67 & 111 & 1.49 & 549 & 0.00 & 31.52 \\ 
 & 2 & 51.44 & 670 & 11.53 & 4013 & 0.00 & 665.60 \\ 
\hline 
 \multirow{1}{*}{w-P400D5$\%\bar{D}$70} & 1 & 0.86 & 190 & 3.48 & 414 & 0.00 & 79.99 \\ 
 & 2 & 50.34 & 480 & 2.74 & 1652 & 0.00 & 891.45 \\ 
\hline 
 \multirow{1}{*}{w-P400D7$\%\bar{D}$50} & 1 & 2.02 & 264 & 3.40 & 974 & 0.00 & 241.08 \\ 
 & 2 & 50.29 & 566 & 3.15 & 2819 & 0.00 & 1557.25 \\ 
\hline 
\hline 
 \multirow{1}{*}{w-P200D5$\%\bar{D}$90} & 1 & 2.96 & 84 & 2.30 & 3427 & 0.00 & 427.78 \\ 
 & 2 & 50.00 & 293 & 3.73 & 12045 & 32.56 & 7200.00 \\ 
 & 3 & *67.03 & 289 & 4.70 & 20671 & *1.09 & 7200.00 \\ 
\hline 
 \multirow{1}{*}{w-P200D7$\%\bar{D}$70} & 1 & 2.69 & 87 & 3.01 & 3736 & 0.00 & 514.63 \\ 
 & 2 & 50.00 & 428 & 3.89 & 6632 & 0.00 & 2647.32 \\ 
 & 3 & *66.82 & 382 & 3.03 & 25358 & *0.46 & 7200.00 \\ 
\hline 
 \multirow{1}{*}{w-P400D5$\%\bar{D}$90} & 1 & 1.26 & 442 & 6.14 & 3048 & 0.00 & 1532.36 \\ 
 & 2 & *50.01 & 381 & 4.95 & 3415 & *48.87 & 7200.00 \\ 
 & 3 & *66.86 & 430 & 10.48 & 2994 & *11.63 & 7200.00 \\ 
\hline 
 \multirow{1}{*}{w-P400D7$\%\bar{D}$70} & 1 & 0.68 & 628 & 6.68 & 9429 & 0.00 & 6031.09 \\ 
 & 2 & *50.00 & 387 & 7.68 & 2605 & *49.52 & 7200.00 \\ 
 & 3 & *66.78 & 650 & 6.11 & 2089 & *33.02 & 7200.00 \\  
\hline 
\end{tabular}
\end{table}

\begin{figure}[!tb]
    \centering   
	\includegraphics[width=3.2in]{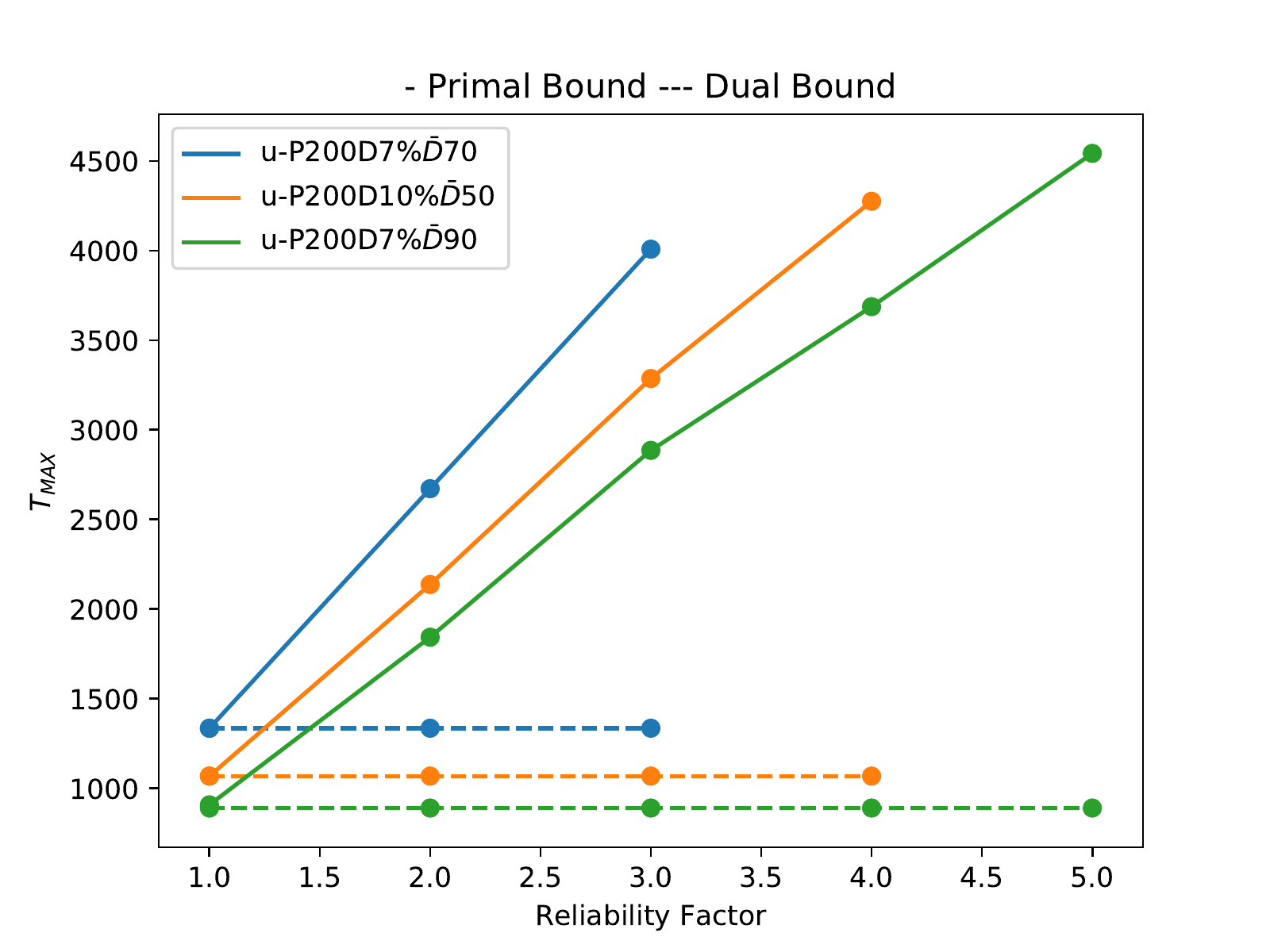}
	\caption{\emph{Primal bound} and \emph{dual bound} on increasing $\sigma$ for unweighted instances.}
	\label{fig:optimums_dualbounds_x_sigma_unweighted}
\end{figure}

\begin{figure}[!tb]
    \centering	    
	\includegraphics[width=3.2in]{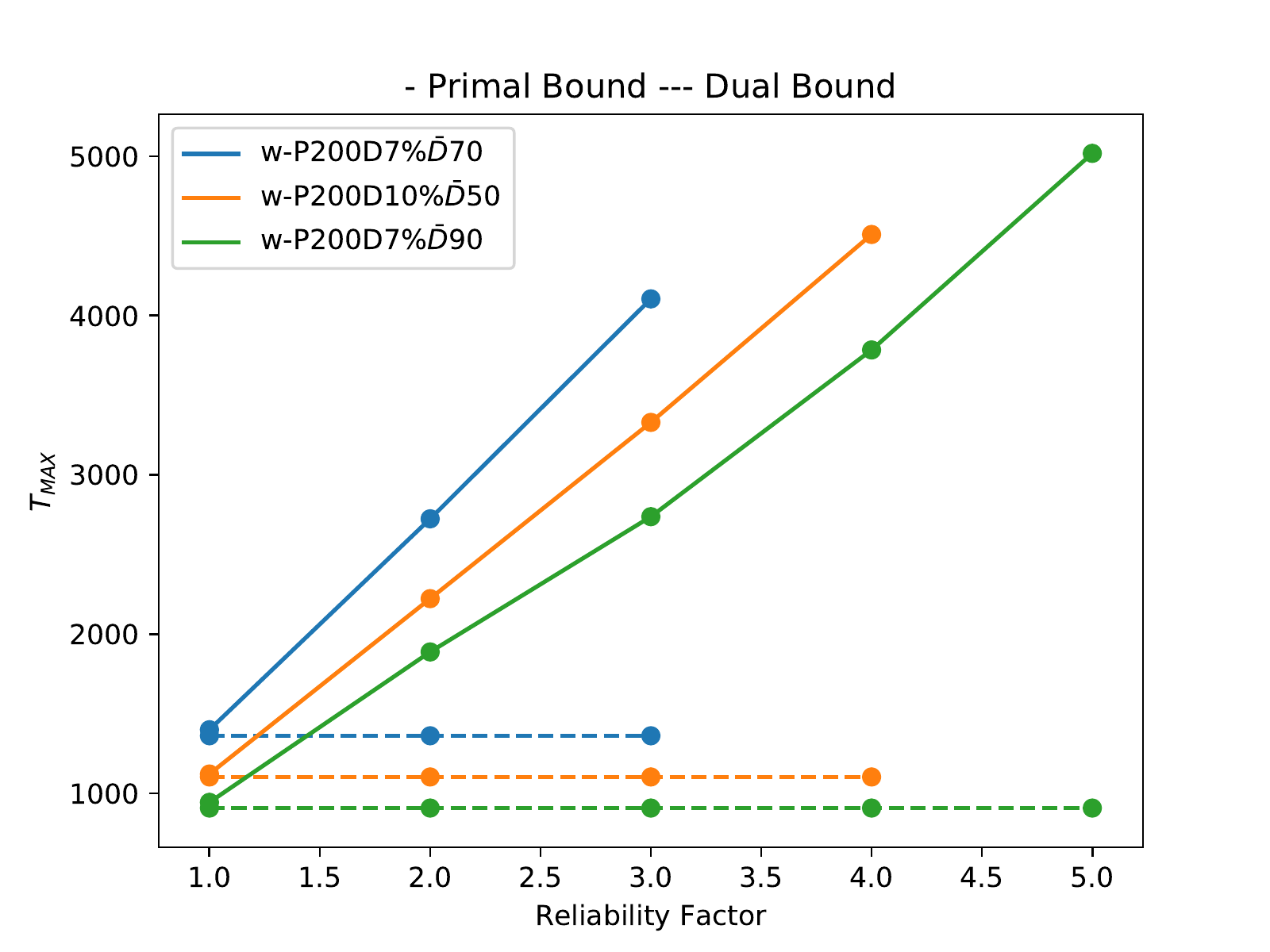}
	\caption{\emph{Primal bound} and \emph{dual bound} on increasing $\sigma$ for weighted instances.}
	\label{fig:optimums_dualbounds_x_sigma_weighted}
\end{figure}

Tables \ref{tab:sensitivity_sigma_unweighted} and \ref{tab:sensitivity_sigma_weighted} show that initial dual gaps are largely affected by $\sigma$. 
Those large dual gaps result from the increase of the optimum solution values with $\sigma$ not accompanied 
by the increase in the dual bounds obtained at the root node.
This is illustrated in Fig.~\ref{fig:optimums_dualbounds_x_sigma_unweighted} and Fig.~\ref{fig:optimums_dualbounds_x_sigma_weighted}, where the changes in the optimal objective function value (named “$T_{\max}$”) given the increase of $\sigma$ are presented for unweighted instances “u-P200D7$\%\bar{D}$70”, “u-P200D10$\%\bar{D}$50”, “u-P200D7$\%\bar{D}$90”, and for weighted instances “w-P200D7$\%\bar{D}$70”, “w-P200D10$\%\bar{D}$50”, “w-P200D7$\%\bar{D}$90”. 
Those instances were selected to include distinct values of $|\bar{D}|$.

\subsubsection{Maximum transmission time sensitivity} 
\label{subsec:sensitivity_hat_T}

\begin{figure}[!tb]
    \centering	      
	\includegraphics[width=3.2in]{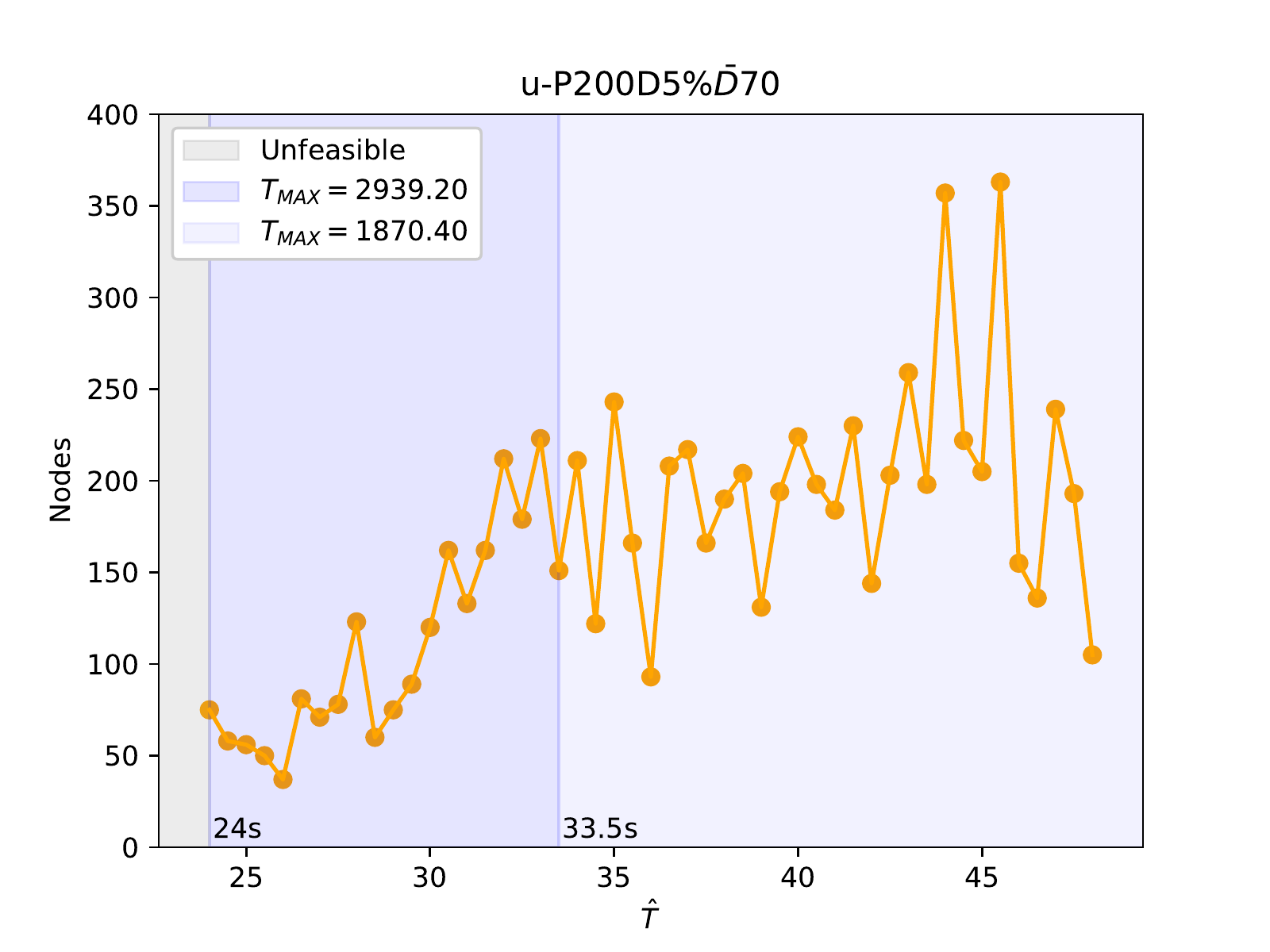}
	\caption{Number of nodes explored by CPLEX on varying $\hat{T}$ for the instance u-P200D5$\%\bar{D}$70.}
	\label{fig:sensitivity_t_hat_unweighted}
\end{figure}

\begin{figure}[!tb]
    \centering	        
	\includegraphics[width=3.2in]{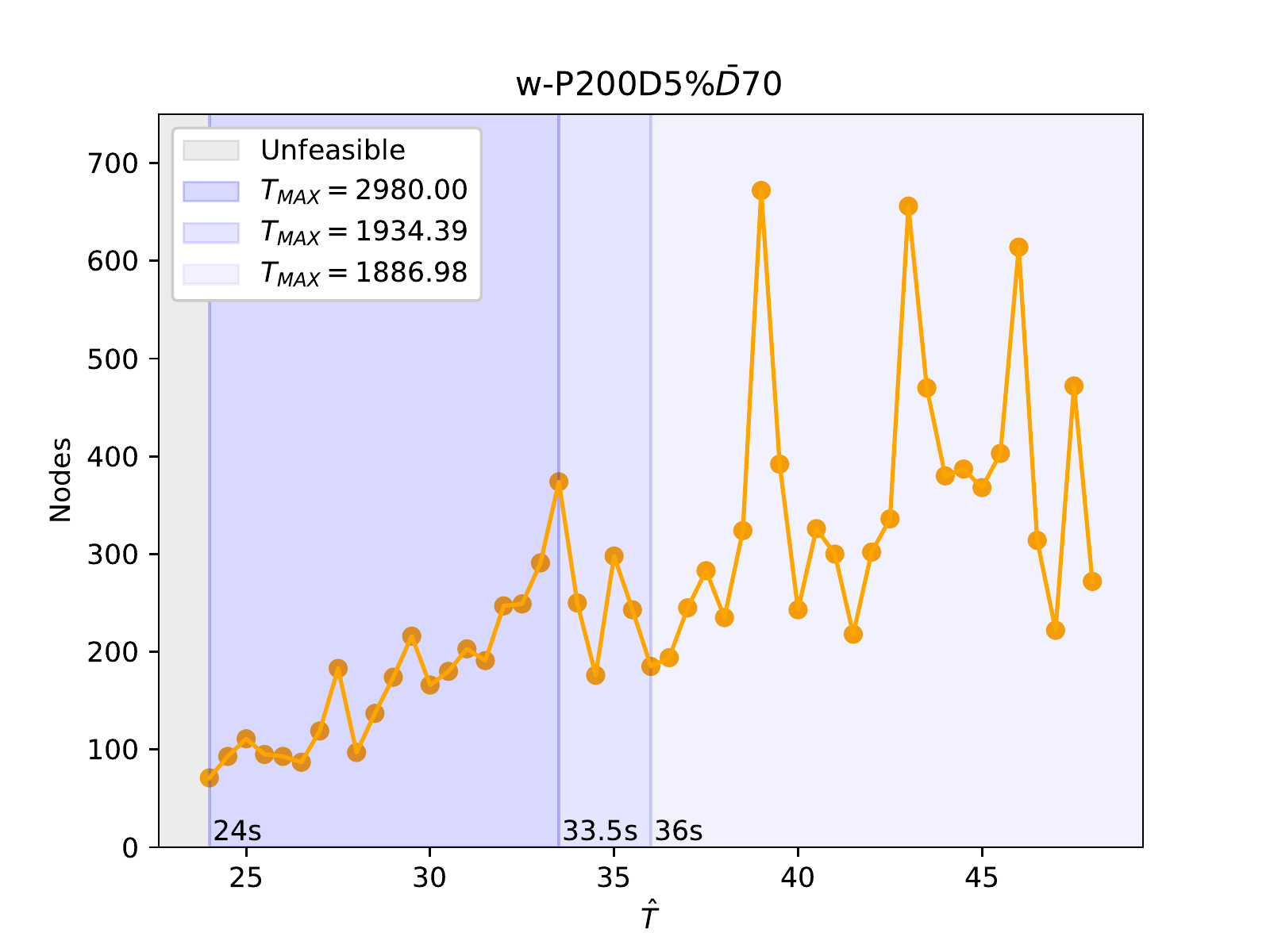}
	\caption{Number of nodes explored by CPLEX on varying $\hat{T}$ for the instance w-P200D5$\%\bar{D}$70.}
	\label{fig:sensitivity_t_hat_weighted}
\end{figure}

The sensitivity analysis of formulation “$PB$” to parameter $\hat{T}$ is performed by decreasing its values progressively (the value of $\sigma$ is fixed to 1 in this set of experiments). 
The first value of $\hat{T}$ tested corresponds to the allocated communication time between the drones when no time limit is imposed for their communication, i.e., $\hat{T} {=}{+}\infty$. 
From that value, $\hat{T}$ is decreased by 0.5 seconds until formulation “$PB$” becomes infeasible. 
Fig.~\ref{fig:sensitivity_t_hat_unweighted} and Fig.~\ref{fig:sensitivity_t_hat_weighted} present results for  instances “u-P200D5$\%\bar{D}$70” and “w-P200D5$\%\bar{D}$70”. 
For them, the first value of $\hat{T}$ is 48 seconds, being decreased down to 23.5 when both problems become infeasible.
The reported number of nodes explored by CPLEX branch-and-cut method and the optimum objective function value  are obtained within 2h of computing time. 

The decreasing of $\hat{T}$ tends to reduce the number of nodes explored in the branch-and-cut enumeration whereas the objective function value increases until the problem becomes infeasible.
For example, in Fig. \ref{fig:sensitivity_t_hat_unweighted}, the optimum objective function value increases from $1870.4$ to $2939.2$ starting at $\hat{T} = 33.5$. The problem becomes infeasible for $\hat{T}$ smaller than $24$s.

\section{Conclusion}
\label{sec:conclusions}
A swarm of drones (UAVs) can be used to automate a wide range of missions, from surveillance to search and rescue, from 3D mapping to telecommunication enhancement. While  UAVs are typically responsible for the mission phases related to data collection --- thanks to their flying capabilities and to the availability of embedded sensors --- most of the data processing is offloaded to dedicated machines (virtual or bare-metal) placed in the cloud. However, when the communication bandwidth between the swarm and the cloud is limited, an ad-hoc cloud established on top of the UAVs' computing resources (and those of other elements available in the area) can be leveraged to replace the cloud and keep data processing local.

For the purpose of optimizing the use of such ad-hoc cloud infrastructure powered by the swarming UAVs, we introduced a new optimization problem, namely the Covering-Assignment Problem for swarm-powered ad-hoc clouds - CAPsac, based on a real-life use-case in the emergency management field: swarm-powered distributed 3D reconstruction for humanitarian emergency response.
After having established  the relationship between the general problem and the specific use-case, we presented the NP-Hardness proof of the CAPsac and described two MILP formulations for it.

Given a set of geo-positioned aerial pictures (data) that are subject to geolocation/clustering constraints, CAPsac minimizes the 3D mapping (data-processing phase) completion time by jointly computing: (i) the optimal covering of photos (workload configuration), and (ii) the optimal assignment of photographed sub-regions (workload assignment) to UAVs (computing elements).
Besides being a way to provide optimal solutions for the problem, our integrated decision model contrasts with the \textit{decompose-then-allocate} and the \textit{allocate-then-decompose} paradigms usually seen in (both the cloud computing optimization and) the multi-robot task allocation literature.
Finally, modeling CAPsac in this way is flexible and amendable to take into account any other additional ground computing elements connected to the swarm itself.

In order to assess the proposed formulations, a series of computational experiments was conducted with a set of unweighted and weighted realistic benchmark instances available online (\url{https://github.com/ds4dm/CAPsac}). 
The experiments revealed that the photo-based formulation “$PB$” was more efficient by using ordering inequalities that remove from the feasible continuous search space those sub-regions whose boundaries are not regular (e.g. left boundary at the right of a right boundary). 
However, the different branching priority strategies and row generation methods have not proven to yield a performance gain while solving “$PB$”.
Column Generation was employed in the region-based formulation “$RB$”, but the presence of highly degenerate optimums led to long execution times.

Finally, the sensitivity analysis of the formulation “$PB$” showed that it becomes more difficult to solve as the reliability factor $\sigma$ increases. 
Tests with varying values for the maximum allowed transmission time $\hat{T}$ also presented a slight gain of performance as $\hat{T}$ approaches a limit when the problem becomes infeasible.

\appendix[Region-based CAPsac]
\label{appdx:set_covering_formulation}

The $CAPsac$ problem can be addressed by explicitly considering the set $\mathcal{S}$ of all feasible rectangular subsets of photos, such that each element of $\mathcal{S}$ corresponds to a possible rectangular sub-region to be 3D-reconstructed. 
It is important to remark that the cardinality of $\mathcal{S}$ is polynomial and bounded by $O(|C|^2 |L|^2)$, which is $O(|P|^4)$ in the worst scenario:
\begin{proposition}
\label{prop:polynomial_S}
\textit{Let $\mathcal{S}$ be the set of all feasible rectangular sub-regions to a CAPsac instance. 
Then, $|\mathcal{S}|$ is bounded by $O(|C|^2 |L|^2)$, which is $O(|P|^4)$ in the worst case.}
\end{proposition}
\begin{IEEEproof}
As in Section \ref{subsec:convex_constraints}, any feasible hyperectangle $S \in \mathcal{S}$ is defined by a tuple ($\alpha^S$, $\beta^S$, $\gamma^S$, $\omega^S$) of latitudes and longitudes corresponding to the left, right, bottom, and top borders of $S$, respectively, with $\alpha^S,\beta^S \in C$ and $\gamma^S, \omega^S \in L$, and such that $\alpha^S \leq \beta^S$ and $\gamma^S \leq \omega^S$.
Therefore, $\mathcal{S} = C \times C \times L \times L$.
Since $1 \leq |C| \leq |P|$ and $1 \leq |L| \leq |P|$, $|\mathcal{S}|$ is bounded by $O(|P|^4)$.
\end{IEEEproof}

In particular, the photos are commonly spread across the target region in a grid pattern to fulfill photo footprint overlapping constraints~\cite{pepe2018planning}. 
Consequently, $|C|$ and $|L|$ are usually far smaller than $|P|$, and hence, $|C|^2 \cdot |L|^2$ is in practice usually significantly smaller than $|P|^4$.

Let $\mathcal{S}_{p}$ be the collection of rectangular subsets $S \in \mathcal{S}$ that cover photo $p \in P$. 
For each set $S \in \mathcal{S}$, denote $t^S$ the photo processing time of $S$, and $\mu^{hd}_S$ the amount of data to transfer from drone $h \in D$ to the drone $d \in \bar{D}$ if $S$ is selected. 
Let $q^{S}_{d}$ be the binary variable equal to 1 if $S$ is allocated to drone $d \in \bar{D}$.
Finally, let us denote $o^{S}$ the auxiliary binary variable that is equal to 1 if $S$ is selected, and 0 otherwise.

The region-based formulation of the $CAPsac$ is expressed as follows:
\begin{align}
    \min\limits_{q, o} &\quad T_{\max} \label{eq:scf_obj_func} \\
    \text{s.t. } &T_{\max} \geq \sum_{S \in \mathcal{S}} t^S q^{S}_{d} && \forall d \in \bar{D} \label{eq:scf_1} && \\
    &\hat{T} \cdot \phi^{hd} \geq \sum_{S \in \mathcal{S}} \mu^{hd}_{S} q^{S}_{d} && \forall (h,d) \in D \times \bar{D} \label{eq:scf_2}&& \\
	&\sum_{d \in \bar{D}} q^{S}_{d} \geq \sigma o^{S} && \forall S \in \mathcal{S} \label{eq:scf_3}&& \\
	&\sum_{S \in \mathcal{S}_{p}} o^{S} \geq 1 && \forall p \in P \label{eq:scf_5}&& \\
	&\sum_{S \in \mathcal{S}} o^{S} = m \label{eq:scf_52} \\
	&z^{hd} \leq \sum_{S \in \mathcal{S}} \mu^{hd}_{S} q^{S}_{d} && \forall (h,d) \in  D \times \bar{D} \label{eq:scf_6}&& \\
	& \phi^{hd} \leq \bar{c}^{hd} z^{hd} && \forall (h,d) \in D \times \bar{D}  &&\\
   & \text{MMF constraints~\cite{amaldi2013network} }  \nonumber \\
	&o^{S}, q^{S}_{d} \in \{ 0, 1 \} && \forall S \in \mathcal{S}, \forall d \in \bar{D} \label{eq:scf_7} && \\
	&w^{hd}_{ij} \in \{0, 1\} && \forall (i,j) \in A, \forall (h,d) \in  D \times \bar{D} \label{eq:scf_8} && \\
	&\phi^{hd} \geq 0, z^{hd} \in \{0, 1\} && \forall (h,d) \in  D \times \bar{D} \label{eq:scf_9} && \\
	&u_{ij} \geq 0 && \forall (i,j) \in A. && \label{eq:scf_10}
\end{align}
The objective function (\ref{eq:scf_obj_func}) minimizes the makespan $T_{\max}$, which is computed by constraints (\ref{eq:scf_1}). 
Constraints (\ref{eq:scf_2}) limit the networking delay for each photo transmission traffic demand. 
Constraints (\ref{eq:scf_3}) impose that the selected subsets in $\mathcal{S}$ are assigned to $\sigma$ drones that can do the 3D reconstruction. 
The set of constraints (\ref{eq:scf_5}) ensures that each photo $p \in P$ is covered at least once, and constraint (\ref{eq:scf_52}) defines the number of selected subsets to $m$, i.e., the number of drones that can perform the 3D reconstruction. 
The transmission rates are defined by (\ref{eq:scf_6}) and the ones in~\cite{amaldi2013network}, following MMF rate allocation, as explained in Section \ref{subsec:pr_formulation}. 
Finally, domain constraints are given in (\ref{eq:scf_7})-(\ref{eq:scf_10}).

The cardinality of $\mathcal{S}$ is bounded by $O(|P|^4)$ (Proposition \ref{prop:polynomial_S}). 
Therefore, the number of constraints in the formulation is bounded by $O(|P|^4)$ due to the amount of constraints (\ref{eq:scf_3}). 
The number of variables is bounded by $O(|D|\cdot|P|^4)$ due to the number of variables $q^{S}_{d}$.

\section*{Acknowledgment}

The authors thank all support from Humanitas Solutions and the Canada Excellence Research Chair in Data Science for Real-Time Decision-Making. 
The authors also would like to thank the referees for relevant and constructive insights.

\ifCLASSOPTIONcaptionsoff
  \newpage
\fi

\bibliographystyle{IEEEtran}
\bibliography{bib}

\begin{thebibliography}{10}
\providecommand{\url}[1]{#1}
\csname url@samestyle\endcsname
\providecommand{\newblock}{\relax}
\providecommand{\bibinfo}[2]{#2}
\providecommand{\BIBentrySTDinterwordspacing}{\spaceskip=0pt\relax}
\providecommand{\BIBentryALTinterwordstretchfactor}{4}
\providecommand{\BIBentryALTinterwordspacing}{\spaceskip=\fontdimen2\font plus
\BIBentryALTinterwordstretchfactor\fontdimen3\font minus
  \fontdimen4\font\relax}
\providecommand{\BIBforeignlanguage}[2]{{%
\expandafter\ifx\csname l@#1\endcsname\relax
\typeout{** WARNING: IEEEtran.bst: No hyphenation pattern has been}%
\typeout{** loaded for the language `#1'. Using the pattern for}%
\typeout{** the default language instead.}%
\else
\language=\csname l@#1\endcsname
\fi
#2}}
\providecommand{\BIBdecl}{\relax}
\BIBdecl

\bibitem{tan2013research}
Y.~Tan and Z.-Y. Zheng, ``Research advance in swarm robotics,'' \emph{Defence
  Technology}, vol.~9, no.~1, pp. 18 -- 39, 2013.

\bibitem{nex2014uav}
F.~Nex and F.~Remondino, ``{UAV} for {3D} mapping applications: a review,''
  \emph{Applied Geomatics}, vol.~6, no.~1, pp. 1--15, 2014.

\bibitem{berni2009thermal}
J.~A.~J. {Berni}, P.~J. {Zarco-Tejada}, L.~{Suarez}, and E.~{Fereres},
  ``Thermal and narrowband multispectral remote sensing for vegetation
  monitoring from an unmanned aerial vehicle,'' \emph{IEEE Transactions on
  Geoscience and Remote Sensing}, vol.~47, no.~3, pp. 722--738, March 2009.

\bibitem{grenzdorffer2008photogrammetric}
G.~Grenzd{\"o}rffer, A.~Engel, and B.~Teichert, ``The photogrammetric potential
  of low-cost {UAV}s in forestry and agriculture,'' \emph{The International
  Archives of the Photogrammetry, Remote Sensing and Spatial Information
  Sciences}, vol. XXXVII, no.~B1, pp. 1207--1214, 2008.

\bibitem{martinez2006experimental}
J.~Mart{\'\i}nez-de Dios, L.~Merino, F.~Caballero, A.~Ollero, and D.~Viegas,
  ``Experimental results of automatic fire detection and monitoring with
  {UAV}s,'' \emph{Forest Ecology and Management}, vol. 234, no.~1, p. S232,
  2006.

\bibitem{chiabrando2011uav}
F.~Chiabrando, F.~Nex, D.~Piatti, and F.~Rinaudo, ``{UAV} and {RPV} systems for
  photogrammetric surveys in archaelogical areas: two tests in the {Piedmont
  region (Italy)},'' \emph{Journal of Archaeological Science}, vol.~38, no.~3,
  pp. 697--710, 2011.

\bibitem{lambers2007combining}
K.~Lambers, H.~Eisenbeiss, M.~Sauerbier, D.~Kupferschmidt, T.~Gaisecker,
  S.~Sotoodeh, and T.~Hanusch, ``Combining photogrammetry and laser scanning
  for the recording and modelling of the late intermediate period site of
  {Pinchango Alto, Palpa, Peru},'' \emph{Journal of Archaeological Science},
  vol.~34, no.~10, pp. 1702--1712, 2007.

\bibitem{oczipka2009small}
M.~Oczipka, J.~Bemmann, H.~Piezonka, J.~Munkabayar, B.~Ahrens, M.~Achtelik, and
  F.~Lehmann, ``Small drones for geo-archaeology in the steppes: locating and
  documenting the archaeological heritage of the {Orkhon Valley in Mongolia},''
  in \emph{Remote Sensing for Environmental Monitoring, GIS Applications, and
  Geology IX}, U.~Michel and D.~L. Civco, Eds., vol. 7478, International
  Society for Optics and Photonics.\hskip 1em plus 0.5em minus 0.4em\relax
  SPIE, 2009, pp. 53 -- 63.

\bibitem{rinaudo2012archaeological}
F.~Rinaudo, F.~Chiabrando, A.~M. Lingua, and A.~Span{\`o}, ``Archaeological
  site monitoring: {UAV} photogrammetry can be an answer,'' \emph{ISPRS -
  International Archives of the Photogrammetry, Remote Sensing and Spatial
  Information Sciences}, vol. XXXIX-B5, pp. 583--588, 2012.

\bibitem{verhoeven2009providing}
G.~J.~J. Verhoeven, ``Providing an archaeological bird's-eye view – an
  overall picture of ground-based means to execute low-altitude aerial
  photography ({LAAP}) in archaeology,'' \emph{Archaeological Prospection},
  vol.~16, no.~4, pp. 233--249, 2009.

\bibitem{hartmann2012determination}
W.~Hartmann, S.~Tilch, H.~Eisenbeiss, and K.~Schindler, ``Determination of the
  {UAV} position by automatic processing of thermal images,'' \emph{ISPRS -
  International Archives of the Photogrammetry, Remote Sensing and Spatial
  Information Sciences}, vol. XXXIX-B6, pp. 111--116, 2012.

\bibitem{manyoky2011unmanned}
M.~Manyoky, P.~Theiler, D.~Steudler, and H.~Eisenbeiss, ``Unmanned aerial
  vehicle in cadastral applications,'' \emph{ISPRS-international archives of
  the photogrammetry, remote sensing and spatial information sciences}, vol.
  XXXVIII-1/C22, pp. 57--62, 2011.

\bibitem{niethammer2010uav}
U.~Niethammer, S.~Rothmund, M.~James, J.~Travelletti, and M.~Joswig,
  ``{UAV}-based remote sensing of landslides,'' \emph{International Archives of
  Photogrammetry, Remote Sensing and Spatial Information Sciences}, vol.~38,
  no. Part 5, pp. 496--501, 2010.

\bibitem{smith2009volcano}
J.~G. Smith, J.~Dehn, R.~P. Hoblitt, R.~G. LaHusen, J.~B. Lowenstern, S.~C.
  Moran, L.~McClelland, K.~A. McGee, M.~Nathenson, P.~G. Okubo, J.~S.
  Pallister, M.~P. Poland, J.~A. Power, D.~J. Schneider, and T.~W. Sisson,
  ``Volcano monitoring,'' in \emph{Geological Monitoring}.\hskip 1em plus 0.5em
  minus 0.4em\relax Geological Society of America, 2009, pp. 273--305.

\bibitem{zhang2008uav}
C.~Zhang, ``An {UAV}-based photogrammetric mapping system for road condition
  assessment,'' \emph{ISPRS - International Archives of the Photogrammetry,
  Remote Sensing and Spatial Information Sciences}, vol. XXXVII-B5, pp.
  627--632, 2008.

\bibitem{chou2010disaster}
T.-Y. Chou, M.-L. Yeh, Y.~C. Chen, and Y.~H. Chen, ``Disaster monitoring and
  management by the unmanned aerial vehicle technology,'' in
  \emph{International Archives of Photogrammetry, Remote Sensing and Spatial
  Information Sciences}, vol.~38, no.~7B, 2010, p. 137–142.

\bibitem{haarbrink2006helicopter}
R.~Haarbrink and E.~Koers, ``Helicopter {UAV} for photogrammetry and rapid
  response,'' in \emph{2nd Int. Workshop “The Future of Remote Sensing”,
  ISPRS Inter-Commission Working Group I/V Autonomous Navigation},
  vol.~1.\hskip 1em plus 0.5em minus 0.4em\relax Citeseer, 2006.

\bibitem{molina2012searching}
P.~Molina, I.~Colomina, T.~Victoria, J.~Skaloud, W.~Kornus, R.~Prades, and
  C.~Aguilera, ``Searching lost people with {UAV}s: The system and results of
  the {CLOSE-SEARCH} project,'' in \emph{International Archives of the
  Photogrammetry, Remote Sensing and Spatial Information Sciences}, vol.~39,
  no. EPFL-CONF-182482, 2012, pp. 441--446.

\bibitem{zarco_tejada2014}
P.~Zarco-Tejada, R.~Diaz-Varela, V.~Angileri, and P.~Loudjani, ``Tree height
  quantification using very high resolution imagery acquired from an unmanned
  aerial vehicle ({UAV}) and automatic {3D} photo-reconstruction methods,''
  \emph{European Journal of Agronomy}, vol.~55, pp. 89 -- 99, 2014.

\bibitem{bendig2014}
J.~Bendig, A.~Bolten, S.~Bennertz, J.~Broscheit, S.~Eichfuss, and G.~Bareth,
  ``Estimating biomass of barley using crop surface models ({CSM}s) derived
  from {UAV}-based {RGB} imaging,'' \emph{Remote Sensing}, vol.~6, no.~11, pp.
  10\,395--10\,412, 2014.

\bibitem{diaz_varela2015}
R.~A. Díaz-Varela, R.~De~la Rosa, L.~León, and P.~J. Zarco-Tejada,
  ``High-resolution airborne {UAV} imagery to assess olive tree crown
  parameters using {3D} photo reconstruction: Application in breeding trials,''
  \emph{Remote Sensing}, vol.~7, no.~4, pp. 4213--4232, 2015.

\bibitem{alena2018optm_uav}
A.~Otto, N.~Agatz, J.~Campbell, B.~Golden, and E.~Pesch, ``Optimization
  approaches for civil applications of unmanned aerial vehicles ({UAV}s) or
  aerial drones: A survey,'' \emph{Networks}, vol.~0, no.~0, 2018.

\bibitem{coutinho2018}
\BIBentryALTinterwordspacing
W.~P. Coutinho, M.~Battarra, and J.~Fliege, ``The unmanned aerial vehicle
  routing and trajectory optimisation problem, a taxonomic review,''
  \emph{Computers \& Industrial Engineering}, vol. 120, pp. 116 -- 128, 2018.
  [Online]. Available:
  \url{http://www.sciencedirect.com/science/article/pii/S0360835218301815}
\BIBentrySTDinterwordspacing

\bibitem{Sujit2004search}
P.~B. Sujit and D.~Ghose, ``Search using multiple {UAV}s with flight time
  constraints,'' \emph{IEEE Transactions on Aerospace and Electronic Systems},
  vol.~40, no.~2, pp. 491--509, April 2004.

\bibitem{oh2014road_search}
H.~Oh, S.~Kim, A.~Tsourdos, and B.~A. White, ``Coordinated road-network search
  route planning by a team of {UAV}s,'' \emph{International Journal of Systems
  Science}, vol.~45, no.~5, pp. 825--840, 2014.

\bibitem{oh2015road_relay}
H.~Oh, H.-S. Shin, S.~Kim, A.~Tsourdos, and B.~A. White, \emph{Cooperative
  Mission and Path Planning for a Team of {UAV}s}.\hskip 1em plus 0.5em minus
  0.4em\relax Dordrecht: Springer Netherlands, 2015, pp. 1509--1545.

\bibitem{lanillos2014}
P.~Lanillos, S.~K. Gan, E.~Besada-Portas, G.~Pajares, and S.~Sukkarieh,
  ``Multi-{UAV} target search using decentralized gradient-based negotiation
  with expected observation,'' \emph{Information Sciences}, vol. 282, pp. 92 --
  110, 2014.

\bibitem{gan2011multi}
S.~K. Gan and S.~Sukkarieh, ``Multi-{UAV} target search using explicit
  decentralized gradient-based negotiation,'' in \emph{Robotics and Automation
  (ICRA), 2011 IEEE International Conference on}.\hskip 1em plus 0.5em minus
  0.4em\relax IEEE, 2011, pp. 751--756.

\bibitem{ji2015cooperative}
X.~Ji, X.~Wang, Y.~Niu, and L.~Shen, ``Cooperative search by multiple unmanned
  aerial vehicles in a nonconvex environment,'' \emph{Mathematical Problems in
  Engineering}, vol. 2015, pp. 1--19, 2015.

\bibitem{Tang2005MMMT}
Z.~Tang and U.~Ozguner, ``Motion planning for multitarget surveillance with
  mobile sensor agents,'' \emph{IEEE Transactions on Robotics}, vol.~21, no.~5,
  pp. 898--908, Oct 2005.

\bibitem{KARAMAN2008CG}
S.~Karaman and G.~Inalhan, ``Large-scale task/target assignment for {UAV}
  fleets using a distributed branch and price optimization scheme,'' \emph{IFAC
  Proceedings Volumes}, vol.~41, no.~2, pp. 13\,310 -- 13\,317, 2008, 17th IFAC
  World Congress.

\bibitem{niccolini2010multiple}
M.~Niccolini, M.~Innocenti, and L.~Pollini, ``Multiple {UAV} task assignment
  using descriptor functions,'' \emph{IFAC Proceedings Volumes}, vol.~43,
  no.~15, pp. 93--98, 2010.

\bibitem{viguria2010distributed}
A.~Viguria, I.~Maza, and A.~Ollero, ``Distributed service-based cooperation in
  aerial/ground robot teams applied to fire detection and extinguishing
  missions,'' \emph{Advanced Robotics}, vol.~24, no. 1-2, pp. 1--23, 2010.

\bibitem{choi2011genetic}
H.~Choi, Y.~Kim, and H.~Kim, ``Genetic algorithm based decentralized task
  assignment for multiple unmanned aerial vehicles in dynamic environments,''
  \emph{International Journal Aeronautical and Space Sciences}, vol.~12, no.~2,
  pp. 163--174, 2011.

\bibitem{barrientos2011aerial}
A.~Barrientos, J.~Colorado, J.~d. Cerro, A.~Martinez, C.~Rossi, D.~Sanz, and
  J.~Valente, ``Aerial remote sensing in agriculture: A practical approach to
  area coverage and path planning for fleets of mini aerial robots,''
  \emph{Journal of Field Robotics}, vol.~28, no.~5, pp. 667--689, 2011.

\bibitem{Moon2013}
S.~Moon, E.~Oh, and D.~H. Shim, ``An integral framework of task assignment and
  path planning for multiple unmanned aerial vehicles in dynamic
  environments,'' \emph{Journal of Intelligent {\&} Robotic Systems}, vol.~70,
  no.~1, pp. 303--313, Apr 2013.

\bibitem{moon2015}
S.~Moon, D.~H. Shim, and E.~Oh, \emph{Cooperative Task Assignment and Path
  Planning for Multiple {UAV}s}.\hskip 1em plus 0.5em minus 0.4em\relax
  Dordrecht: Springer Netherlands, 2015, pp. 1547--1576.

\bibitem{turpin2014capt}
M.~Turpin, N.~Michael, and V.~Kumar, ``Capt: Concurrent assignment and planning
  of trajectories for multiple robots,'' \emph{The International Journal of
  Robotics Research}, vol.~33, no.~1, pp. 98--112, 2014.

\bibitem{enright2015handbook}
J.~J. Enright, E.~Frazzoli, M.~Pavone, and K.~Savla, \emph{{UAV} Routing and
  Coordination in Stochastic, Dynamic Environments}.\hskip 1em plus 0.5em minus
  0.4em\relax Dordrecht: Springer Netherlands, 2015, pp. 2079--2109.

\bibitem{sadeghi2017heterogeneous}
A.~Sadeghi and S.~L. Smith, ``Heterogeneous task allocation and sequencing via
  decentralized large neighborhood search,'' \emph{Unmanned Systems}, vol.~5,
  no.~02, pp. 79--95, 2017.

\bibitem{ZORBAS201616}
D.~Zorbas, L.~D.~P. Pugliese, T.~Razafindralambo, and F.~Guerriero, ``Optimal
  drone placement and cost-efficient target coverage,'' \emph{Journal of
  Network and Computer Applications}, vol.~75, pp. 16 -- 31, 2016.

\bibitem{Ladosz2018}
P.~Ladosz, H.~Oh, and W.-H. Chen, ``Trajectory planning for communication relay
  unmanned aerial vehicles in urban dynamic environments,'' \emph{Journal of
  Intelligent {\&} Robotic Systems}, vol.~89, no.~1, pp. 7--25, Jan 2018.

\bibitem{CARABALLO2017725}
L.~Caraballo, J.~Díaz-Báñez, I.~Maza, and A.~Ollero, ``The
  block-information-sharing strategy for task allocation: A case study for
  structure assembly with aerial robots,'' \emph{European Journal of
  Operational Research}, vol. 260, no.~2, pp. 725 -- 738, 2017.

\bibitem{grancharova2015uavs}
A.~Grancharova, E.~I. Gr{\o}tli, D.-T. Ho, and T.~A. Johansen, ``{UAV}s
  trajectory planning by distributed {MPC} under radio communication path loss
  constraints,'' \emph{Journal of Intelligent \& Robotic Systems}, vol.~79,
  no.~1, pp. 115--134, 2015.

\bibitem{koulali2016green}
S.~Koulali, E.~Sabir, T.~Taleb, and M.~Azizi, ``A green strategic activity
  scheduling for {UAV} networks: A sub-modular game perspective,'' \emph{IEEE
  Communications Magazine}, vol.~54, no.~5, pp. 58--64, May 2016.

\bibitem{xu2017distributed}
S.~Xu, K.~Do{\u{g}}an{\c{c}}ay, and H.~Hmam, ``Distributed pseudolinear
  estimation and {UAV} path optimization for {3D} {AOA} target tracking,''
  \emph{Signal Processing}, vol. 133, pp. 64--78, 2017.

\bibitem{Gomarasca2009}
M.~A. Gomarasca, \emph{Elements of Photogrammetry}.\hskip 1em plus 0.5em minus
  0.4em\relax Dordrecht: Springer Netherlands, 2009, ch.~3, pp. 79--121.

\bibitem{Linder2016}
W.~Linder, \emph{Digital Photogrammetry}.\hskip 1em plus 0.5em minus
  0.4em\relax Springer Berlin Heidelberg, 2016.

\bibitem{doherty2016collaborative}
P.~Doherty, J.~Kvarnstr{\"o}m, P.~Rudol, M.~Wzorek, G.~Conte, C.~Berger,
  T.~Hinzmann, and T.~Stastny, ``A collaborative framework for {3D} mapping
  using unmanned aerial vehicles,'' in \emph{International Conference on
  Principles and Practice of Multi-Agent Systems}.\hskip 1em plus 0.5em minus
  0.4em\relax Springer, 2016, pp. 110--130.

\bibitem{loianno2016swarm}
G.~Loianno, Y.~Mulgaonkar, C.~Brunner, D.~Ahuja, A.~Ramanandan, M.~Chari,
  S.~Diaz, and V.~Kumar, ``A swarm of flying smartphones,'' in
  \emph{Intelligent Robots and Systems (IROS), 2016 IEEE/RSJ International
  Conference on}.\hskip 1em plus 0.5em minus 0.4em\relax IEEE, 2016, pp.
  1681--1688.

\bibitem{schmiemann2017distributed}
J.~Schmiemann, H.~Harms, J.~Schattenberg, M.~Becker, S.~Batzdorfer, and
  L.~Frerichs, ``A distributed online {3D}-lidar mappin system,'' \emph{ISPRS -
  International Archives of the Photogrammetry, Remote Sensing and Spatial
  Information Sciences}, vol. XLII-2/W6, pp. 339--346, 2017.

\bibitem{kobayashi2017method}
K.~Kobayashi, H.~Shishido, Y.~Kameda, and I.~Kitahara, ``Method to generate
  disaster-damage map using {3D} photometry and crowd sourcing,'' in \emph{Big
  Data (Big Data), 2017 IEEE International Conference on}.\hskip 1em plus 0.5em
  minus 0.4em\relax IEEE, 2017, pp. 4397--4399.

\bibitem{golodetz2018collaborative}
S.~Golodetz, T.~Cavallari, N.~A. Lord, V.~A. Prisacariu, D.~W. Murray, and
  P.~H. Torr, ``Collaborative large-scale dense {3D} reconstruction with online
  inter-agent pose optimisation,'' \emph{arXiv preprint arXiv:1801.08361},
  2018.

\bibitem{milani2016impact}
S.~Milani and A.~Memo, ``Impact of drone swarm formations in {3D} scene
  reconstruction,'' in \emph{2016 IEEE International Conference on Image
  Processing (ICIP)}.\hskip 1em plus 0.5em minus 0.4em\relax IEEE, Sep. 2016,
  pp. 2598--2602.

\bibitem{zanella2014}
A.~{Zanella}, N.~{Bui}, A.~{Castellani}, L.~{Vangelista}, and M.~{Zorzi},
  ``Internet of things for smart cities,'' \emph{IEEE Internet of Things
  Journal}, vol.~1, no.~1, pp. 22--32, 2014.

\bibitem{abbas2018}
N.~{Abbas}, Y.~{Zhang}, A.~{Taherkordi}, and T.~{Skeie}, ``Mobile edge
  computing: A survey,'' \emph{IEEE Internet of Things Journal}, vol.~5, no.~1,
  pp. 450--465, 2018.

\bibitem{chen2019}
\BIBentryALTinterwordspacing
W.~Chen, B.~Liu, H.~Huang, S.~Guo, and Z.~Zheng, ``{When {UAV} Swarm Meets
  Edge-Cloud Computing: The QoS Perspective},'' \emph{IEEE Network}, vol.~33,
  no.~2, pp. 36--43, mar 2019. [Online]. Available:
  \url{https://ieeexplore.ieee.org/document/8675170/}
\BIBentrySTDinterwordspacing

\bibitem{yu2020}
Z.~{Yu}, Y.~{Gong}, S.~{Gong}, and Y.~{Guo}, ``Joint task offloading and
  resource allocation in {UAV}-enabled mobile edge computing,'' \emph{IEEE
  Internet of Things Journal}, vol.~7, no.~4, pp. 3147--3159, 2020.

\bibitem{asheralieva2019}
A.~{Asheralieva} and D.~{Niyato}, ``Hierarchical game-theoretic and
  reinforcement learning framework for computational offloading in
  {UAV}-enabled mobile edge computing networks with multiple service
  providers,'' \emph{IEEE Internet of Things Journal}, vol.~6, no.~5, pp.
  8753--8769, 2019.

\bibitem{mei2020}
H.~{Mei}, K.~{Yang}, Q.~{Liu}, and K.~{Wang}, ``Joint trajectory-resource
  optimization in {UAV}-enabled edge-cloud system with virtualized mobile
  clone,'' \emph{IEEE Internet of Things Journal}, vol.~7, no.~7, pp.
  5906--5921, 2020.

\bibitem{wei2019}
X.~{Wei}, C.~{Tang}, J.~{Fan}, and S.~{Subramaniam}, ``Joint optimization of
  energy consumption and delay in cloud-to-thing continuum,'' \emph{IEEE
  Internet of Things Journal}, vol.~6, no.~2, pp. 2325--2337, 2019.

\bibitem{wan2020}
S.~{Wan}, J.~{Lu}, P.~{Fan}, and K.~B. {Letaief}, ``Toward big data processing
  in {IoT}: Path planning and resource management of {UAV} base stations in
  mobile-edge computing system,'' \emph{IEEE Internet of Things Journal},
  vol.~7, no.~7, pp. 5995--6009, 2020.

\bibitem{hamdaqa2015}
\BIBentryALTinterwordspacing
M.~Hamdaqa, M.~M. Sabri, A.~Singh, and L.~Tahvildari, ``{Adoop: MapReduce for
  Ad-hoc Cloud Computing},'' \emph{Proceedings of the 25th Annual International
  Conference on Computer Science and Software Engineering}, pp. 26--34, 2015.
  [Online]. Available: \url{http://dl.acm.org/citation.cfm?id=2886444.2886449}
\BIBentrySTDinterwordspacing

\bibitem{malhotra2014}
\BIBentryALTinterwordspacing
A.~Malhotra, S.~K. Dhurandher, and B.~Kumar, ``{Resource allocation in
  multi-hop Mobile Ad hoc cloud},'' in \emph{2014 Recent Advances in
  Engineering and Computational Sciences ({RAECS})}.\hskip 1em plus 0.5em minus
  0.4em\relax IEEE, mar 2014, pp. 1--6. [Online]. Available:
  \url{http://ieeexplore.ieee.org/lpdocs/epic03/wrapper.htm?arnumber=6799578}
\BIBentrySTDinterwordspacing

\bibitem{yaqoob2016}
\BIBentryALTinterwordspacing
I.~Yaqoob, E.~Ahmed, A.~Gani, S.~Mokhtar, M.~Imran, and S.~Guizani, ``{Mobile
  ad hoc cloud: A survey},'' \emph{Wireless Communications and Mobile
  Computing}, pp. 421--430, 2016. [Online]. Available:
  \url{http://doi.wiley.com/10.1002/wcm.2709}
\BIBentrySTDinterwordspacing

\bibitem{hu2019}
Q.~{Hu}, Y.~{Cai}, G.~{Yu}, Z.~{Qin}, M.~{Zhao}, and G.~Y. {Li}, ``Joint
  offloading and trajectory design for {UAV}-enabled mobile edge computing
  systems,'' \emph{IEEE Internet of Things Journal}, vol.~6, no.~2, pp.
  1879--1892, 2019.

\bibitem{zhang2019}
J.~{Zhang}, L.~{Zhou}, Q.~{Tang}, E.~C.-H. {Ngai}, X.~{Hu}, H.~{Zhao}, and
  J.~{Wei}, ``Stochastic computation offloading and trajectory scheduling for
  {UAV}-assisted mobile edge computing,'' \emph{IEEE Internet of Things
  Journal}, vol.~6, no.~2, pp. 3688--3699, 2019.

\bibitem{tan_qu2020}
Z.~{Tan}, H.~{Qu}, J.~{Zhao}, S.~{Zhou}, and W.~{Wang}, ``{UAV}-aided edge/fog
  computing in smart {IoT} community for social augmented reality,'' \emph{IEEE
  Internet of Things Journal}, vol.~7, no.~6, pp. 4872--4884, 2020.

\bibitem{yu_wang2019}
T.~{Yu}, X.~{Wang}, and A.~{Shami}, ``{UAV}-enabled spatial data sampling in
  large-scale {IoT} systems using denoising autoencoder neural network,''
  \emph{IEEE Internet of Things Journal}, vol.~6, no.~2, pp. 1856--1865, 2019.

\bibitem{Humanitas}
\BIBentryALTinterwordspacing
CBC/Radio-Canada. (2020, April) Émission découverte: Drones humanitaires.
  [Online]. Available:
  \url{https://ici.radio-canada.ca/tele/decouverte/site/segments/reportage/139538/drones-humanitaires}
\BIBentrySTDinterwordspacing

\bibitem{RaspberryPi}
W.~Gay, \emph{Raspberry Pi Hardware Reference}, 1st~ed.\hskip 1em plus 0.5em
  minus 0.4em\relax USA: Apress, 2014.

\bibitem{luca2020}
L.~G. Gianoli, April 2020, private communication.

\bibitem{Khamis2015}
A.~Khamis, A.~Hussein, and A.~Elmogy, ``Multi-robot task allocation: A review
  of the state-of-the-art,'' in \emph{Cooperative Robots and Sensor Networks
  2015}.\hskip 1em plus 0.5em minus 0.4em\relax Springer, 2015, pp. 31--51.

\bibitem{huang2020}
P.~{Huang}, Y.~{Wang}, K.~{Wang}, and K.~{Yang}, ``Differential evolution with
  a variable population size for deployment optimization in a {UAV}-assisted
  {IoT} data collection system,'' \emph{IEEE Transactions on Emerging Topics in
  Computational Intelligence}, vol.~4, no.~3, pp. 324--335, 2020.

\bibitem{wang2020}
Y.~{Wang}, Z.~Y. {Ru}, K.~{Wang}, and P.~Q. {Huang}, ``Joint deployment and
  task scheduling optimization for large-scale mobile users in
  multi-{UAV}-enabled mobile edge computing,'' \emph{IEEE Transactions on
  Cybernetics}, vol.~50, no.~9, pp. 3984--3997, 2020.

\bibitem{Bekmezci2013}
I.~Bekmezci, O.~K. Sahingoz, and S.~Temel, ``{Flying Ad-Hoc Networks (FANETs):
  A survey},'' \emph{Ad Hoc Networks}, vol.~11, no.~3, pp. 1254--1270, may
  2013.

\bibitem{massoulie1999bandwidth}
L.~Massouli{\'e} and J.~Roberts, ``Bandwidth sharing: Objectives and
  algorithms,'' in \emph{INFOCOM'99. Eighteenth Annual Joint Conference of the
  IEEE Computer and Communications Societies. Proceedings. IEEE}, vol.~3.\hskip
  1em plus 0.5em minus 0.4em\relax IEEE, 1999, pp. 1395--1403.

\bibitem{bertsekas1992data}
D.~P. Bertsekas, R.~G. Gallager, and P.~Humblet, \emph{Data networks}.\hskip
  1em plus 0.5em minus 0.4em\relax Prentice-Hall International New Jersey,
  1992, vol.~2.

\bibitem{raniwala2007}
A.~Raniwala, P.~De, S.~Sharma, R.~Krishnan, and T.~Chiueh, ``End-to-end flow
  fairness over {IEEE 802.11}-based wireless mesh networks,'' in \emph{IEEE
  INFOCOM 2007-26th IEEE International Conference on Computer
  Communications}.\hskip 1em plus 0.5em minus 0.4em\relax IEEE, 2007, pp.
  2361--2365.

\bibitem{coniglio2020}
S.~Coniglio, L.~G. Gianoli, E.~Amaldi, and A.~Capone, ``Elastic traffic
  engineering subject to a fair bandwidth allocation via bilevel programming,''
  \emph{IEEE/ACM Transactions on Networking}, 2020.

\bibitem{bertsimas2003}
\BIBentryALTinterwordspacing
D.~Bertsimas and M.~Sim, ``{Robust discrete optimization and network flows},''
  \emph{Mathematical Programming}, vol.~98, no. 1-3, pp. 49--71, sep 2003.
  [Online]. Available:
  \url{http://www.springerlink.com/openurl.asp?genre=article{\&}id=doi:10.1007/s10107-003-0396-4}
\BIBentrySTDinterwordspacing

\bibitem{delage2018}
E.~Delage, L.~G. Gianoli, and B.~Sans{\`o}, ``A practicable robust counterpart
  formulation for decomposable functions: A network congestion case study,''
  \emph{Operations Research}, vol.~66, no.~2, pp. 535--567, 2018.

\bibitem{mccormick1976}
G.~P. McCormick, ``Computability of global solutions to factorable nonconvex
  programs: Part i—convex underestimating problems,'' \emph{Mathematical
  programming}, vol.~10, no.~1, pp. 147--175, 1976.

\bibitem{tang2020}
\BIBentryALTinterwordspacing
P.~Tang, S.~Vick, J.~Chen, and S.~{German Paal}, ``Chapter 2 - surveying,
  geomatics, and {3D} reconstruction,'' in \emph{Infrastructure Computer
  Vision}, I.~Brilakis and C.~Haas, Eds.\hskip 1em plus 0.5em minus 0.4em\relax
  Butterworth-Heinemann, 2020, pp. 13 -- 64. [Online]. Available:
  \url{http://www.sciencedirect.com/science/article/pii/B9780128155035000024}
\BIBentrySTDinterwordspacing

\bibitem{pepe2018planning}
M.~Pepe, L.~Fregonese, and M.~Scaioni, ``Planning airborne photogrammetry and
  remote-sensing missions with modern platforms and sensors,'' \emph{European
  Journal of Remote Sensing}, vol.~51, no.~1, pp. 412--436, 2018.

\bibitem{nace2008max}
D.~Nace and M.~Pi{\'o}ro, ``Max-min fairness and its applications to routing
  and load-balancing in communication networks: a tutorial,'' \emph{IEEE
  Communications Surveys \& Tutorials}, vol.~10, no.~4, 2008.

\bibitem{amaldi2013network}
E.~Amaldi, A.~Capone, S.~Coniglio, and L.~G. Gianoli, ``Network optimization
  problems subject to max-min fair flow allocation,'' \emph{IEEE Communications
  Letters}, vol.~17, no.~7, pp. 1463--1466, 2013.

\bibitem{fowler1981boxcover}
R.~J. Fowler, M.~S. Paterson, and S.~L. Tanimoto, ``Optimal packing and
  covering in the plane are {NP}-complete,'' \emph{Information Processing
  Letters}, vol.~12, no.~3, pp. 133 -- 137, 1981.

\bibitem{johnson1982np}
D.~S. Johnson, ``The {NP}-completeness column: An ongoing guide,''
  \emph{Journal of Algorithms}, vol.~3, no.~2, pp. 182 -- 195, 1982.

\bibitem{column2005}
\BIBentryALTinterwordspacing
G.~Desaulniers, J.~Desrosiers, and M.~M. Solomon, Eds., \emph{Column
  Generation}.\hskip 1em plus 0.5em minus 0.4em\relax Springer {US}, 2005.
  [Online]. Available: \url{https://doi.org/10.1007/b135457}
\BIBentrySTDinterwordspacing

\bibitem{lodi2010mixed}
A.~Lodi, ``Mixed integer programming computation,'' in \emph{50 Years of
  Integer Programming 1958-2008}.\hskip 1em plus 0.5em minus 0.4em\relax
  Springer, Berlin, Heidelberg, 2010, pp. 619--645.

\bibitem{dolan2002}
\BIBentryALTinterwordspacing
E.~D. Dolan and J.~J. Mor{\'{e}}, ``Benchmarking optimization software with
  performance profiles,'' \emph{Mathematical Programming}, vol.~91, no.~2, pp.
  201--213, Jan. 2002. [Online]. Available:
  \url{https://doi.org/10.1007/s101070100263}
\BIBentrySTDinterwordspacing

\bibitem{jain1990art}
R.~Jain, \emph{The art of computer systems performance analysis: techniques for
  experimental design, measurement, simulation, and modeling}.\hskip 1em plus
  0.5em minus 0.4em\relax New York: Wiley, 1991.

\end{thebibliography}

\newpage
\begin{IEEEbiography}[{\includegraphics[width=1in,height=1.25in,clip,keepaspectratio]{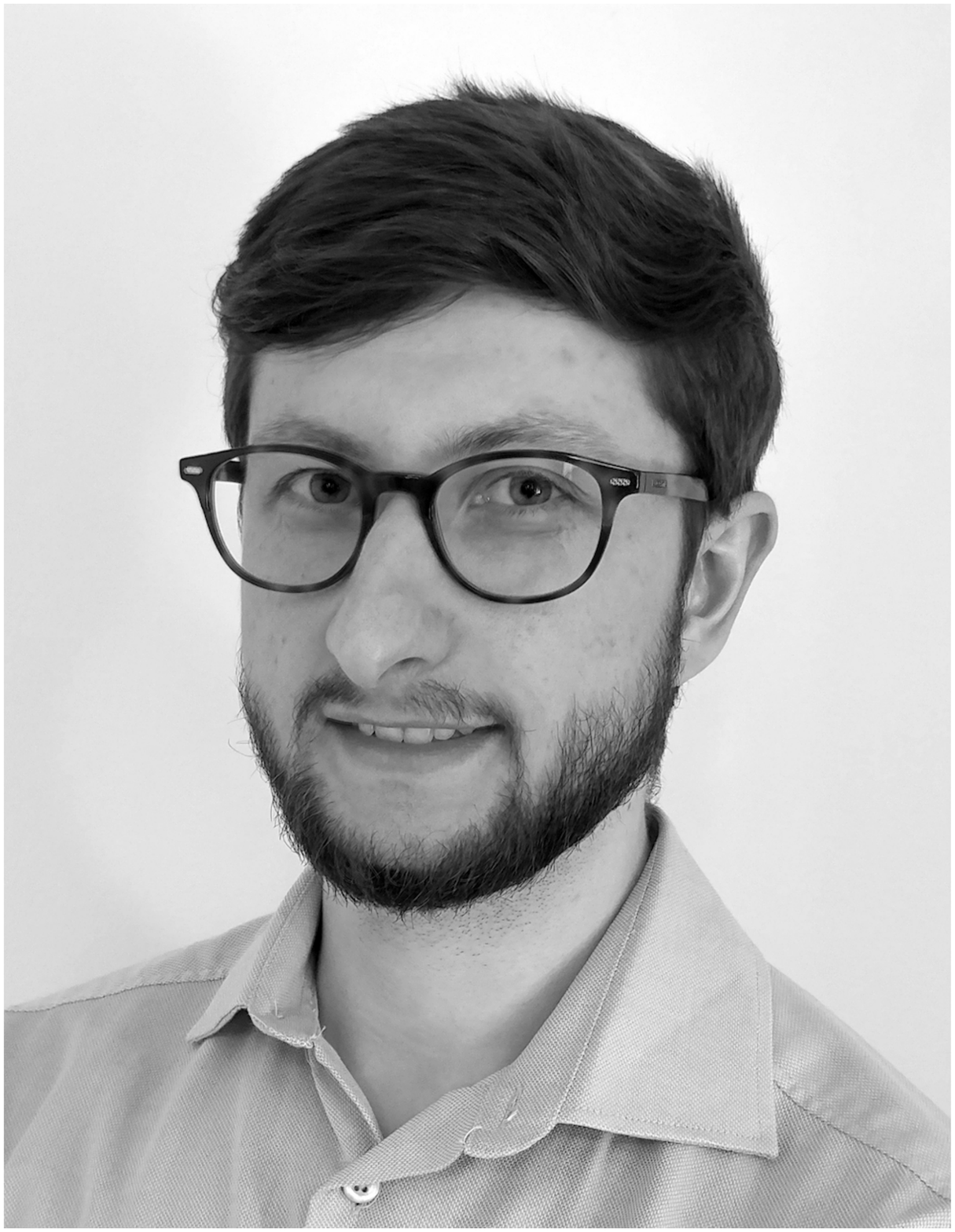}}]{Leandro R. Costa}
received the B.S. and M.S. degrees in computer engineering from the Universidade Federal do Rio Grande do Norte, Natal, Brazil, in 2014 and 2016. respectively.
He is currently pursuing a Ph.D degree in Applied Mathematics with the Canada Excellence Research Chair in Data Science for Real-Time Decision-Making and the GERAD (Group for Research in Decision Analysis), Department of Mathematical and Industrial Engineering, Polytechnique Montréal, Montreal, Canada.
His research interests include relevant real-life applications exploiting combinatorial optimization, mathematical programming, and data science.
\end{IEEEbiography}

\vspace*{-2\baselineskip}
\begin{IEEEbiography}[{\includegraphics[width=1in,height=1.25in,clip,keepaspectratio]{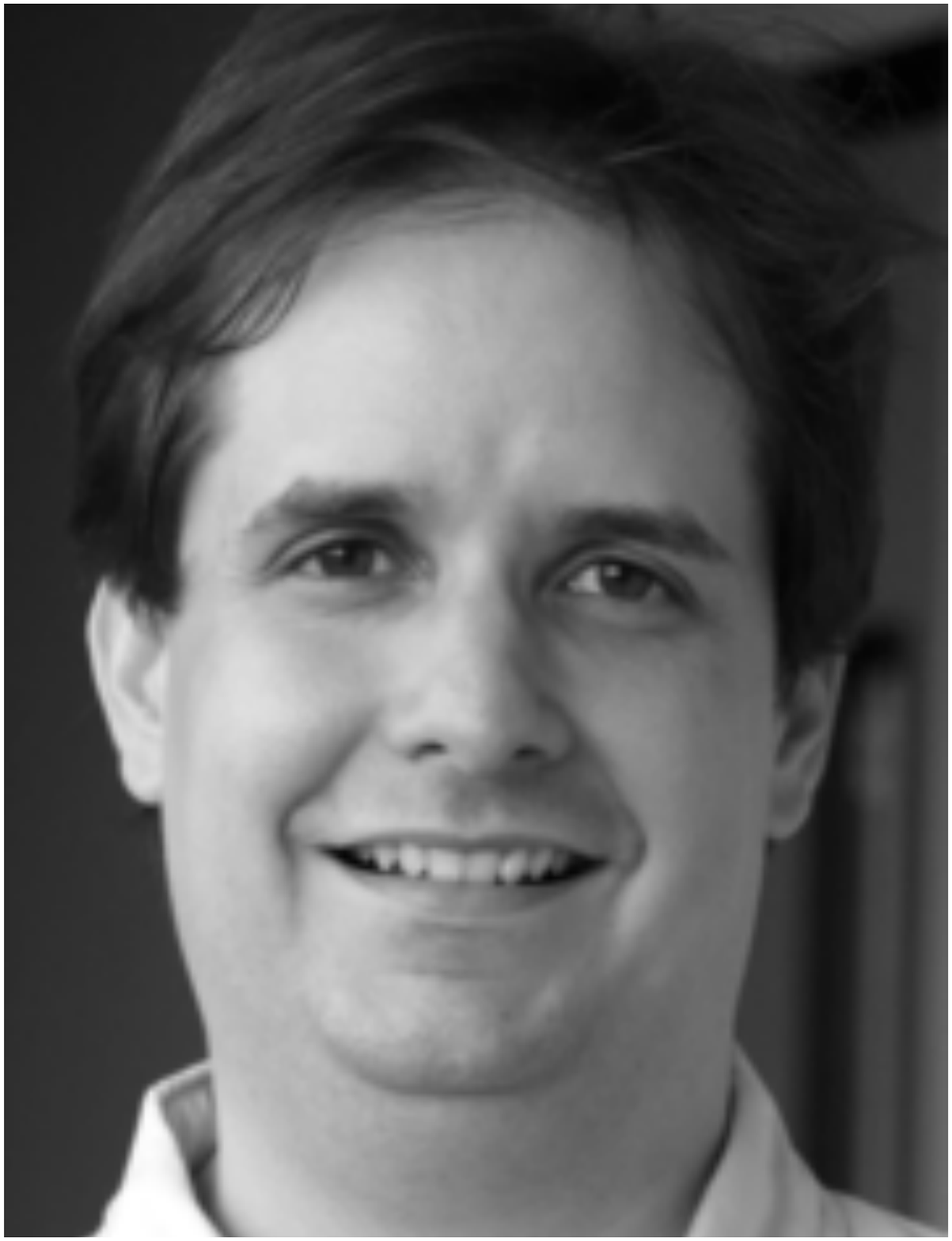}}]{Daniel Aloise}
is associate professor of the Computer and Software Engineering Department at Polytechnique Montréal. 
He obtained his PhD in Applied Mathematics from Polytechnique Montréal in 2009. 
He is member of the GERAD (Group for Research in Decision Analysis) and fellow of the Canada Excellence Research Chair in Data Science for Real-Time Decision-Making. 
His research interests include data mining, optimization, mathematical programming and how these disciplines interact to tackle problems in the Big Data era. 
He has published papers in leading data mining and operations research journals during his career.
\end{IEEEbiography}

\vspace*{-2\baselineskip}
\begin{IEEEbiography}[{\includegraphics[width=1in,height=1.25in,clip,keepaspectratio]{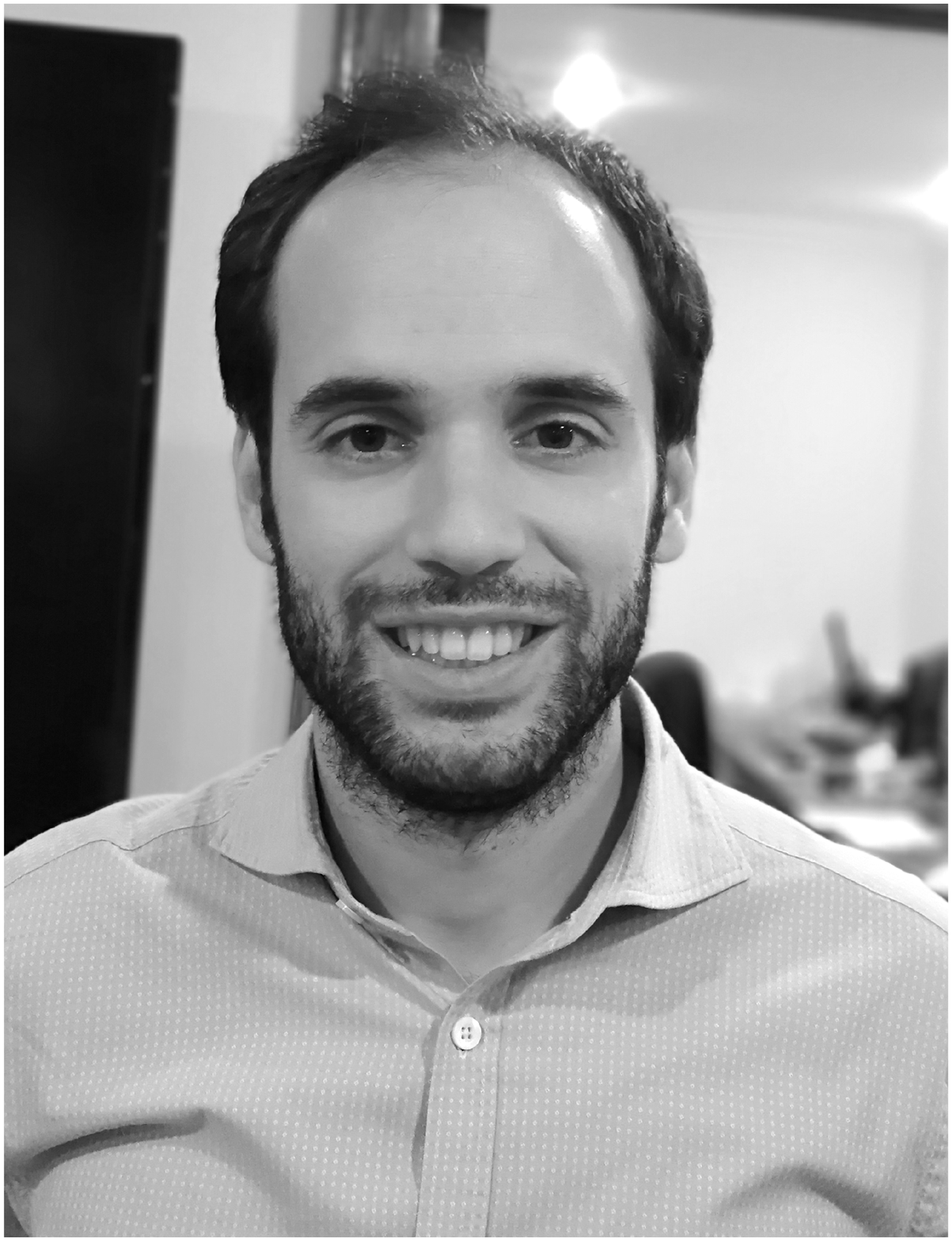}}]{Luca G. Gianoli}
obtained a Ph.D. in electrical engineering at both Politecnico di Milano and Polytechnique Montreal, Canada, in 2014. 
After two years as a postdoc at Polytechnique, where he conducted research on cloud optimization and communication for swarm robotics, he is now CTO at Humanitas Solutions, Montreal.
His research interests include network optimization, ad-hoc networking, cloud/edge computing and collaborative robotics.
\end{IEEEbiography}

\vspace*{-2\baselineskip}
\begin{IEEEbiography}[{\includegraphics[width=1in,height=1.25in,clip,keepaspectratio]{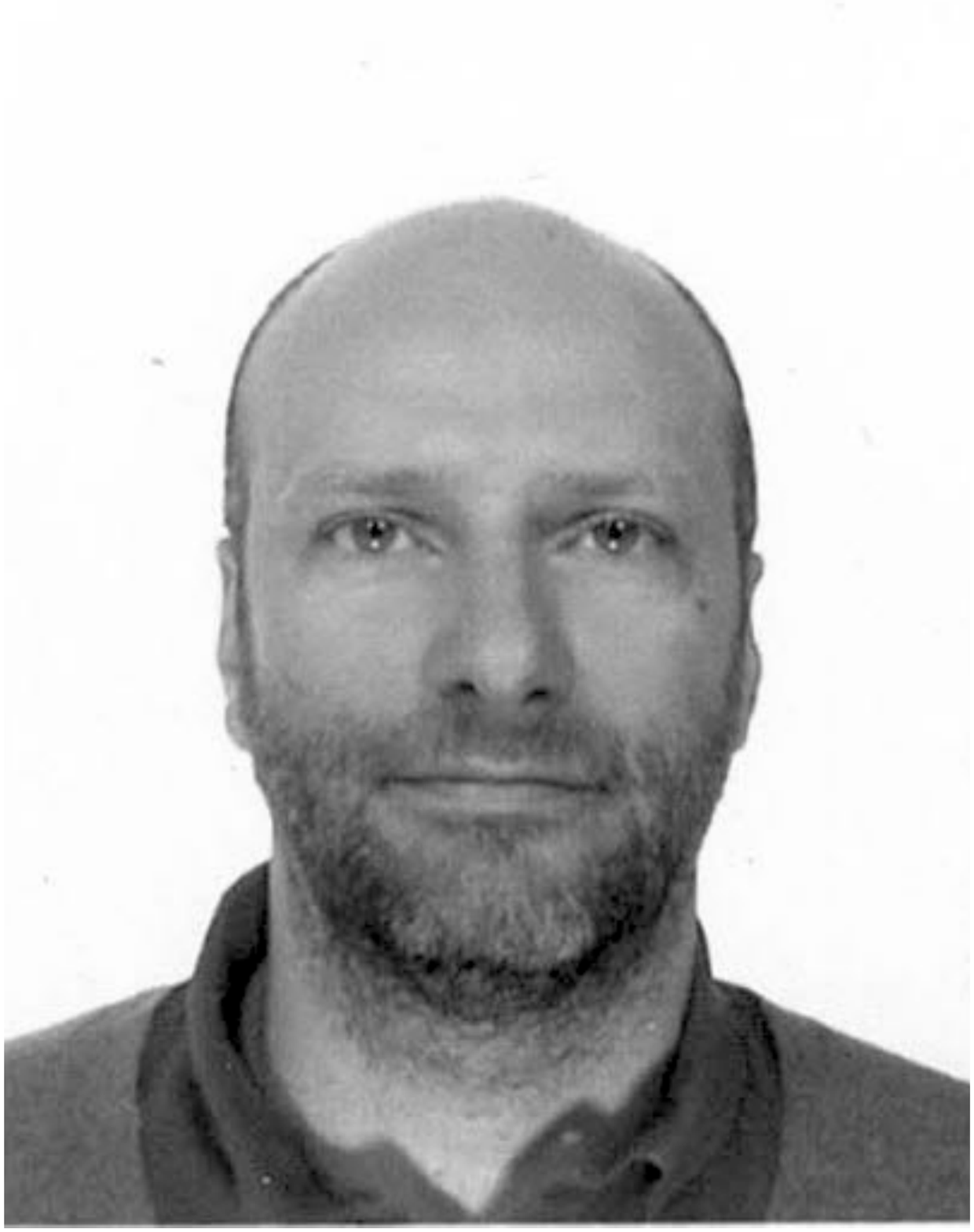}}]{Andrea Lodi}
is a full professor in the Department of Mathematical and Industrial Engineering at Polytechnique Montréal and holds the Canada Excellence Research Chair in Data Science for Real-Time Decision-Making. 
His research interests are mixed-integer linear and nonlinear programming and data science. 
He received several awards including the IBM Herman Goldstine Fellowship, the Google Faculty Research award and the IBM Faculty Award.
\end{IEEEbiography}
\vfill

\end{document}